\documentclass[fleqn,usenatbib]{mnras_mod}
\RequirePackage{rotating}

\usepackage[T1]{fontenc}
\usepackage{ae,aecompl}

\usepackage{graphicx}	%
\usepackage{amsmath}	%
\usepackage{amssymb}	%

\usepackage[flushleft]{threeparttable}
\usepackage{float}
\usepackage{datetime}
\usepackage{textpos}
\usepackage{booktabs}
\usepackage{rotating}
\usepackage{adjustbox}
\usepackage{epsfig}
\usepackage{textcomp}
\usepackage{color}
\usepackage{bm}

\usepackage[all]{hypcap}

\usepackage{breqn}

\providecommand{\adsurl}[1]{\href{#1}{ADS}}

\providecommand{\url}[1]{\href{#1}{#1}}
\usepackage{url}

\usepackage{placeins}

\def \aj{Astronom.~J.}

\def \aap{A\&A }
\def \apjl{ApJ}
\def \apjs{ApJS}
\def \apj{ApJ}
\def \jcap{JCAP}

\def \mnras{MNRAS}

\def\alt{\raise0.3ex\hbox{$\;<$\kern-0.75em\raise-1.1ex\hbox{$\sim\;$}}}
\def\agt{\raise0.3ex\hbox{$\;>$\kern-0.75em\raise-1.1ex\hbox{$\sim\;$}}}

\newcommand{\bw}{\begin{widetext}}
	\newcommand{\ew}{\end{widetext}}

\newcommand{\lsim}{\,\rlap{\raise 0.35ex\hbox{$<$}}{\lower 0.7ex\hbox{$\sim$}}\,}
\newcommand{\gsim}{\,\rlap{\raise 0.35ex\hbox{$>$}}{\lower 0.7ex\hbox{$\sim$}}\,}

\def\lesssim{\mathrel{\hbox{\rlap{\hbox{\lower3pt\hbox{$\sim$}}}\hbox{\raise2pt\hbox{$<$}}}}}
\def\gtrsim{\mathrel{\hbox{\rlap{\hbox{\lower3pt\hbox{$\sim$}}}\hbox{\raise2pt\hbox{$>$}}}}}

\def\xlinkspace#1 #2{%
	\ifx\relax#2%
	\xlinkdash#1-\relax
	\else
	\xlinkdash#1 -\relax
	\expandafter\xlinkspace\expandafter#2%
	\fi}

\def\xlinkdash#1-#2{%
	\ifx\relax#2%
	\tmp{#1}%
	\else
	\tmp{#1-}%
	\expandafter\xlinkdash\expandafter#2%
	\fi}

\newcommand{\newtext}[1]{\textcolor{black}{#1}}

\usepackage{natbib}

\title[Maximum Entropy Estimation of the Galactic Bulge Morphology]{Maximum Entropy Estimation of the Galactic Bulge Morphology via the VVV Red Clump}

\author[Coleman et al.]{
		\mbox{
		B.~Coleman$^1$\thanks{E-mail: bjc174@uclive.ac.nz},
		D.~Paterson$^1$\thanks{E-mail: dnp16@uclive.ac.nz},
		  C.~Gordon$^1$\thanks{E-mail: chris.gordon@canterbury.ac.nz}, O. Macias$^{2,3}$\thanks{oscar.macias@ipmu.jp} and	H.~Ploeg$^1$\thanks{hzp10@uclive.ac.nz}} 
\\
1) School of Physical and Chemical Sciences, University of Canterbury, Christchurch, New Zealand\\
2) Kavli Institute for the Physics and Mathematics of the
Universe (WPI), University of Tokyo, Kashiwa, Chiba 277-8583, Japan\\
3) GRAPPA Institute, University of Amsterdam, 1098 XH Amsterdam, The Netherlands\\
}

\date{}

\pubyear{}

\begin{document}
\label{firstpage}
\pagerange{\pageref{firstpage}--\pageref{lastpage}}
\maketitle

\begin{abstract}
The abundance and narrow magnitude dispersion of Red Clump (RC) stars make them a popular candidate for mapping the morphology of the bulge region of the Milky Way.
Using an estimate of the RC's intrinsic luminosity function, we extracted the three-dimensional density distribution of the RC from deep photometric catalogues of the VISTA Variables in the Via Lactea (VVV) survey. 
We used maximum entropy based deconvolution to extract the spatial distribution of the bulge from $K_s$--band star counts.
We obtained our \newtext{extrapolated} non-parametric model of the bulge over the inner $40^\circ \times 40^\circ$ region of the Galactic centre. Our reconstruction also naturally matches onto a parametric fit to the
bulge outside the VVV region and inpaints overcrowded and high extinction regions.
We found a range of bulge properties consistent with other recent investigations based on the VVV data. \newtext{In particular, we estimated the bulge mass to be in the range $[1.3,1.7]\times 10^{10} M_\odot$,
 the X-component to be between 18\% and 25\% of the bulge mass, and the bulge angle  with respect to the Sun-Galactic centre line to be between  $18^\circ$ and $32^\circ$.}
Studies of the {\em Fermi  Large Area Telescope\/} (LAT) gamma-ray  Galactic centre excess suggest that the excess may be traced by Galactic bulge distributed sources.
We applied our deconvolved density in a template fitting analysis of this {\em Fermi--LAT\/} GeV excess and found an improvement in the fit compared to previous parametric based templates.

\end{abstract}


\section{Introduction}

Since the advent of near infrared  surveys, we have begun to view the Milky Way centre behind dust reddening obscuration  \citep{Bland-Hawthorn2016}.
Through the COBE/DIRBE survey the presence of a Galactic bulge/bar was %
established  \citep{BinneyUnderstandingkinematicsGalactic1991,WeilandCOBEDiffuseInfrared1994}.
Models fitted to the DIRBE data typically found a \newtext{triaxial bar with its major axis rotated at an angle in the range between 10 and 45 degrees to the Sun-Galactic centre line} \citep{Bissantzmicrolensingopticaldepth1997, FreudenreichCOBEModelGalactic1998, DwekMorphologynearinfraredluminosity1995,BissantzSpiralarmsbar2002}.
Subsequently, surveys such as OGLE, 2MASS, and VVV have provided us with increasingly sensitive observations of the stellar distribution in the Galactic centre.
The main observational dataset of interest to this study is the VISTA Variables in the Via Lactea (VVV) survey \citep{MinnitiVISTAVariablesLactea2010}, in particular, the stars occupying the Red Clump (RC) region of the Colour Magnitude Diagram (CMD).

The narrow dispersion of the RC \newtext{\citep{ChanemphGaiaDR2parallax2019,HallTestingasteroseismologyGaia2019}} combined with the photometric star catalogues in the near infrared regime enables estimates of the distance to stars based on their apparent magnitudes, though this comes with some caveats (see \cite{GirardiRedClumpStars2016}).
The RC has been the focus of several studies characterising the three-dimensional density structure of the Galactic bulge.
Many studies have exploited this property of the RC to fit triaxial models to the bulge \citep{StanekModellingGalacticBar1997,RattenburyModellingGalacticbar2007,Caonewphotometricmodel2013,Simionparametricdescription3D2017}.
Non-parametric methods have also been used in viewing the RC distribution, initially with an assumed constant intrinsic RC magnitude \cite{SaitoMappingXshapedMilky2011}, then later accounting for its dispersion in works such as \cite{WeggMappingthreedimensionaldensity2013} (from here on WG13).
\newtext{The Galactic RC magnitude distribution was found to produce a double peak} by \cite{NatafSplitRedClump2010} using  OGLE-III data 
and \cite{McWilliamTwoRedClumps2010} using 2MASS.
This has been interpreted as being the result of an X-shaped structure
which is characteristic of the 
boxy/peanut like morphology seen in
extragalactic studies of barred galaxies  \citep[e.g.][]{LaurikainenMilkyWaymass2014,CiamburGraham2016}
and N-body simulations \citep[e.g.][]{GardnerNbodysimulationinsights2014}.
However, some works have disputed the physical separation of the RC, positing population effects in the luminosity function account for the photometric split in the RC peaks \citep{Lopez-CorredoiracaseXshapedstructure2016,JooNewInsightOrigin2017,
LeeAssemblingMilkyWay2018}.
However, the cross matching of VVV RC stars with \emph{Gaia} in \cite{SandersTransversekinematicsGalactic2019} and \cite{ClarkeMilkyWaybar2019} found proper motions of the VVV RC stars which indicate a spatial separation in the split RC peak.

Triaxial symmetry has often been assumed in morphological studies, 
such as the analytic models used by \cite{Simionparametricdescription3D2017} (from here on S17).  The models used by S17 represent only a subset of the  broader class of triaxial bulge models \citep{DwekMorphologynearinfraredluminosity1995}. \newtext{Triaxial symmetry has also been enforced for non-parametric studies such as that of WG13 (hereafter, eight-fold symmetry for this context) to overcome gaps in the data and improve signal to noise when producing their final model.} 
In this article, we use maximum entropy and smoothness regularisation 
\citep{JaynesInformationTheoryStatistical1957,StormSkyFACTHighdimensionalmodeling2017a}
to help estimate the bulge morphology. This allows us to make fewer symmetry assumptions and it also provides a natural way of inpainting masked regions and matching onto parametric fits outside the region of interest covered by the data.

\citeauthor{PaperI} (submitted), hereafter P19,  modelled the VVV data without any symmetry requirements, which exposed features adjacent to the bulge. In this paper, we made a \newtext{mirror} symmetry assumption about the Galactic plane to enable a constrained extension of the non-parametric RC bulge model to the inner \newtext{$40^\circ \times 40^\circ$} region, which is important for our intended application described below. In addition, we absorb into our background known features outside the bulge that may otherwise be picked up by the deconvolution. We also performed systematic checks of this bulge analysis pipeline.
%
%

%
%

Knowledge of the Galactic bulge density distribution can provide useful information when modelling the {\em Fermi\/} Galactic Centre Excess (GCE) \citep{AckermannFermiGalacticCenter2017} observed in the {\em Fermi\/} Large Area Telescope (LAT) data \citep{AtwoodLargeAreaTelescope2009}.
The GCE was identified early on as a possible dark matter self-annihilation signal \citep{GoodenoughPossibleEvidenceDark2009,AbazajianDetectiongammaraysource2012,Gordon:2013vta}
due to its apparent diffuse spherical nature, and soon after as possibly due to a Millisecond Pulsar (MSP) population in the Galactic centre \citep{AbazajianconsistencyFermiLATobservations2011}.
More recently, the non-spherical nature of the GCE came to the foreground in importance, interpreted as strongly in favour of gamma-ray emission tracing a Galactic bulge morphology rather than the more spherically distributed dark matter self-annihilation case \citep{MaciasGalacticBulgePreferred2018,BartelsFermiLATGeVexcess2018}.
However, there is some debate about whether
the resolved MSPs are consistent with 
the needed bulge population~\citep{Cholis:2014lta,Hooper:2015jlu,PloegConsistencyLuminosityFunction2017,BartelsBayesianmodelcomparison2018a}.
In this work, we employ the same template fitting procedure described in \cite{MaciasStrongEvidencethat2019} and compare our non-parametrically deconvolved bulge model to the bulge models of past works.

Our article is arranged as follows:
In Section \ref{sec:Method} we provide an overview of our VVV dataset preparation and our non-parametric deconvolution method for inverting stellar statistics to recover the three-dimensional RC density distribution.
%
%
We also motivate our choice of parametric model  as a prior distribution and as a simple geometric model of the bulge with a peanut/X-shape morphology.
In Section \ref{sec:simulation}, we test our deconvolution pipeline against simulations.
We present our results and discuss them in Sections \ref{sec:maxentVVV} and \ref{sec:Systematics}.
In Section \ref{sec:applications}, we estimate various properties of the bulge and analyse the impact of our non-parametric model on the GCE template fitting.

%

%

%
%
%

%

%

%

%
%
%
%
%

%

%
%

%
%
%
%
%
%
%
%
%
%
%
%
%
%
%

%
%
%

%
%
%
%
%
%
%
%
%

\section{Method}\label{sec:Method}

\subsection{VVV Data Preparation}
This paper employs the MW-BULGE-PSFPHOT ultra deep photometric catalogue of \cite{SurotMappingstellarage2019}, corrected for known calibration issues discussed by \cite{Hajdu2019OptimalVVVCalibration} through cross matching sources with 2MASS \citep{SkrutskieTwoMicronAll2006}.
\newtext{Note that the $K_s$ and $J$ apparent magnitudes in the catalogue have been extinction corrected \citep{SurotMappingstellarage2019}.}
A standard colour cut of $0.4 < J - K_s < 1.0$ was applied to restrict sources to the predominantly RC region of the CMD \newtext{ and exclude any bluer foreground stars}.
\newtext{We bin the extinction corrected stellar catalogue with resolution $(0.05\, \mathrm{mag} \times 0.2^\circ \times  0.2^\circ)$ in magnitude $(K_s$), Galactic latitude ($l$), and Galactic longitude ($b$). The grid was bounded by the ranges: 11 < $K_s$ < 15, $-10^{\circ} < l <  10^{\circ}$, and  $-10^{\circ} < b < 5^{\circ}$. This binned dataset was corrected for completeness by dividing by the mean of the completeness values for stars in each voxel, utilising the effective completeness value assigned to each star in the catalogue of \cite{SurotMappingstellarage2019}.}
Due to crowding and known photometric error effects, we argue for a mask \newtext{of our gridded line of sight data} based on the mean
$K_{s}$ uncertainty ($\sigma$)
of the binned stars in the catalogue rather than  a colour excess based mask.
A boundary of $\sigma  = 0.06$ was chosen.
\newtext{This value causes the new mask to approximately match the $E(J-K)=0.9$ boundary in the less crowded regions of $|l|>5^{\circ}$.}
A systematic check of this method is investigated in Section \ref{sec:Systematics}.
The data preparation is discussed in further detail in P19.

\subsection{Luminosity Function}
We utilise the semi-analytic luminosity function constructed in P19 using the PARSEC+COLIBRI isochrone sets of \cite{MarigonewgenerationPARSECCOLIBRI2017} and a Chabrier log-normal Initial Mass Function (IMF) \cite{ChabrierGalacticStellarSubstellar2003}.
Using the evolutionary stage flags in the isochrones, the semi-analytic luminosity function is divided into 3 components: a red giant branch, an RC, and, an asymptotic giant branch. An exponential function was fitted to the red giant branch, excluding the absolute magnitude range $-1.75<M_{K_s}<-0.75$, to extract the Red Giant Branch Bump (RGBB) component.
We assumed a bulge age of 10 Gyr and a metal content normally distributed with solar mean metallicity $\mu_{\text{\lbrack Fe/H \rbrack}} = 0.0 $ and dispersion $\sigma_{\left \lbrack \text{Fe/H} \right \rbrack} = 0.4$ \citep{Zoccali2008BulgeMetalContent}.

\subsection{Deconvolution Procedure}
The RC+RGBB stellar density ($\rho$)
of the Galactic bulge  can be reconstructed by inverting the equation of stellar statistics 
\begin{multline}\label{eq:stellarstatistics}
    N\left(K_{s}, l, b\right) = B\left(K_s, l, b\right) \\ + \Delta \Omega \Delta K_s \int^{13\, \mathrm{kpc}}_{4\, \mathrm{kpc}} \rho\left(s, l,b\right) \Phi\left(K_{s} - 5\log s - 10\right) s^2 \, {\rm d}s,
\end{multline}
where $N$ is the predicted number of stars in a voxel centred at $(K_s,l, b)$ and 
 $B$ is the number of smooth background stars in the voxel that are neither RC or RGBB stars. The
 $\Delta \Omega$ denotes the solid angle subtended by the line-of-sight, $\Delta K_s$ is the width of the $K_s$ magnitude bin, and $s$ (measured in kpc) is the distance from the Sun.
The luminosity function $\Phi$ is the sum of the bulge RC and bulge RGBB luminosity function components.
Note that as the RGBB is a much smaller component than the RC, we sometimes refer to our obtained density in terms of the RC only, but more precisely it does contain both the RC and RGBB.
As the Galactic bulge density tends to become negligible beyond several kpc, we only integrate the range  $4~{\rm kpc}\leq s\leq 13~{\rm kpc}$ when computing the bulge contribution in modelling stellar counts.

As in P19, our analysis uses penalised likelihoods with penalties which come in two general forms: the first is maximum entropy regularisation, \newtext{inspired by its application in \cite{StormSkyFACTHighdimensionalmodeling2017a}}, which is defined for a \newtext{3-D grid of numbers} $q$, \begin{equation}\label{eq:maxentdef}
    -2 \ln \mathcal{L}_{MEM} = 2 \lambda \sum_{i,j,k} \left( 1 - q_{i,j,k} + q_{i,j,k} \ln q_{i,j,k} \right)
\end{equation}
where $i$, $j$, and $k$ are the grid points for $K_s$, $b$, and $l$ respectively. 
The maximum entropy regularisation has a minimum at $q_i = 1$, so for our application we will use a parameterisation where $q$ is the ratio between a modelled quantity of interest and a smooth prior estimation of the quantity.
As shown in Appendix A of P19, the prior relative standard deviation of the reconstructed density from the prior density is of order $1/\sqrt{\lambda}$.
So, the larger the value of $\lambda$ chosen, the smaller the prior uncertainty assumed and so the more regularisation of the solution is applied.

The second form of likelihood penalty we use is the $\ell_2$-norm regularisation of the second derivative of the logarithm of some quantity \newtext{(also inspired by its application in \cite{StormSkyFACTHighdimensionalmodeling2017a})}. For a \newtext{3-D grid of numbers}, $F$, which varies over one dimension, we use the second order central difference equation approximation of curvature:
\begin{equation}\label{eq:l2normdef}
  -2 \ln \mathcal{L}_{\rm smooth} = 
\eta \sum_{i}  \left( \ln F_{i-1} + \ln F_{i+1} - 2\ln F_i  \right)^2.
\end{equation}
This penalty has a minimum when $F$ is the exponential of a linear function of grid coordinates.
As shown in Appendix A of P19, the prior relative standard deviation from an exponential of a linear function is approximately $1/\sqrt{6\eta}$.
So, the larger the value chosen for $\eta$, the more smoothness regularisation is applied.

\subsection{Background}\label{subsec:background}
We modelled the background ($B$) non-parametrically as a free parameter for each $(K_s,l,b)$ voxel.
Without regularisation we would have a Poisson likelihood for data $n_{i,j,k}$ with expected counts $B_{i,j,k}$ where $i,j,k$ are the grid points for $(K_s,l,b)$ respectively.
With maximum entropy and smoothness regularisation, we have the following formula for the natural log of the penalised likelihood ($\cal L$):
\begin{equation}\label{eq:likelihoodbg}
   \begin{aligned}
         \ln \mathcal{L} = &\sum_{\{i,j,k\}\in \{K_s,l,b\}}  \left[ \vphantom{-\eta_{b}  \left( \ln B_{i,j,k-1} + \ln B_{i,j,k+1} - 2\ln B_{i,j,k}  \right)^2/2}
         n_{i,j,k} \ln B_{i,j,k} - B_{i,j,k}\right.
         \\ 
         &\left.
          -\lambda \left( 1 - q_{i,j,k} +q_{i,j,k} \ln q_{i,j,k} \right)\right. \\ 
         &\left. -\eta_{K_s}  \left( \ln B_{i-1,j,k} + \ln B_{i+1,j,k} - 2\ln B_{i,j,k}  \right)^2/2\right.
         \\ 
         &\left. -\eta_{l}  \left( \ln B_{i,j-1,k} + \ln B_{i,j+1,k} - 2\ln B_{i,j,k}  \right)^2/2
         \right.
         \\ 
         &\left. -\eta_{b}  \left( \ln B_{i,j,k-1} + \ln B_{i,j,k+1} - 2\ln B_{i,j,k}  \right)^2/2
         \right],
   \end{aligned}
\end{equation}%
where $q$ is the ratio between our background model and a smooth prior estimation of the background:
\begin{equation}\label{eq:entropydefbg}
    q \equiv \frac{B}{B_{\rm prior}}.
\end{equation}
The first line on the RHS of  Eq.~\ref{eq:likelihoodbg} is from the usual Poisson likelihood distribution. The second line is an
entropy regularisation of the form of Eq.~\ref{eq:maxentdef} and 
the third, fourth, and fifth lines
are smoothness regularisations of the form given in Eq.~\ref{eq:l2normdef} 
for $K_s$, $l$, and $b$ respectively.
The regularisation parameter values we used are listed in Table~\ref{tab:regularisationparameters} and we discuss their choice  in Section~\ref{sec:simulation}.
We maximised Eq.~\ref{eq:likelihoodbg} using the magnitude ranges $11<K_s<11.7$ and $14.3<K_s<15$,
see Section~\ref{sec:simulation} for more details. This means  the behaviour in $11.7\leq K_s\leq 14.3$ is determined entirely by the prior, maximum entropy, and smoothness regularisation. 

\begin{table}\label{tab:regularisationparameters}
    \begin{center}
    \caption{Regularisation parameters used when fitting to the simulated population and the VVV sample.
    }
    \begin{tabular}{ |c|c|c|c|c| }
        \hline
          & $\lambda$ & $\eta_{s}$ & $\eta_{l}$ & $\eta_{b}$ \\
         \hline
         Background & 1.0 & 1000.0 & 100.0 & 100.0\\  
         \hline
         3-D Deconvolution & 0.01  & 400.0 & 200.0  & 100.0 \\
         \hline
    \end{tabular}
    \end{center}
\end{table}

\begin{figure}\label{fig:structurebehindbar}
    \centering
    \includegraphics[width=\columnwidth]{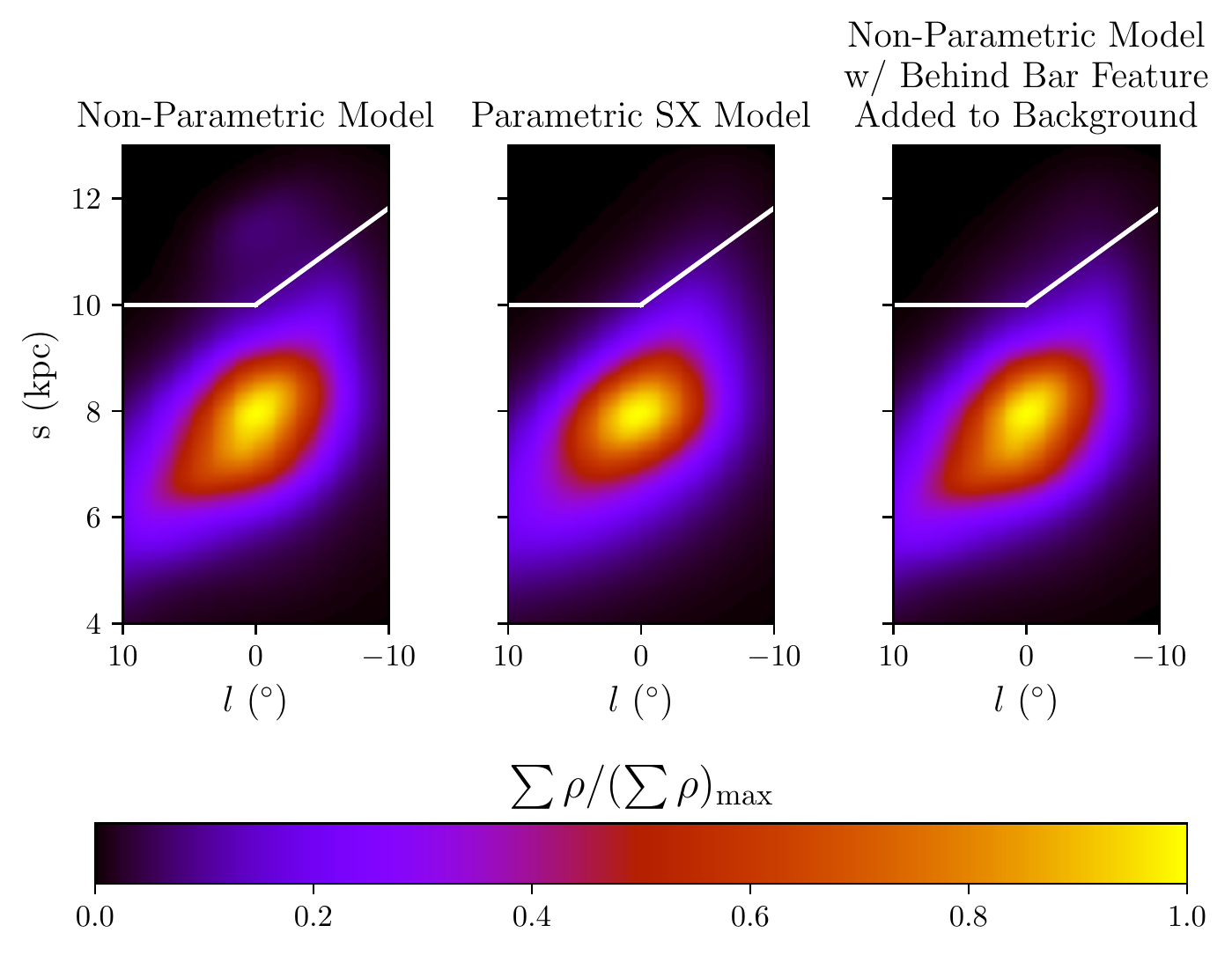}
    \caption{Apparent structure behind the bar in the VVV data, visible in the left panel, was added to the background of our model. We remove any density which is significantly greater than the fitted parametric model (middle panel) and at distances greater than indicated by the white line. In these figures, the density has been summed in the range $|b| < 10^{\circ}$. }
\end{figure}

\begin{figure}\label{fig:MEMmodellosbg}
    \centering
    \includegraphics[width=0.8\columnwidth]{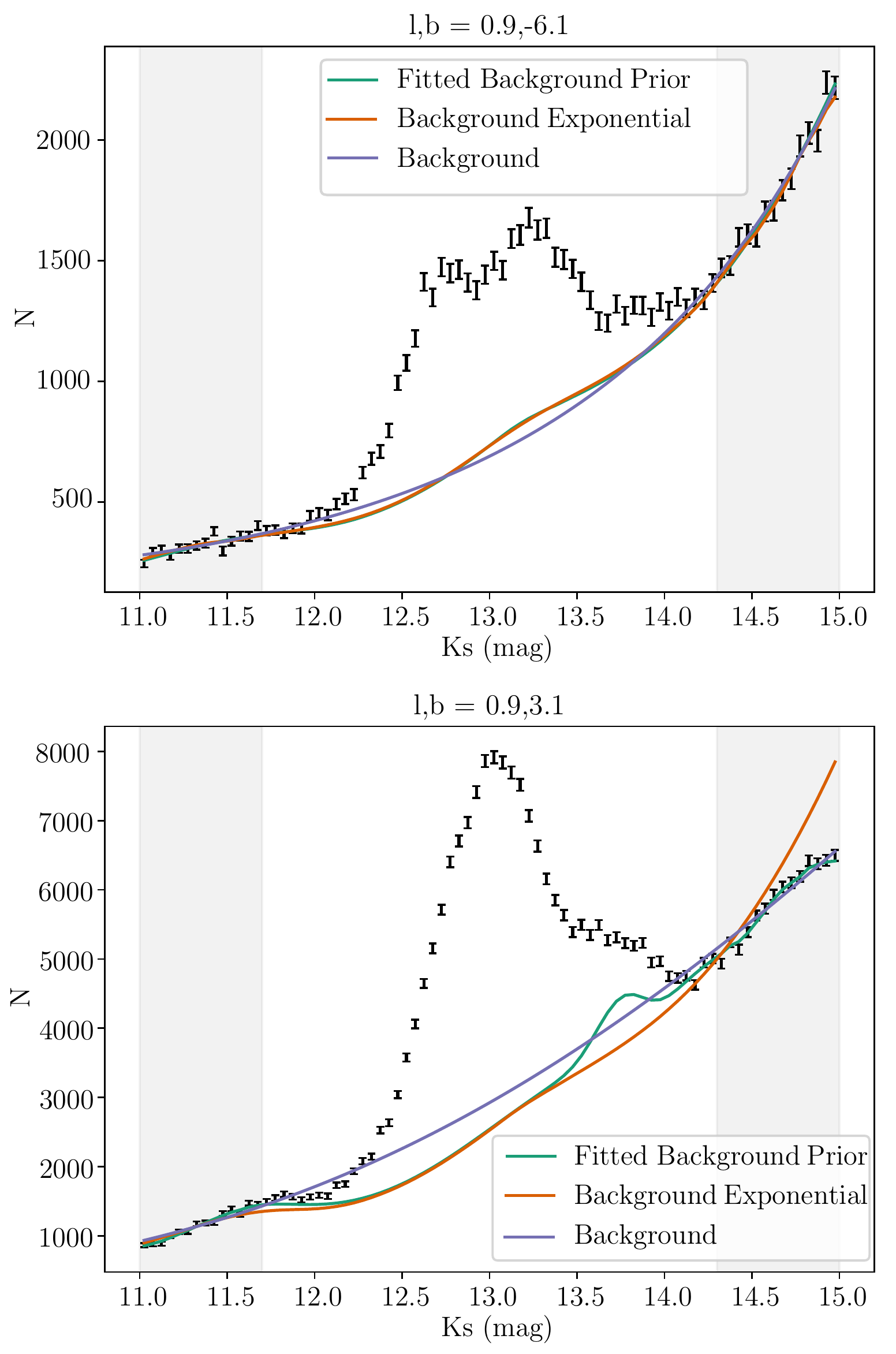}
    \caption{Demonstration of the maximum entropy background fitting in two $1^{\circ}\times1^{\circ}$ regions. The background has been fitted in the grey shaded regions using the maximum entropy method. The prior background was calculated using the S17 S-model+discs, which has been scaled to match the VVV observations between $11.0 < K_s < 11.5$~mag.
    The bump in the bottom panel  ``fitted background'' at $K_s\sim 13.8$~mag is from a feature behind the bar, see text in Subsection~\ref{subsec:background} for more details.
    The exponential background is described in Section \ref{sec:Richardson-Lucy}.}
\end{figure}

The background is mainly composed of red giant stars in the bulge and foreground disc stars, so for the prior background ($B_{\rm prior}$) we used the S-model+discs fitted by S17 with the RC and RGBB  components subtracted.  Only the asymptotic giant branch and red giant branch (excluding the RGBB) components of the semi-analytic luminosity function are used for the bulge component in determining the background. Included in the S-model+discs are thin and thick disc components of the Besan\c{c}on galaxy model of \cite{Robinsyntheticviewstructure2003}, where we have used the updated thin disc parameters from \cite{Robin2012thindiskupdate} and the updated thick disc parameters from \cite{Robin2014thickdiskupdate}. The S-model+discs of S17 was fitted to aperture photometry of the VVV DR2 data in the range $12 < K_s < 14$, so the background was underestimated for some lines of sight. To compensate for this, 
we multiplied
each pixel (line-of-sight) of the prior background by a constant, so that its mean matched the mean of our data in the range $11 < K_s < 11.5$~mag.
Initial tests of our deconvolution method on the VVV data showed that our method was finding a feature in the density consistent with the structure behind the bar reported in \cite{Gonzalez2018StructureBehindBar} and P19.
 As we are trying to determine the bulge component, we decided to add this feature to our background, by first estimating our density using our maximum entropy background, then adding the star counts associated with any density significantly greater than our prior parametric density (see SX model of Section~\ref{sec:parametricmodel}) to the maximum entropy background. We considered any density which was beyond the limits 
\begin{equation}\label{eq:behindbardensitylimits}
    \begin{array}{lr}
    s > 10 \textrm{ kpc}                & l \geq 0^{\circ} \\
    s > (10 -0.1818 \, l) \textrm{ kpc}   & l < 0^{\circ} \\
    \end{array}
\end{equation}
and at least $2.6 \times 10^{-5}$ stars pc$^{-3}$ sr$^{-1}$ above the parametric model density to be part of the structure behind bar. In Fig.~\ref{fig:structurebehindbar} we display the density summed over |b| < 10$^{\circ}$, where the feature behind the bar is visible in the model fitted using our maximum entropy method. The contribution of the feature behind the bar to the background is visible in the bottom panel of Fig. \ref{fig:MEMmodellosbg} as a bump in the fitted background at $K_s \sim$13.8 mag. When using the updated background, the feature behind the bar is no longer present in the density, as seen in the right panel of Fig. \ref{fig:structurebehindbar}.

Shown in the top panel of Fig. \ref{fig:MEMmodellosbg} is the fitted background for a $1^{\circ}\times1^{\circ}$ box around $(l,b) = (0.9^{\circ},-6.1^{\circ})$, where we can see that the fitted background is only slightly deviating from the prior background. In the bottom panel, the background fitted in a $1^{\circ}\times1^{\circ}$ box around $(l,b) = (0.9^{\circ},3.1^{\circ})$ fits the data well in the shaded regions. However, the background needs to deviate significantly from the prior background at $K_s > 14.7$~mag, where the data may have  residual extinction and completeness issues.  In the unshaded region, apart
from the added feature behind the bar,
the background closely follows the shape of the prior solution. The background also  smoothly trends back to passing through the data in the shaded regions.

\subsection{Maximum Entropy Deconvolution}\label{subsec:maxentdeconv}
Our maximum entropy method provides a non-parametric estimate of the stellar density which predicts the binned star counts of a stellar catalogue.
It maximises the same $\ln{\cal L}$ as Eq.~\ref{eq:likelihoodbg}, except that $B$ is replaced with the total expected star counts ($N$) and $q$ is replaced by 
$\kappa$ which is the ratio between the bulge density model and a prior estimation of the density \newtext{such as a parametric bulge model like that of Section~\ref{sec:parametricmodel}}:
\begin{equation}\label{eq:entropydef}
    \kappa \equiv \frac{\rho}{\rho_{\rm prior}}.
\end{equation}
Also, as we are estimating $\rho$ on a grid of $(s,l,b)$, we need a separate sum for the regularisation terms in contrast to  Eq.~\ref{eq:likelihoodbg} where we could use one sum as we estimated the background ($B$) on a $(K_s,l,b)$ grid. This gives
\begin{equation}\label{eq:likelihoodmaxent}
   \begin{aligned}
         \ln \mathcal{L} = & \sum_{\{i,j,k\}\in \{K_s,l,b\}} \left( n_{i,j,k} \ln N_{i,j,k} - N_{i,j,k}\right)
         \\ 
         &
          -\sum_{\{h,j,k\}\in \{s,l,b\}} \left[\vphantom{\left( \ln \rho_{h,j,k-1} + \ln \rho_{h,j,k+1} - 2\ln \rho_{h,j,k}  \right)^2}
         \lambda \left( 1 - \kappa_{h,j,k} +\kappa_{h,j,k} \ln \kappa_{h,j,k} \right)\right. \\ 
         &\left. +\eta_{s}  \left( \ln \rho_{h-1,j,k} + \ln \rho_{h+1,j,k} - 2\ln \rho_{h,j,k}  \right)^2/2\right.
         \\ 
         &\left. +\eta_{l}  \left( \ln \rho_{h,j-1,k} + \ln \rho_{h,j+1,k} - 2\ln \rho_{h,j,k}  \right)^2/2
         \right.
         \\ 
         &\left. +\eta_{b}  \left( \ln \rho_{h,j,k-1} + \ln \rho_{h,j,k+1} - 2\ln \rho_{h,j,k}  \right)^2/2
         \right ].
   \end{aligned}
\end{equation}
Including the maximum entropy term in the likelihood discourages the modelled density from over-fitting to regions of the data that are dominated by noise, where it will instead favour the smooth prior density. In practice this is important in the regions where the background makes up a significant part of the model ($K_{s}$ near 12.0 and 14.0), where the density should be tending towards zero. Addition of the smoothness terms discourages spurious high frequency variations in the modelled density by minimising curvature in the logarithm of the density. The smoothness term also has the added benefit of inpainting the density in lines of sight which have been masked out. 
For Eq.~\ref{eq:likelihoodmaxent}, we set $\lambda=0$ in masked regions so as they are only affected by the smoothness term and the values of the model at the edge of the mask.

\subsection{Parametric Model of the  X-Bulge}\label{sec:parametricmodel}

In light of the X-shape apparent in the eight-fold symmetrised WG13 style deconvolution, we consider a closed form parametric base case that allows for a X-bulge perturbation.
We characterise its potential pathologies in fitting to data and simulations.
The parametric density models fitted in this section are used as prior estimates for the density ($\rho_{\rm prior}$) with the maximum entropy deconvolution in Section \ref{sec:maxentVVV}. Our base case parametric-model fit was subsequently applied in a template fitting analysis of the {\em Fermi\/} GCE for comparison with our base non-parametric model result (see Section \ref{sec:fermiresults}).

Triaxial models of the bulge have been investigated by \cite{DwekMorphologynearinfraredluminosity1995}, \cite{Athanassoulashapebarsearlytype1990}, and \cite{ FreudenreichCOBEModelGalactic1998}.
We selected the S-model, which proved successful for bulge modelling in \cite{FreudenreichCOBEModelGalactic1998} and S17, as our base distribution. 
Inspired by the X-bulge parametric form of \cite{Lopez-CorredoiracaseXshapedstructure2016}, we perturb the S-model with a X-like shape.
We use a right-handed, Galactic Centre origin,
Cartesian grid $(X,Y,Z)$ aligned with the bulge axes of symmetry. 
The coordinates
are chosen so that the $X$-axis lies along the major axis of
the bulge and the $Z$-axis points towards the north Galactic pole. We refer to the arms of the X-bulge as the X-arms but these are not necessarily aligned with our $X$ coordinate.
The perturbation shape was freed in $X$ and $Y$  to accommodate non-circular X-arm shapes. \newtext{The X-arms in this model part linearly along the bar-aligned $Z$-axis with gradient $C$}. We also allowed the density of the X-arms to trail off as an exponential of a power-law with exponent $n$ rather than assuming an exponential or Gaussian distribution.
We label this parametric form the SX model, with its components defined as follows:
\begin{equation}
\label{eq:s_x_model}
\rho _{{\rm SX}}(X,Y,Z)=\rho_{0} \mathrm{sech}^{2}(r_{1})
\end{equation}\[
\ \ \ \ \times \left[1+A\times (\exp\left(-r_2^n\right)
+ \exp\left(-r_3^n\right) \right)]
,\]\[
r_{1}^{c_{\| }} = \bigg[ \bigg(\frac{|X|}{x_{0}} \bigg)^{c_{\perp}} +
\bigg(\frac{|Y|}{y_{0}}\bigg)^{c_{\perp}} \bigg]^{ \frac{c_{\|}}{c_{\perp}} }
+ \bigg(\frac{|Z|}{z_{0}}\bigg)^{c_{\|}} \quad 
,\]\[
r_2=\bigg[ \bigg(\frac{|X-CZ|}{x_{1}} \bigg)^{2} +
\bigg(\frac{|Y|}{y_{1}}\bigg)^{2} \bigg]^{ \frac{1}{2} }
,\]\[
r_3=\bigg[ \bigg(\frac{|X+CZ|}{x_{1}} \bigg)^{2} +
\bigg(\frac{|Y|}{y_{1}}\bigg)^{2} \bigg]^{ \frac{1}{2} }
\]
using a generalised ellipsoid distribution for the bulge and a simple ellipsoidal X-shape aligned with the bulge that tapers off with the the same $Z$ distribution. 
The parameters $\vartheta=(\rho_0,A,n,x_0,y_0,z_0,c_\perp,c_\parallel,C,x_1,y_1)$
all need to be fit to the data.
We used this parametric fit as a prior ($\rho_{\rm prior}$) for the maximum entropy non-parametric fit  which did not enforce eight-fold symmetry.
Eq.~\ref{eq:s_x_model}
will provide us with an intermediary model between the S and non-parametric models in the {\em Fermi\/} template fitting analysis to gauge the correlation between an improved VVV fit and an improved gamma-ray distribution fit. If the GCE is tracing a bulge and there are no additional unexpected features, we might expect that a model that increasingly traces the morphological features of the bulge will improve the fit.

Investigating the parting rate of the X-arms by fitting a power-law rather than the simple $X \pm CZ$ form, we found the split was still well approximated as a linear function. To avoid convergence issues from excessive parameters, the RC split was left in the linear form.

A tapering of the density at cylindrical radii greater than a cutoff radius, $R_c$, was applied to the density distribution via $\exp(-2(R-R_c)^2)$ with $R_c$ fixed to 4.5~kpc in all fits, following the preferred choice in S17.
We also fit the deviation from an 8 kpc distance from the Sun to the Galactic centre 
so that the new distance is 
$8~\rm{kpc} +\Delta R_0$.
Additionally, we fitted  $\alpha$ which is the angle between the bulge major axis and the line connecting
the Sun to the Galactic centre.

We optimise our parametric models for parameter set $\vartheta$ using the \textsc{scipy BFGS} routine\footnote{\url{https://www.scipy.org/}}, minimising the Poisson likelihood statistic:
\begin{flalign}
         \ln \mathcal{L} = \sum_{\{i,j,k\}\in \{K_s,l,b\}}  \left(
         n_{i,j,k} \ln N_{i,j,k} - N_{i,j,k}\right)+ \mathrm{constant} 
         \label{poissprob_simion}
\end{flalign}
where 
$N_{i}$ is the corresponding model, obtained by integrating the equation of stellar statistics (Eq.~\ref{eq:stellarstatistics}) for parametric density $\rho_{\rm SX}$.
Our best fit likelihoods and uncertainties are listed in our tables of results (Tables \ref{tab:TSmaxentandparam}, \ref{tab:Par_Data_Table}, and \ref{tab:Par_Sims_Table}).
The uncertainties are derived from the corresponding square root of diagonal elements of the inverse Hessian  matrix produced by this routine.
The SX model fit was initialised by randomly picking a starting point somewhere between qualitatively different boundaries that produce physically possible densities for the X perturbation parameters and choosing the initial S parameters from within 10\% of the best fit values from the S-model.

\section{Testing The Deconvolution Against A Simulation}\label{sec:simulation}
%
%
%
%
%
%
%
%

%

%
%
%
%
%
%
%
%
%
%
%
%
%
%
%
%
%
%
%
%


We constructed a simulated Milky Way population comprised of a thin disc, thick disc, and a bulge, as is modelled in S17. To generate the synthetic population, we used
\begin{multline}
    \label{eq:stellarstatisticssim}
    N\left(K_{s}, l, b\right) = \Delta \Omega\Delta K_s\\ \times \sum_{i}\int^{\infty}_{0} \rho_{i}\left(s, l, b\right) \Phi_{i}\left(K_{s} - 5\log s - 10\right) s^2  ds
\end{multline}
where $\rho$ is the density and $\Phi$ is the luminosity function and the sum is over the three model components, to predict the combined star counts in each $(K_s,l,b)$ voxel. We then simulated a population of stars by drawing a Poisson random value from the binned simulation model. 
As in P19, the thin and thick discs were generated from the updated Besan\c{c}on model parameters of \cite{Robin2012thindiskupdate} and \cite{Robin2014thickdiskupdate} respectively. The S-bulge model is given by Eq.~\ref{eq:s_x_model} with $A=0$. The simulation parameters used for this model  are listed in Table \ref{tab:bulgedensity}.

The normalisations we used for each of the three components have been multiplied by the same constant chosen so that  the total number of stars in the unmasked region and in $12 < K_s < 14$ matches the number of stars in the VVV PSF catalogue. The luminosity function we used for the bulge in the simulation is the same as the one we used in our fitting procedure to the VVV data.

\begin{table}\label{tab:bulgedensity}
    \begin{center}
    \caption{
    Density distribution parameters for the bulge component used for our simulation. The second row gives the total number of stars in the unmasked regions of the simulation in the range $12 < K_s < 14 $. In cylindrical coordinates, centred at the maximum density of the bulge, the Sun in located at ($R_{\odot}$,$Z_{\odot}$) = (8.0~kpc,15.0~pc).}
    \begin{tabular}{ |c|c|c|c|c|c| }
        \hline
         $x_0$(kpc) & $y_0$(kpc) & $z_0$(kpc) & $\alpha$($^{\circ}$) & $c_{\parallel}$ & $c_{\perp}$ \\
         \hline
         1.61 & 0.69 & 0.48 & 19.16 & 2.50 & 1.86 \\  
    \end{tabular}
    \begin{tabular}{ |c|c|c| }
        \hline
         $N_{\text{thin}}(\times10^{6})$ & $N_{\text{thick}}(\times10^{6})$ & $N_{\text{bulge}}(\times10^{6})$ \\ 
         \hline
         1.35 & 1.87 & 17.04  \\
        \hline
    \end{tabular}
    \end{center}
\end{table}

\begin{figure*}
    \centering
    \includegraphics[width=0.7\textwidth]{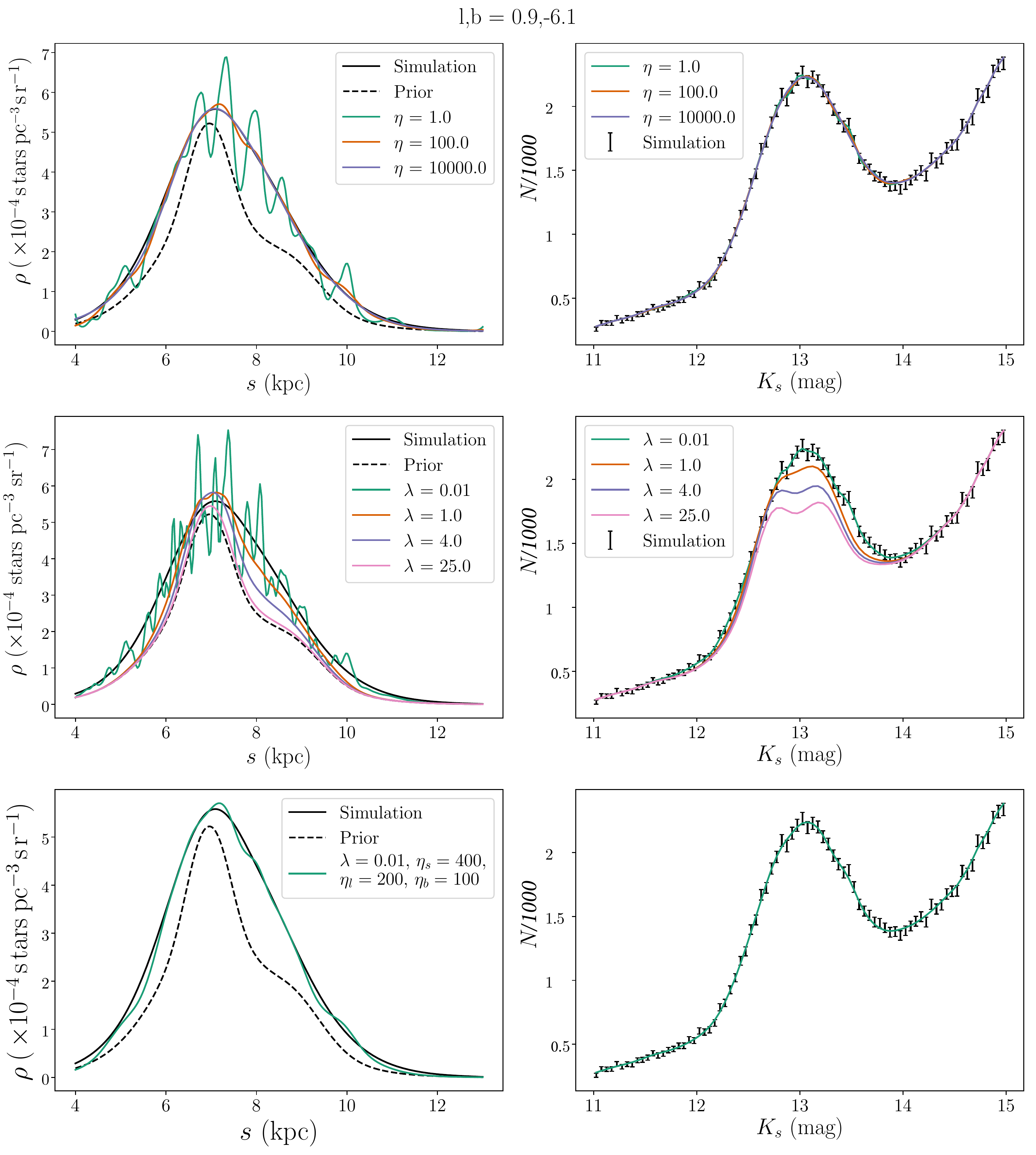}
    \caption{
    Testing the choice of regularisation parameters. We perform our maximum entropy deconvolution to a $1^{\circ}\times1^{\circ}$ region of our simulated population, centred  on $(l,b) = (0.9,-6.1)$. In the top panels, the maximum entropy regularisation is set to zero, and a range spatial smoothness parameter values are tested. The middle panels have the spatial smoothness regularisation set to zero, and a range of maximum entropy regularisation values are tested. The bottom panels have the regularisation parameters used in our final analysis. The left panels show the deconvolved density compared to the true density in the simulation, the right show the  model star counts ($N$) compared to the simulated population ($n$). For small values of $\eta$, the deconvolved density has many spurious features, which get smaller in amplitude as $\eta$ is increased. The predicted star counts is not significantly sensitive to the choice of $\eta$ in the range tested here. For all $\lambda \ge 1.0$, the predicted star counts do not match the simulation, where it is clear that the prior density distribution is not a good estimate of the true density. }
    \label{fig:regularisationparameters}
\end{figure*}

To choose the values of the regularisation parameters we tested a range of choices in a $1^{\circ} \times 1^{\circ}$ region centred on $\left( l,b \right) = (0.9,-6.1)$. 
\newtext{The test region was subdivided into  our usual voxel size of $(0.05\, \mathrm{mag} \times 0.2^\circ \times  0.2^\circ)$.}
For this test, we did not want to use a prior that was too close to the true value, so we used the base SX (Eq. \ref{eq:s_x_model}) model that had been fitted to the VVV data.
We first fixed the maximum entropy regularisation parameter, $\lambda$ from Eq.~\ref{eq:likelihoodmaxent}, to zero and applied our maximum entropy deconvolution method with a range of smoothing regularisations, $\eta$. We repeated this for $\eta = 0$ and a range of $\lambda$ values.
In Fig. \ref{fig:regularisationparameters} the deconvolved density for all choices of $\eta$ follow the general shape of the true density. Small values of $\eta$ give spurious oscillatory deviations from the true density, which decrease in amplitude as $\eta$ increases. There is not a significant difference in the predicted star counts between the choices of $\eta$. For $\lambda \geq 1.0$, the predicted star counts deviate significantly from the simulation, which is also seen in the deconvolved density where it overestimates at distances less than 6 kpc, and underestimates from 6-8 kpc.
This is because the prior density is not a good estimate of the true density for the current case. When $\lambda = 0.01$, the deconvolved density is scattered around the simulated density, and the predicted star counts are over-fitting. The results of this test suggested that a small value of $\lambda$ and a large value of $\eta$ would give the most accurate density deconvolution. Therefore, we used a value of $\lambda=0.01$ and $\eta = 100-1000$. For the background modelling, a simulation is not needed to determine an optimal set of regularisation parameters, as the effectiveness can be determined by directly comparing to the data. 
Also, the prior background from the S17 model gives a good description of the background. This means we expect less deviation from the prior and so
a larger value of
$\lambda$ can be used.
The regularisation parameters used for the background determination are  presented in Table \ref{tab:regularisationparameters}.

\begin{figure}
    \includegraphics[width=0.8\columnwidth]{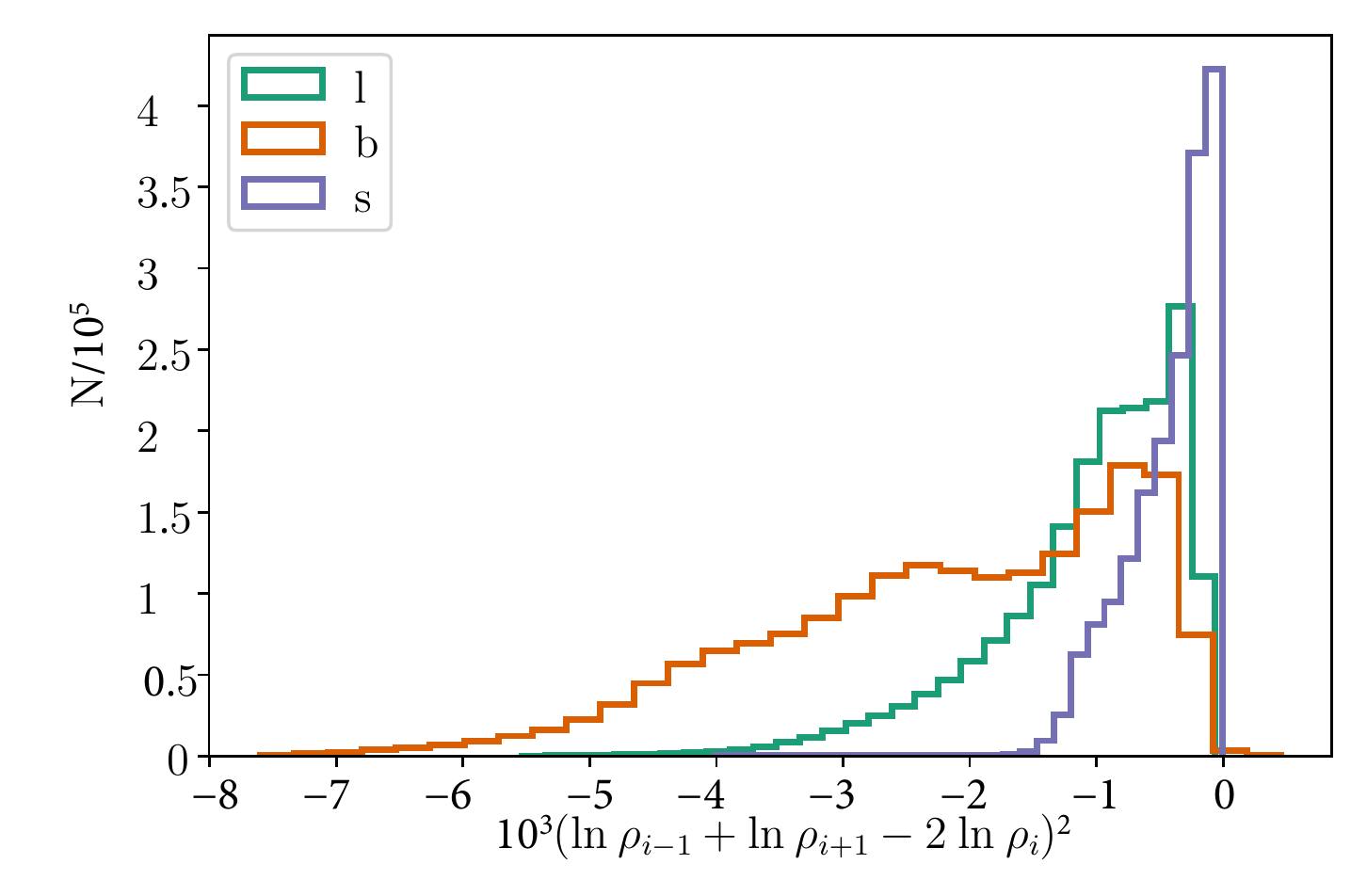}
    \caption{Distributions of the curvature in log-density (Eq. \ref{eq:l2normdef}) along respective density model coordinates of the simulated bulge.}
    \label{fig:curvaturepenaltyhist}
\end{figure}

The distribution of curvature in log-density (Eq.~\ref{eq:l2normdef}) for the simulated bulge in Fig. \ref{fig:curvaturepenaltyhist} is strictly negative. It is broadest in $b$, second broadest in $l$ and narrowest in $s$. 
The $\ell_2$-norm regularisation gives a minimum penalty to the likelihood %
when the log of the fitted density has zero curvature.
\newtext{We chose $\eta_s$, $\eta_l$, and $\eta_b$
such that the overall curvature penalty term in 
  Eq.~\ref{eq:likelihoodmaxent} was of similar magnitude.
 From the distributions of the curvature term in Fig.~\ref{fig:curvaturepenaltyhist} we chose the 
  }
 regularisation parameters used for fitting the simulated population as listed in Table \ref{tab:regularisationparameters}.

We applied the maximum entropy deconvolution process to the simulated star counts, first by fitting the background including the feature behind the bar, then by fitting a parametric density model to determine a prior density estimation for the full 3-D density deconvolution. The parameters of the fitted prior density are presented in Table \ref{tab:Par_Sims_Table}, labelled case A. The maximization of the $\ln\mathcal{L}$ in Eq. \ref{eq:likelihoodmaxent}
and $\ln\mathcal{L}$ in Eq. \ref{eq:likelihoodbg} were both
performed using the python implementation \textsc{pylbfgs}\footnote{\url{https://github.com/dedupeio/pylbfgs}} of the Limited Memory Broyden-Fletcher-Goldfarb-Shanno (L-BFGS) algorithm.

The density was modelled non-parametrically on a (257, 100, 50) grid of $(s, l, b)$, in the range $4 < (s/{\rm kpc}) < 13$, $-10^\circ < b < 0^\circ$ and $-10^\circ < l < 10^\circ$, for a total of 1.285$\times 10^{6}$ free parameters. The grid spacing is ($\Delta s, \Delta l, \Delta b$) = (35 pc, 0.2$^{\circ}$, 0.2$^{\circ}$).
To make the optimization of so many parameters feasible, we evaluated the gradients of  $\ln\mathcal{L}$ in Eq.~\ref{eq:likelihoodmaxent} and  $\ln\mathcal{L}$ in Eq.~\ref{eq:likelihoodbg} analytically, see Appendix A of P19 for more details.
We assumed symmetry about the Galactic mid-plane so that we could reliably extend our non-parametric density model to latitudes $b>5^\circ$, where there are no observations in the VVV sample. Making the \newtext{mirror} symmetry assumption forced us to position the Sun in the Galactic mid-plane ($Z_\odot = 0$~kpc). 
We fixed the reconstructed density just outside the region of interest to the prior density by setting $\lambda=1$ in those regions. This meant that the smoothness regularisation forced  the reconstructed density to smoothly transition to the parametric prior density at $|l| > 10^\circ$ and $|b| > 10^\circ$.   

\begin{figure}
    \centering
    \includegraphics[width=0.7\columnwidth]{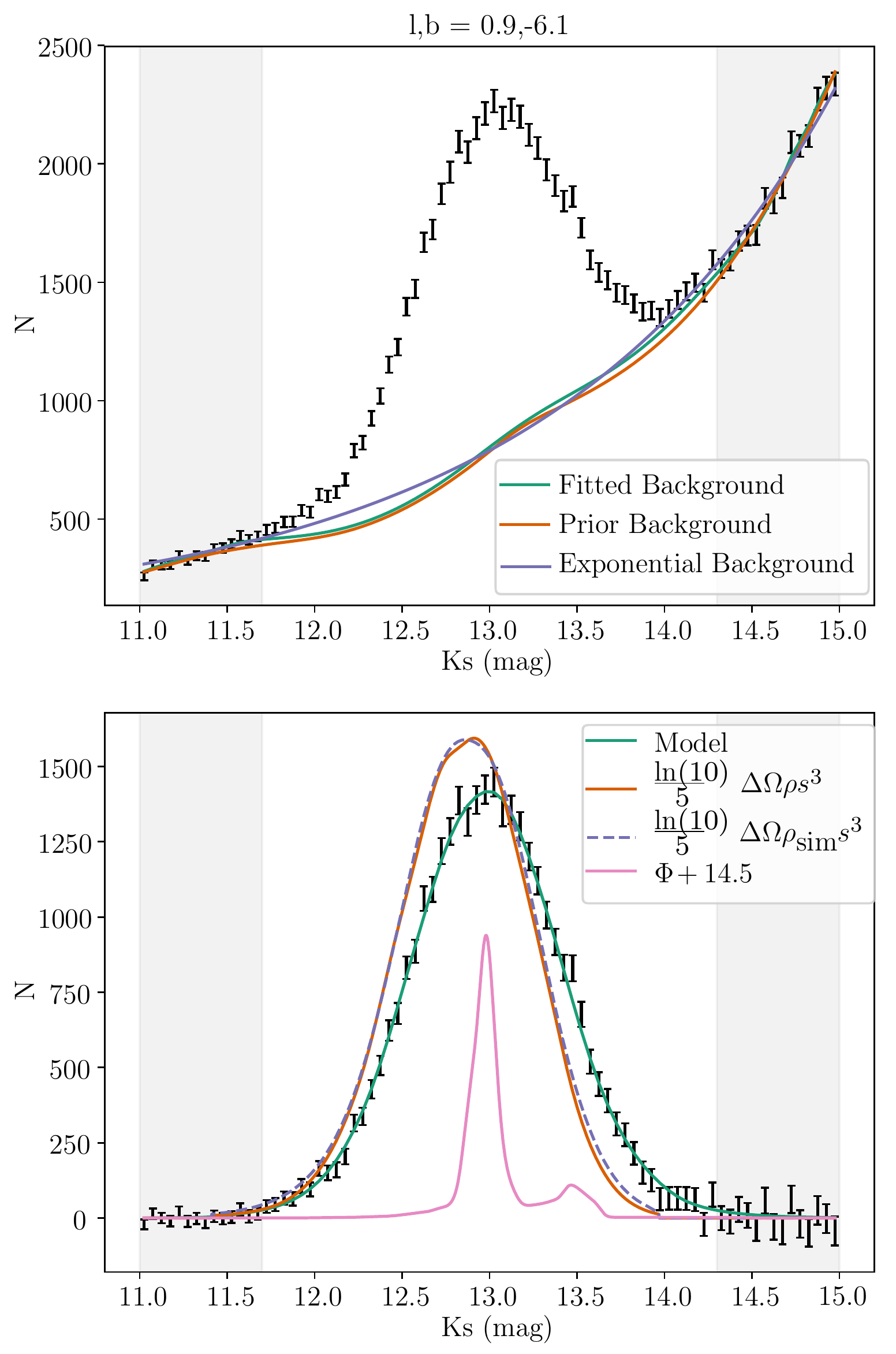}
\caption{Demonstration of the maximum entropy deconvolution method to a simulated population for a $1^{\circ}\times1^{\circ}$ region. \textit{Top}: The background has been fitted in the grey shaded regions using the maximum entropy method. The prior background is the background of the model used to generate the simulation. \textit{Bottom}: Maximum entropy deconvolution of the line-of-sight star count distribution. Shown in green is the predicted number of RC star counts from the convolution of the fitted density (orange) and assumed luminosity function (pink). The density used to produce the simulation is shown as a dashed purple line. \newtext{The luminosity function has been scaled and shifted for display, where 14.5 is the distance modulus added to the absolute magnitudes, $M_{Ks}$.}  
}
    \label{fig:MEMmodellossim}
\end{figure}

Shown in the top panel of Fig. \ref{fig:MEMmodellossim} is the background fitted to the simulation. From the deconvolution of the VVV data shown in Fig. \ref{fig:MEMmodellossim},
we can see the simulated population lacks a splitting of the RC peak that is present in the VVV observations case shown in Fig.~\ref{fig:MEMmodellos}.
In Fig. \ref{fig:MEMsimdensity} we compare the 3-D deconvolved density to the density used in simulating the population. The deconvolved density using the maximum entropy method compares well to the density used in our simulation, even inside of the masked regions where there is no data influencing the deconvolution. However, the reconstruction displays some discrepancy at around $s=4$~kpc. Note that this is due to the low star counts in the bulge at this radius which makes an accurate reconstruction difficult. Note that Fig.~\ref{fig:MEMsimdensity} correctly does not show the X-bulge morphology that is seen in the VVV data which is displayed in Fig.~\ref{fig:MEMdensity}.

\section{Deconvolution of VVV}\label{sec:maxentVVV}
\begin{figure}
    \centering
    \includegraphics[width=0.7\columnwidth]{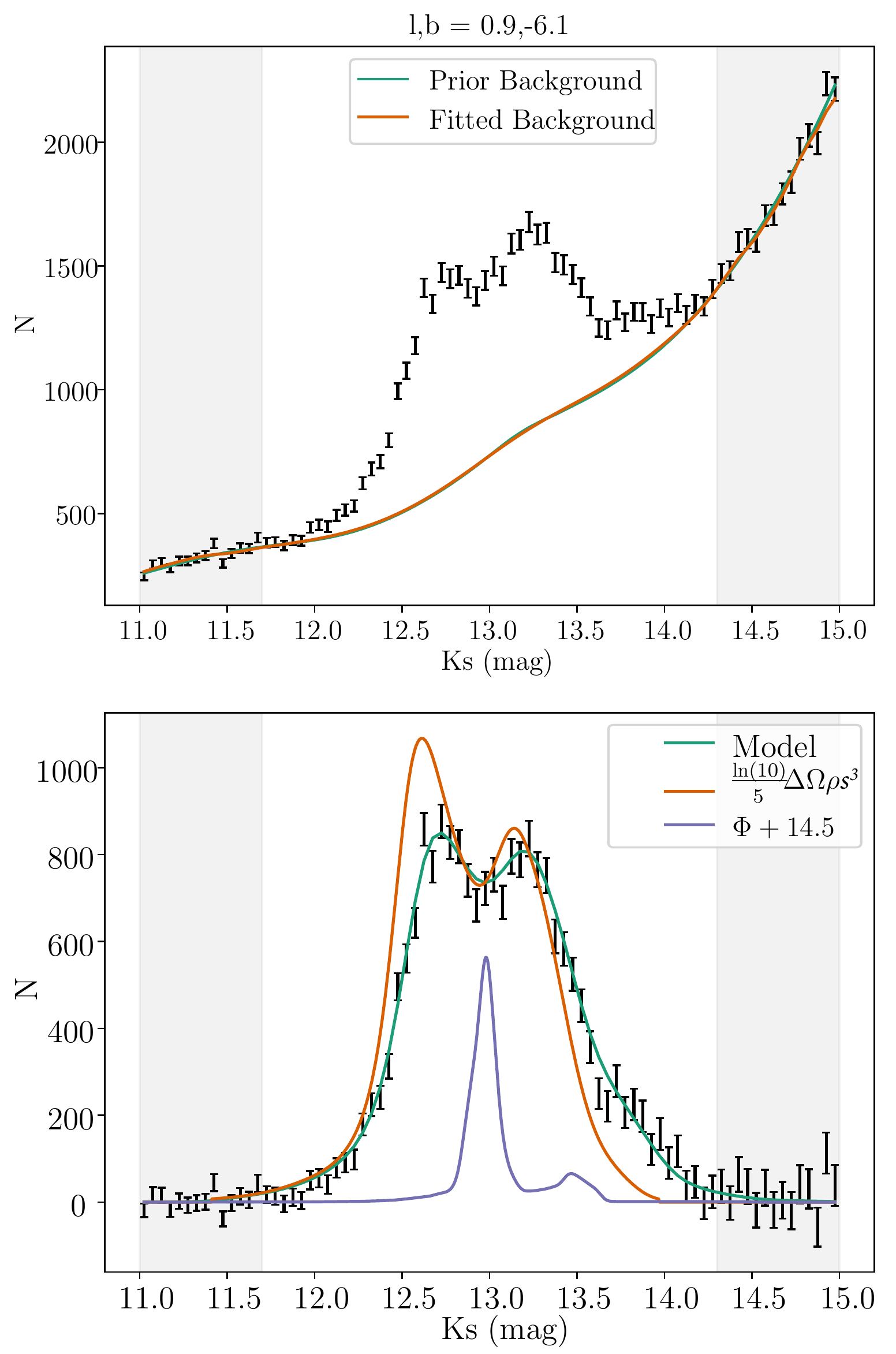}
    \caption{Demonstration of the maximum entropy deconvolution method in a $1^{\circ}\times 1^{\circ}$ ($5 \times 5$ pixels) region.\textit{Top}: The background has been fitted in the grey shaded regions using the maximum entropy method. The prior background was calculated using the S17 S-model+discs, which has been scaled to match the VVV observations between $11.0 < K_s < 11.5$. \textit{Bottom}: Maximum entropy deconvolution of the line-of-sight background subtracted star count distribution. Shown in green is the predicted number of RC star counts from the convolution of the fitted density (orange) and assumed luminosity function (purple). The luminosity function has been scaled for display.} 
    \label{fig:MEMmodellos}
\end{figure}

In this section, we discuss how we applied our maximum entropy deconvolution method to the VVV data sample for our base model which we label as case A. We used a fit of the parametric SX model 
as the prior density distribution and the values for the regularisation parameters in Table~\ref{tab:regularisationparameters}. The background was fitted using the maximum entropy method of Section \ref{subsec:background}.
In Fig. \ref{fig:MEMmodellos} we present a breakdown of the maximum entropy deconvolution model components along a single line-of-sight through the region the photometric split clump has been observed. 

Displayed in Fig. \ref{fig:MEMmodelcontours} is a comparison between the predicted star counts by our maximum entropy deconvolution, the fitted parametric model we used as the prior, and the VVV data. For compactness, we show every tenth magnitude bin.
At $K_s < 12$ and $K_s > 14$ the RC+RGBB stars contribute negligibly to the total star counts, so both the parametric model and maximum entropy deconvolution are dominated by the background. By construction, these regions are well described by the background model, though perhaps there is slight over-fitting in the $K_s=14.975$ bin.
The non-parametric model reproduced the data well and has smaller deviations in comparison to the parametric model, especially notable in the $K_s = 12.525$ bin at $l = 5^{\circ}$ where the X-bulge is prominent. The assumption of symmetry about the Galactic mid-plane seems to be reasonable, as there is no visible bias in fitting to the mirrored contours above and below the plane. 

\begin{figure}
    \centering
    \includegraphics[width=\columnwidth]{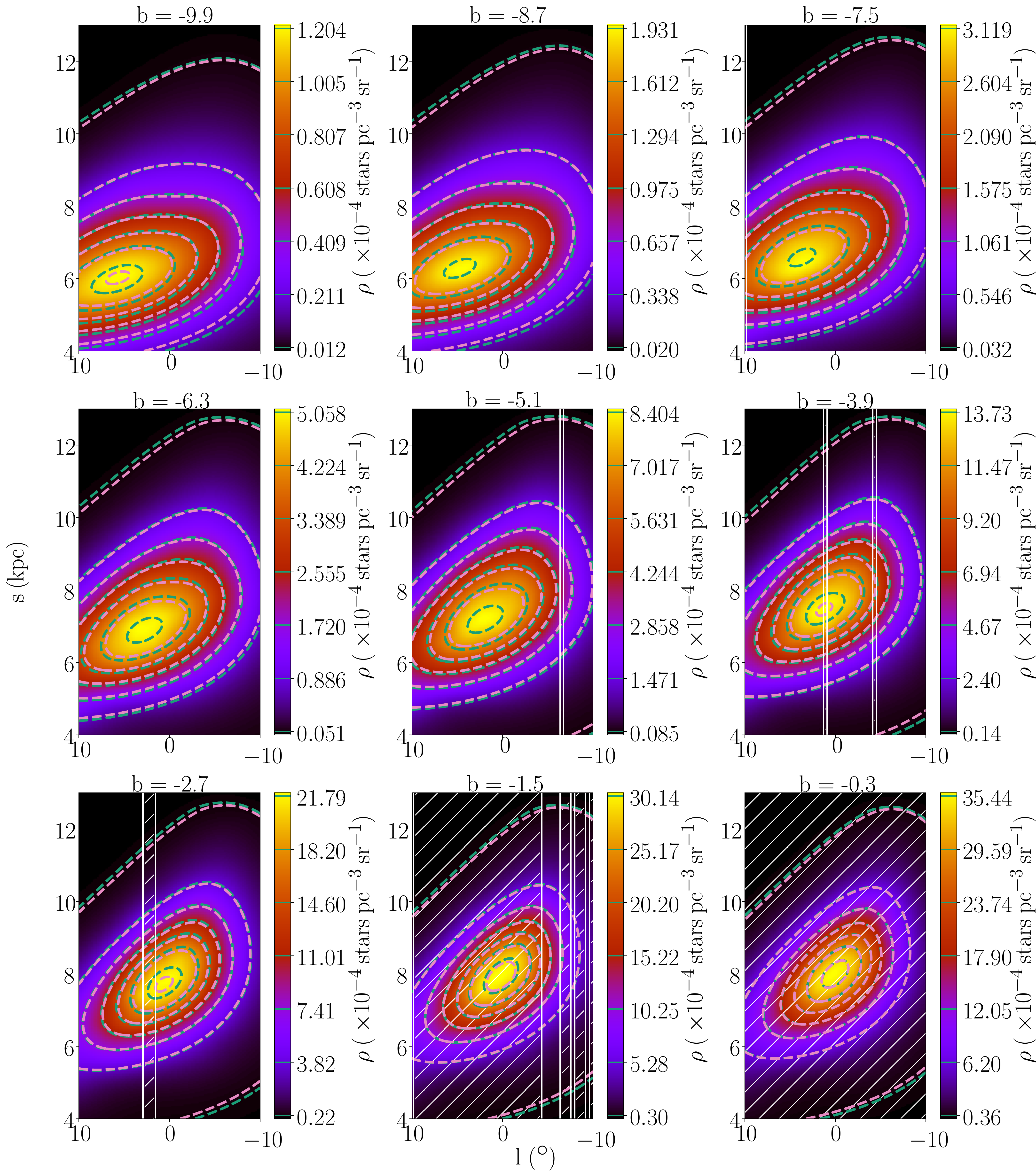}
    \caption{Deconvolved RC+RGBB star density using the maximum entropy method for a simulated 10 Gyr S-model. White hatched regions were masked during the analysis, and were inpainted naturally as part of the deconvolution. Green dashed contours show the true density used in simulating the S-bulge. Pink show the parametric SX model used as the prior density.  }
    \label{fig:MEMsimdensity}
\end{figure}

\begin{figure*}
    \centering
    \includegraphics[width=\textwidth]{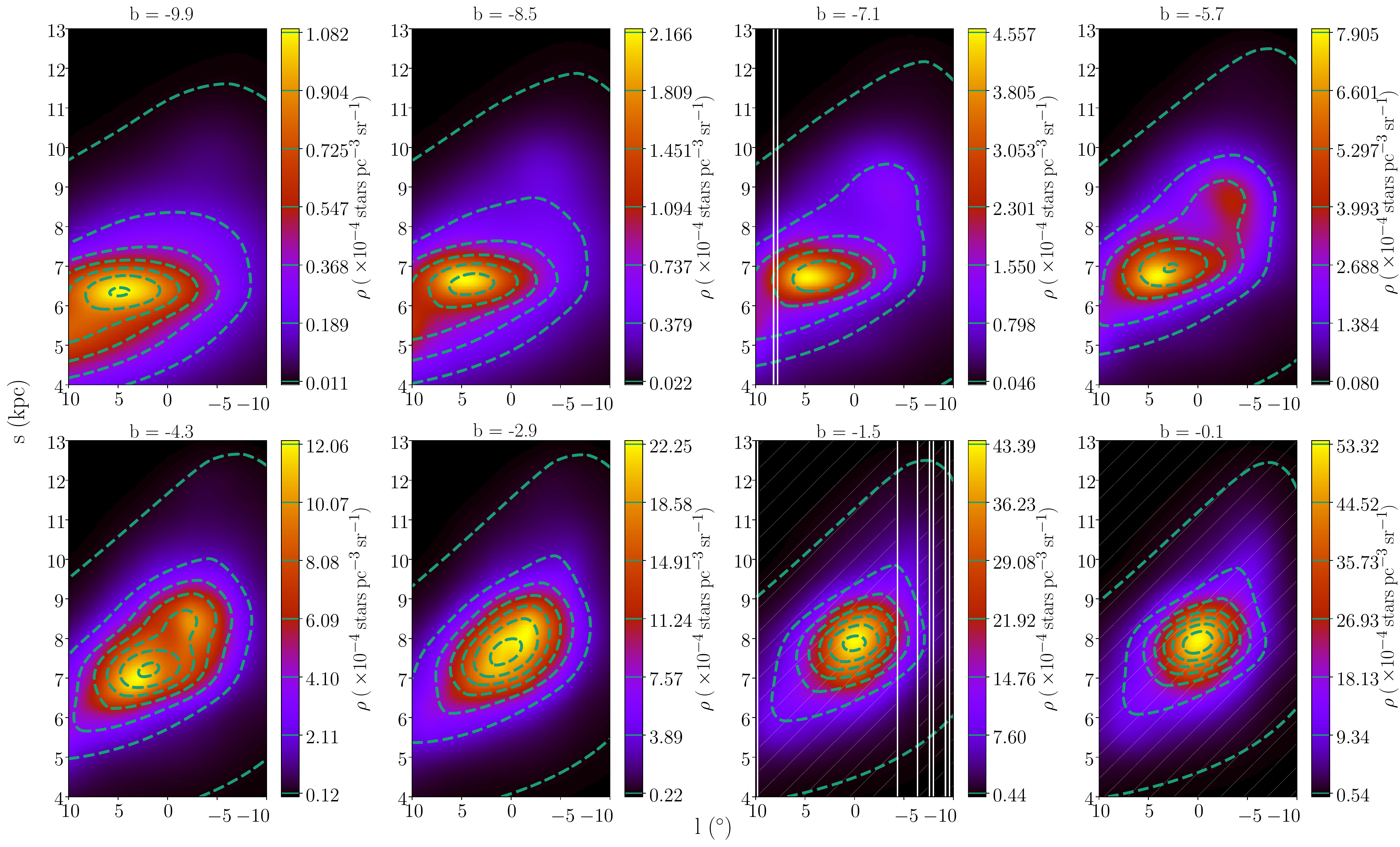}
    \caption{ Deconvolved RC+RGBB star density using the maximum entropy method. White hatched regions were masked during the analysis, and were inpainted as part of the deconvolution. The prior density model is shown in green dashed contours. 
   }
    \label{fig:MEMdensity}
\end{figure*}

\begin{figure*}
    \centering
    \includegraphics[width=0.9\textwidth]{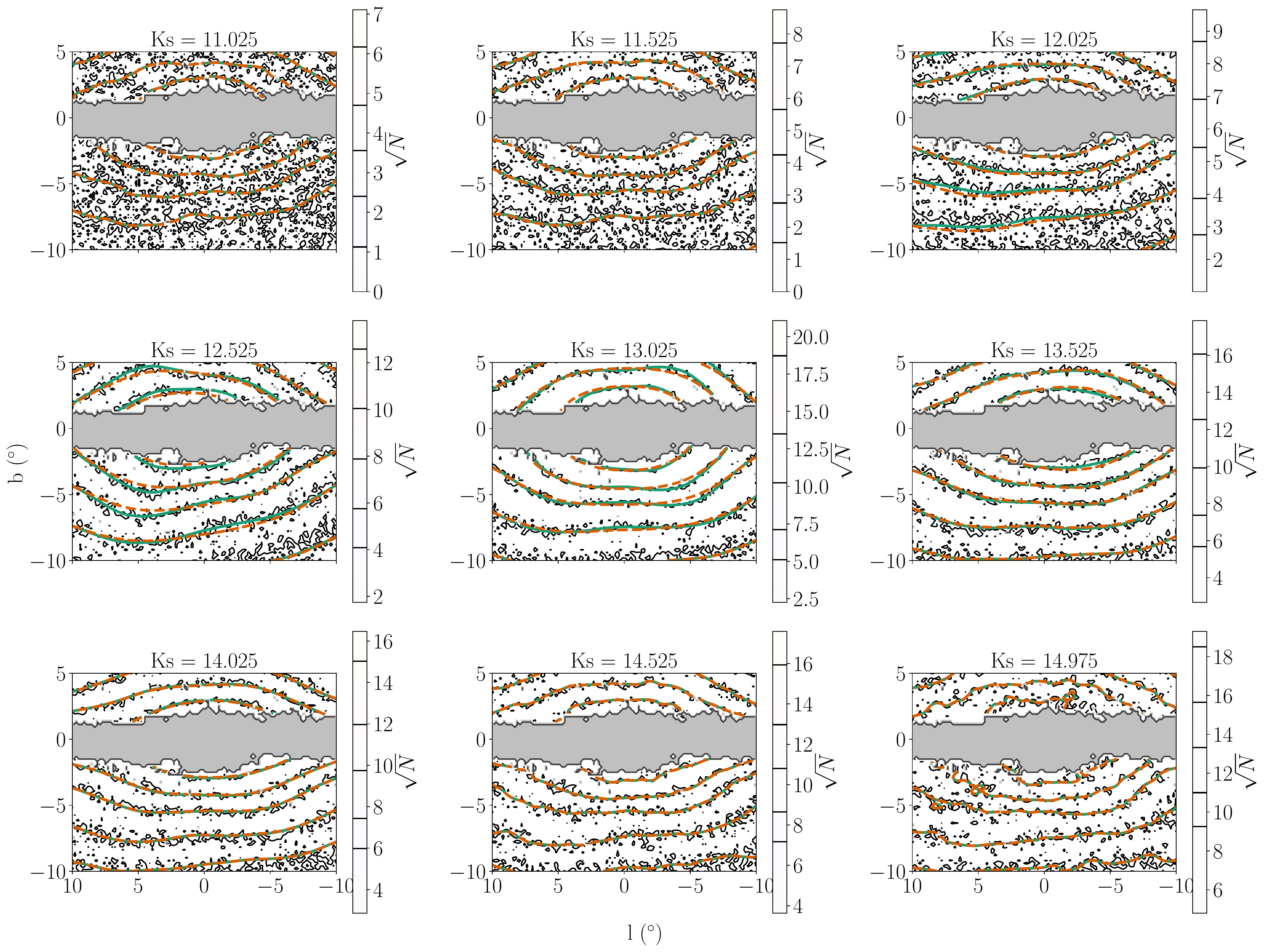}
    \caption{Predicted star counts for our maximum entropy deconvolution method. Black contours show the VVV star counts, where the levels of the contours are indicated by black lines on the colour bar. Green contours show the star counts predicted by the non-parametric model, where the levels match the black contours. The orange dashed line is the parametric model used as the prior. \newtext{Contours are produced using the same resolution as the data, i.e.\ $0.2^\circ \times 0.2^\circ$ degree grid with $K_s$ bins of width 0.5 mag.}}
    \label{fig:MEMmodelcontours}
\end{figure*}

\begin{figure*}\label{fig:datasystematics}
\centering
    \includegraphics[width=0.7\textwidth]{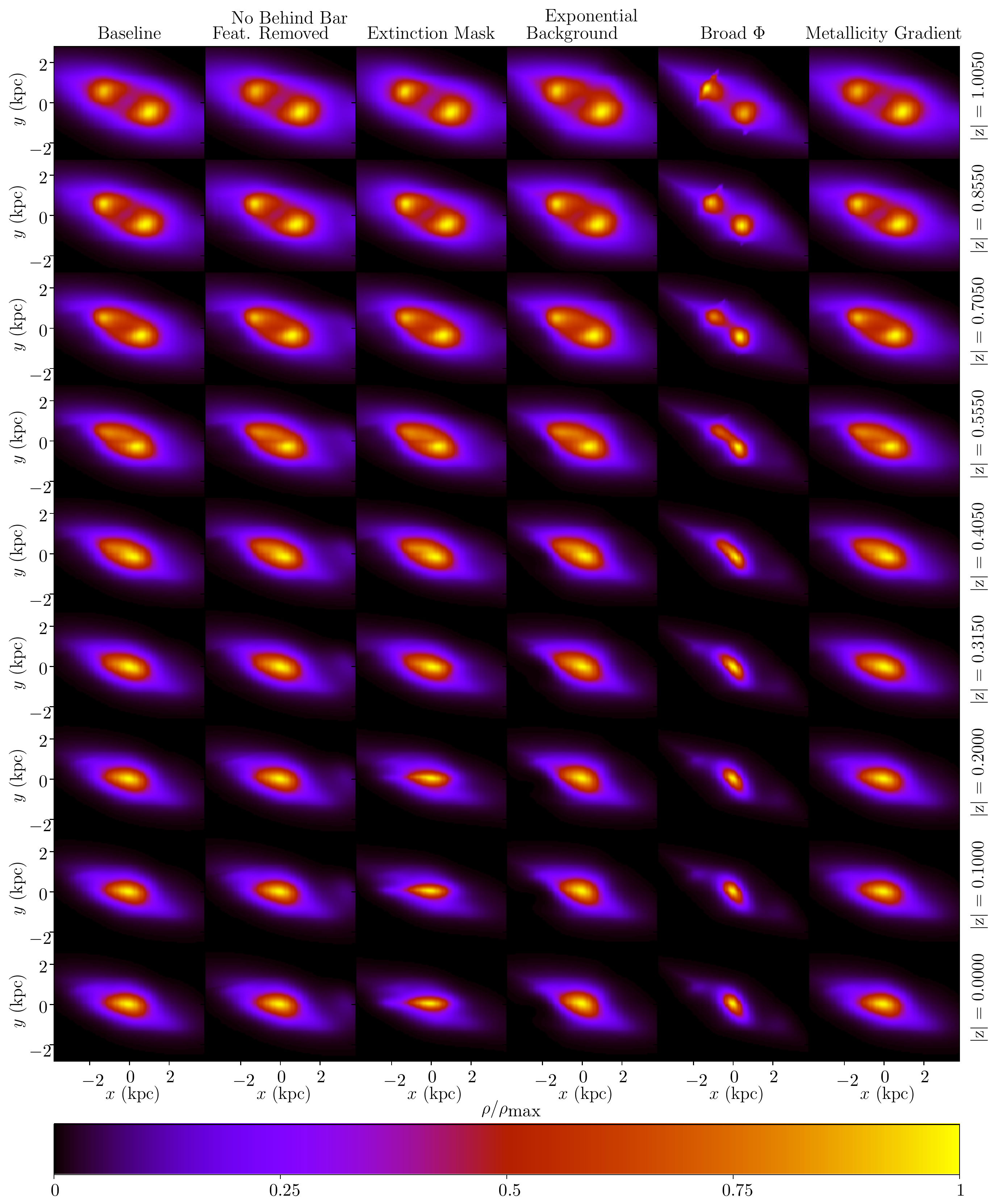}
    \caption{Cartesian projections of the deconstructed density of the VVV RC stars in the bulge, for several systematic test cases. The $x$ axis is aligned with the Sun-Galactic centre line and the $z$-axis is perpendicular to the Galactic plane and measured in kpc. The Galactic centre is located at the maximum bulge density. The significance of each test case is discussed in the text in Section \ref{sec:Systematics}}
\end{figure*}

\begin{figure*}\label{fig:simulationsystematics}
\centering
    \includegraphics[width=0.7\textwidth]{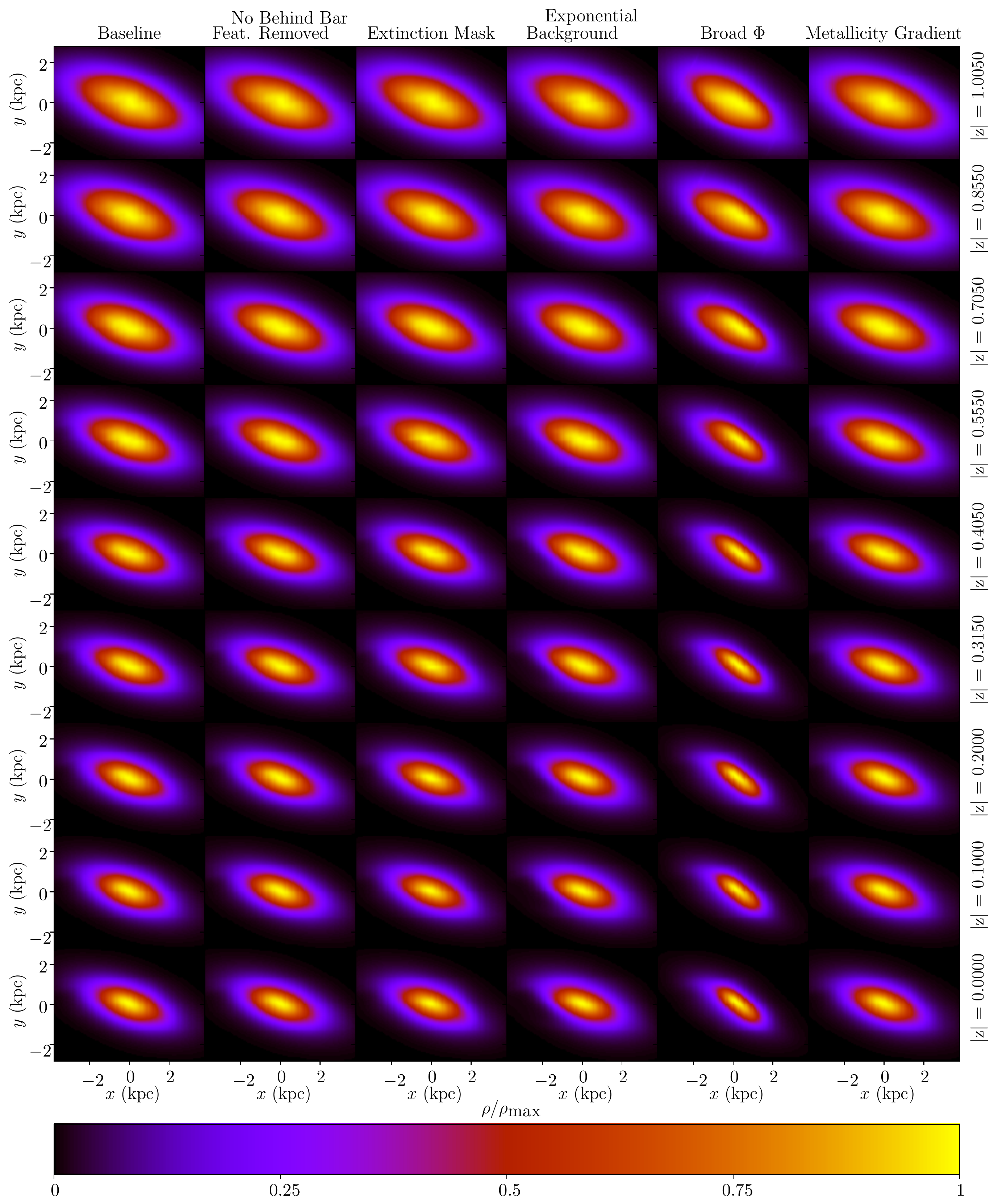}
    \caption{Cartesian projections of the deconstructed density of the simulated bulge population, for several systematic test cases. The $x$ axis is aligned with the Sun-Galactic centre line and the $z$-axis is perpendicular to the Galactic plane and measured in kpc. Nearly all of the cases give a qualitatively similar density to the base case.  However, the exponential background gives densities that are too low at $(x,y) = (-2.5\mathrm{ kpc},0\mathrm{ kpc})$, especially at low latitudes.
    Also, the broadened luminosity function gives a larger bar angle than the base case. The two exceptions noted here are also seen in the VVV data (Fig. \ref{fig:datasystematics}).
    }
\end{figure*}

The deconvolved density and the fitted parametric density, for fixed latitude bins, are shown in Fig. \ref{fig:MEMdensity}. For compactness, only 9 of the 50 bins are displayed and only for $b < 0^\circ$, as the density is symmetric about $b$. Unlike the simulated bulge shown in Fig.~\ref{fig:MEMsimdensity}, the density from deconvolution of the VVV data shows the arms of the X-bulge, first noticeable at $b=-8.7^\circ$ for $(l,s) =(4.7^{\circ}, 6.6$ kpc) and $(l,s) =(-3.3^{\circ}, 9~{\rm kpc})$. As latitude decreases, the arms get closer until they merge at $b=-2.7^\circ$. The maximum density at $b=-2.7^\circ$, where the arms merge, is at longitude $l=-0.7^{\circ}$. The maximum density of the X-bulge arms in the parametric model do not align with the maximum density in the non-parametric model, which is also evident in the star counts. Cartesian versions of the reconstructed bulge from the VVV data and the simulation are shown in the first columns of Figures~\ref{fig:datasystematics} and \ref{fig:simulationsystematics} respectively.


\section{Systematic Tests}\label{sec:Systematics}
In order to gain a better understanding of the robustness of our results we test systematics based on the following:
\begin{itemize}
\item Adding the feature behind the bar to the background (case B).
\item The VVV data mask (case J).
\item The determination of the background component (case C).
\item The semi-analytic luminosity function (case D and I).
\item The metallicity distribution (case E).
\item The position of the Sun (case F, G, H, I).
\item The deconvolution method used (Appendix \ref{sec:Richardson-Lucy}).
\end{itemize}
 We tested the significance of these assumptions by systematically changing one, then repeating the maximum entropy deconvolution, including the background fitting and parametric prior density model fitting. We also repeated the deconvolution with the new assumptions on the simulated population.


\begin{figure}\label{fig:TS}
\begin{center}
    \includegraphics[width=0.9\columnwidth]{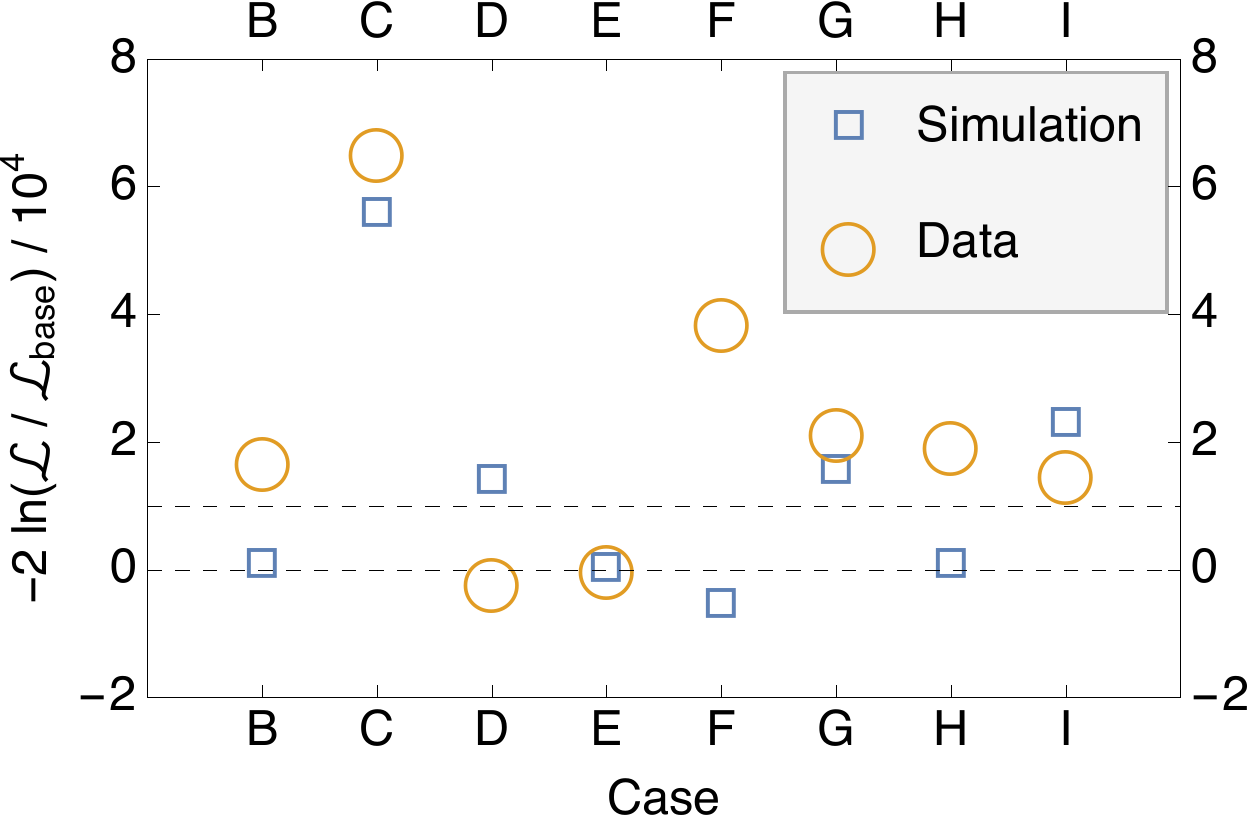}
    \includegraphics[width=0.9\columnwidth]{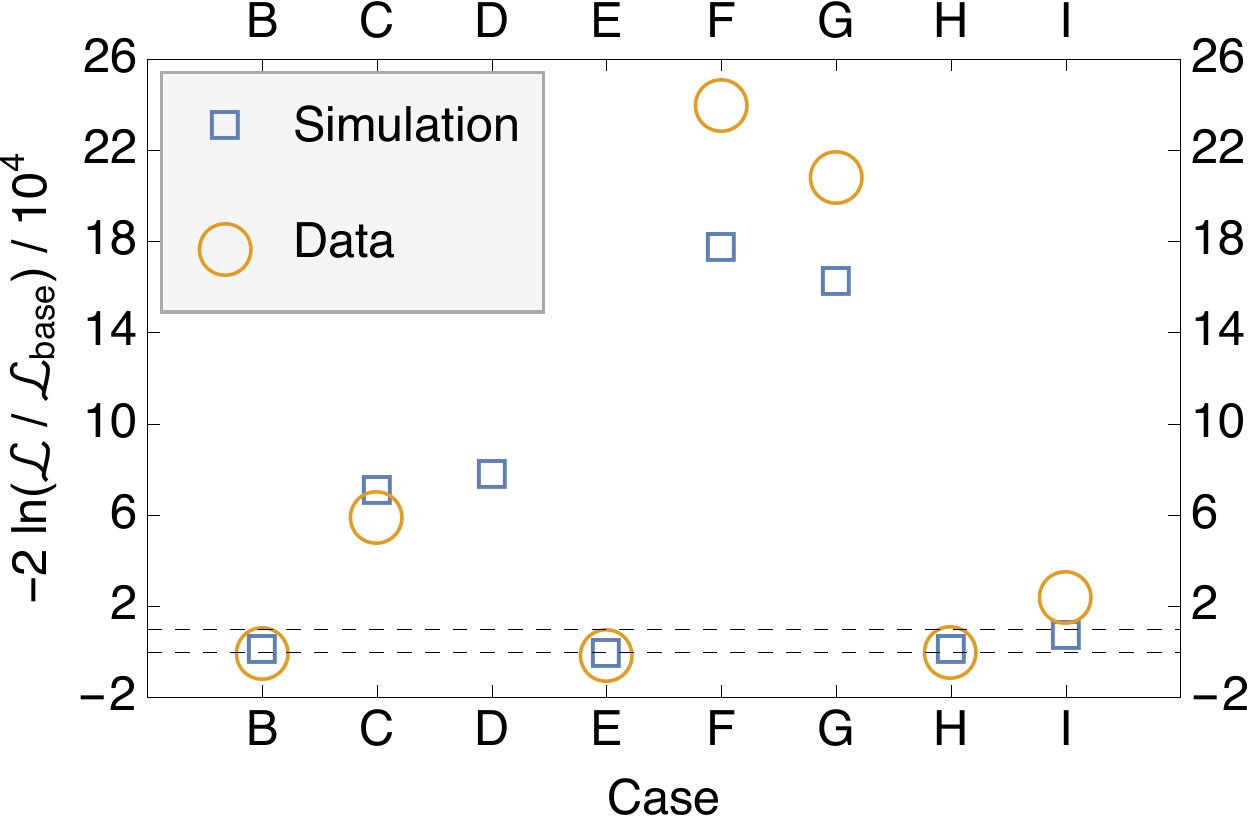}
\end{center}
    \caption{The parametric (top panel) and  non-parametric (bottom panel)  likelihood ($\cal L$)  for the different cases considered.
    The base case's non-parametric likelihood is ${\cal L}_{\rm base}$.
    Results are shown for both the simulations and the data. In the simulation case the base case and labelled case are both fit to the simulated data. In the bottom panel, Case D's data symbol is not shown due to it's very low likelihood value not being in the range of the plot for the non-parametric fit. 
    See Table~\ref{tab:TSmaxentandparam} for  numerical values \newtext{including the one for Case D in the non-parametric fit.
    The dashed lines are for  ${\rm TS}=0$  and the median of the simulations for both the parametric and non-parametric case which is ${\rm TS}\approx 10^4$.
    }
    The cases considered  are: no behind-the-bar feature subtraction (B), exponential background (C), broad luminosity function (D), metallicity gradient (E),   S-model prior with $Z_{\odot}=15$~pc (F), S-model prior and broad luminosity function with $Z_{\odot}=15$~pc (G),  S-model prior with $Z_{\odot}=0$~pc (H), S-model prior with $Z_{\odot}=0$~pc with a broad luminosity function (I).
    \label{fig:results}
    }
\end{figure}

\begin{figure*}
	\centering
	\includegraphics[width=\linewidth]{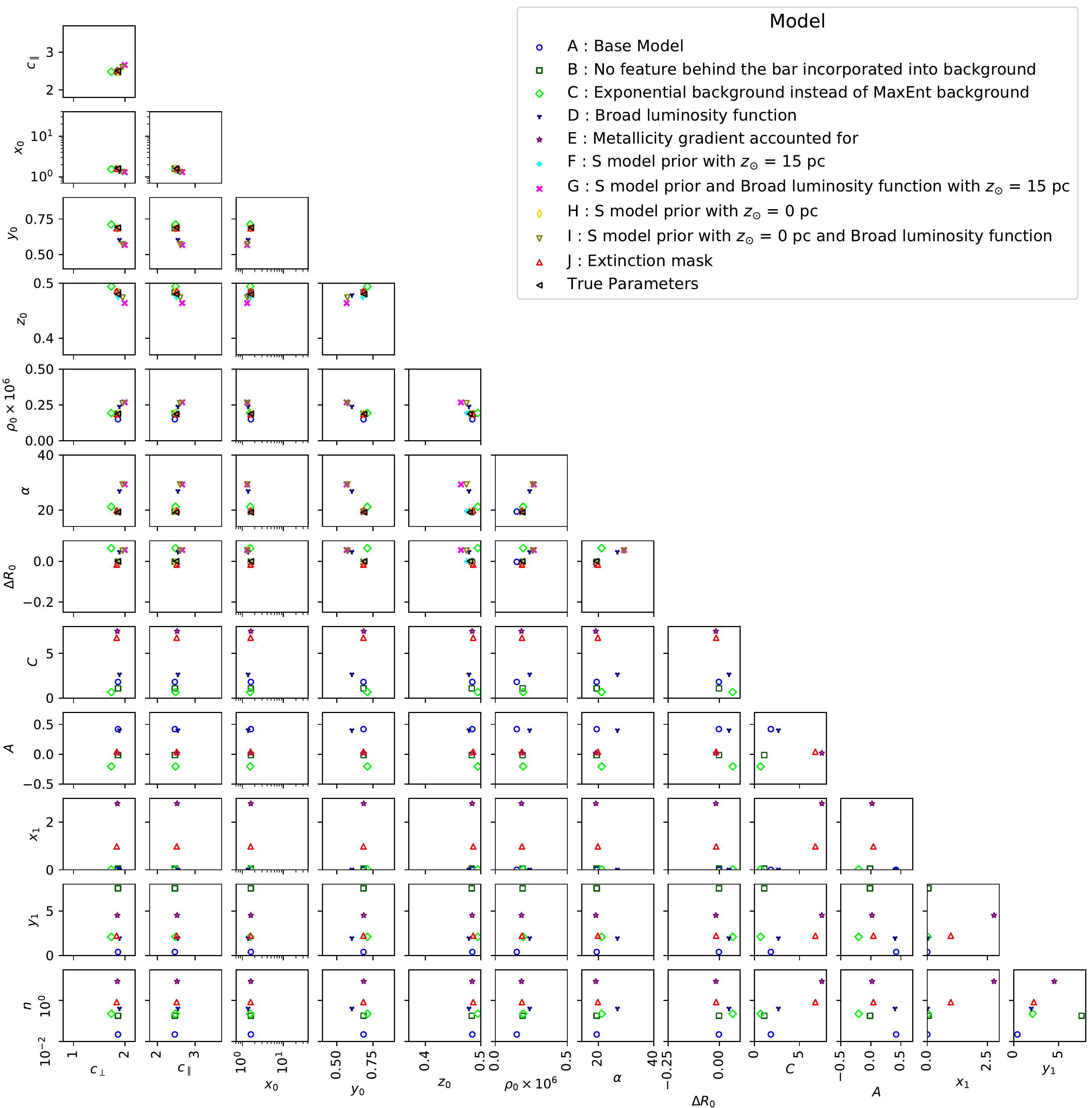}
	\caption{Pair plot of parametric model parameters fitted for the base case and systematics to simulations. See Table~\ref{tab:Par_Sims_Table}.
	The $n$ parameter has been plotted on a logarithmic scale.
	}
	\label{fig:parampairplotsims}
\end{figure*}

The results of fitting the SX model to data and simulations are listed in Tables \ref{tab:Par_Data_Table} and \ref{tab:Par_Sims_Table}, and are plotted in Figures \ref{fig:TS}, \ref{fig:parampairplotsims}, and  \ref{fig:parampairplotdata}. Except where specified, the parametric model has been fitted twice, following the prescription of the deconvolution method in Section \ref{sec:Method}, in which the feature behind the bar is subsumed into the background.
By fitting to the S-model simulation generated by the parameters in Table \ref{tab:bulgedensity}, we hoped to gauge the impact on the likelihood of different background and parametric model cases used in bulge modelling.
Note that in the simulation, we chose $Z_{\odot}=15$~pc.
As can be seen in Fig.~\ref{fig:parampairplotsims}, the range of fitted model parameters is much greater than the error bars in Table~\ref{tab:Par_Sims_Table}. This indicates the main cause of the variation is due to model assumptions rather than statistical error.
 We used the following test statistic (TS) to compare the different cases:
\begin{equation}
    {\rm TS}\equiv-2\ln({\cal L}/{\cal L}_{\rm base})
\end{equation}
As most of the variation between cases was due to systematic  error rather than statistical error, we did not use Wilks' theorem \citep{Wilks1938}
which is also only suited for nested models.
\newtext{This means we cannot associate the TS value with a p-value in the usual way.
%
We can get a rough measure of what a significant TS value is by comparing to the corresponding TS values seen in simulations. The median value of the simulation TSs for the combined top and bottom panels of Fig.~\ref{fig:results} was $TS\approx 10^4$. We take this as our threshold above which the TS value is regarded to be significant.}


 

\subsection{Feature Behind the Bar}
As can be seen both in the top and bottom panels of Fig.~\ref{fig:TS},
the simulation has a negligible TS
when testing against case B which does not account for a feature behind the bar.
This is to be expected as this feature was not present in the simulation. In contrast, for the parametric fit (top panel), the data has a high TS for case B. This indicates that the feature behind the bulge is significant. In P19 we analysed this case in more detail. Also, in that paper we used a parametric background which then also revealed a feature in front of the bulge. The non-parametric background in this paper has absorbed the feature in front of the bar.

However, In the non-parametric case (bottom panel) we do not find a significant change in our penalized likelihood when not removing the feature behind the bar. This can be seen in the bottom panel of Fig.~\ref{fig:results} where case B has a TS very close to zero for both the data and simulation. This is to be expected as the flexibility of the non-parametric method can easily incorporate the feature behind the bar as being part of the bulge as seen by comparing column 1 and 2 in Fig.~\ref{fig:datasystematics}. While for the simulation, where there should be no feature behind the bar, the corresponding columns are virtually indistinguishable as seen in Fig.~\ref{fig:simulationsystematics}.

\subsection{Background Systematics}
We changed the background in case C to one that is common in the literature, a second order polynomial in $\log(N)$, described in Section \ref{sec:Richardson-Lucy}. We have already displayed this background for a couple of lines of sight in Fig. \ref{fig:MEMmodellosbg}. At high latitudes (top panel), this background tends to estimate higher counts than the maximum entropy background for $12 < K_s < 12.5$ and estimate fewer counts at $13 < K_s < 14$. At lower latitudes, this background tends to overestimate at all $K_s$, especially at around $K_s =12.0$. On the simulation, the exponential background significantly over estimates in the range, $11.7 < K_s < 13.0$, as shown in the top panel of Fig. \ref{fig:MEMmodellossim}. As a result, the density is underestimated on the near side ($x<0$) of the bulge at low latitudes when using the exponential background rather than the maximum entropy background in both the VVV data (Fig. \ref{fig:datasystematics}) and simulated population (Fig. \ref{fig:simulationsystematics}). 

In Fig.~\ref{fig:TS}, for the parametric fit (top panel), the exponential background (case C) has the worst TS both for the data and simulation, out of all of the cases considered in that panel. 
The TS was also high for both the data and simulation in the non-parametric case as shown in the bottom panel of Fig.~\ref{fig:results}. This provides further evidence that the maximum entropy method is providing a better background than exponential model approach.

\subsection{Luminosity Function Systematics}
\begin{figure*}
	\centering
	\includegraphics[width=\linewidth]{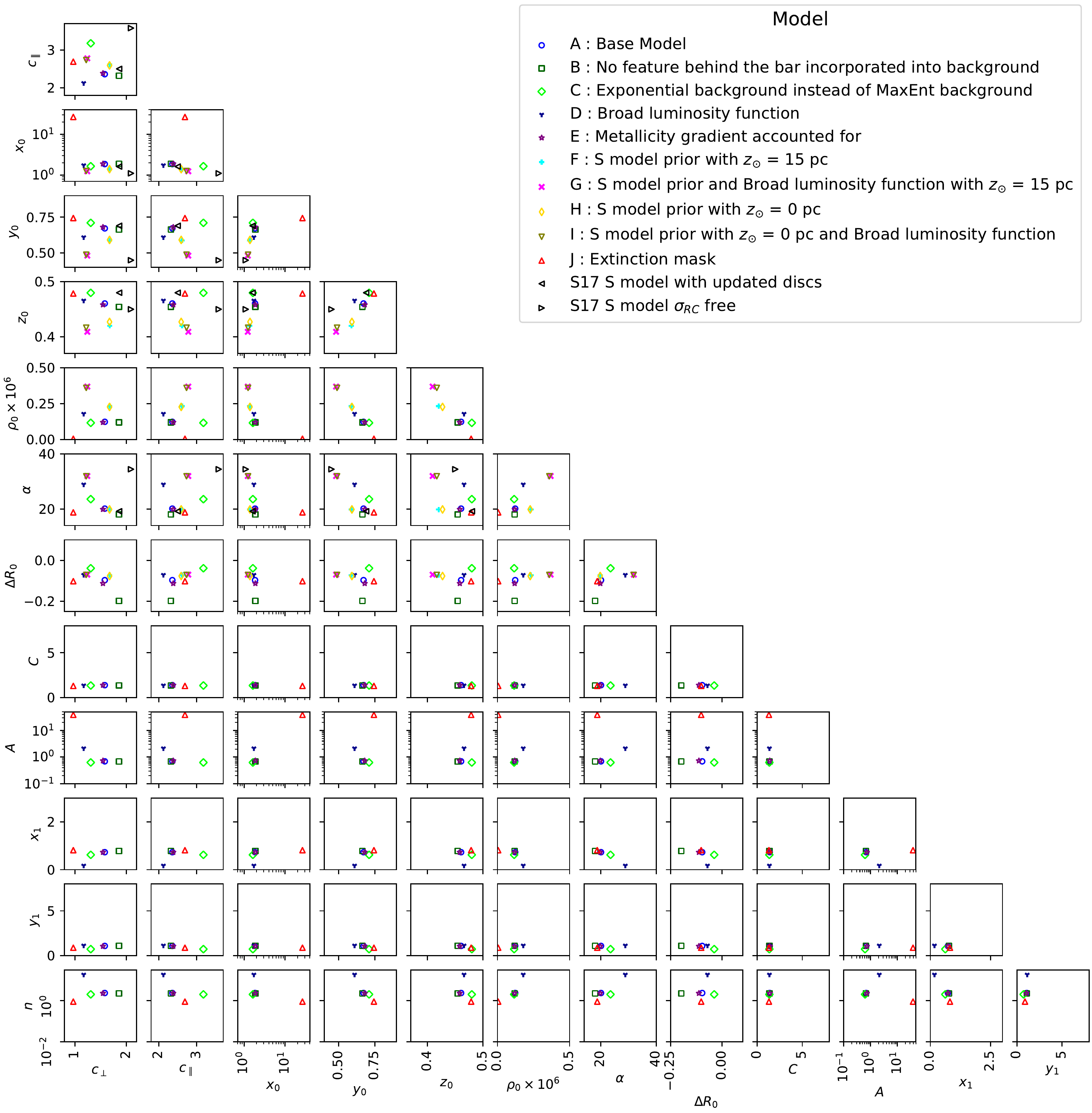} 
	\caption{Pair plot of parametric model parameters fitted for the base case and systematics on the VVV data. Note that the axis scaling for parameters $x_0$, $A$, and $n$ are logarithmic. See Table~\ref{tab:Par_Data_Table}. 
}
	\label{fig:parampairplotdata}
\end{figure*}

\begin{figure}\label{fig:broadluminosity functionsystematics}
\centering
    \includegraphics[width=0.8\columnwidth]{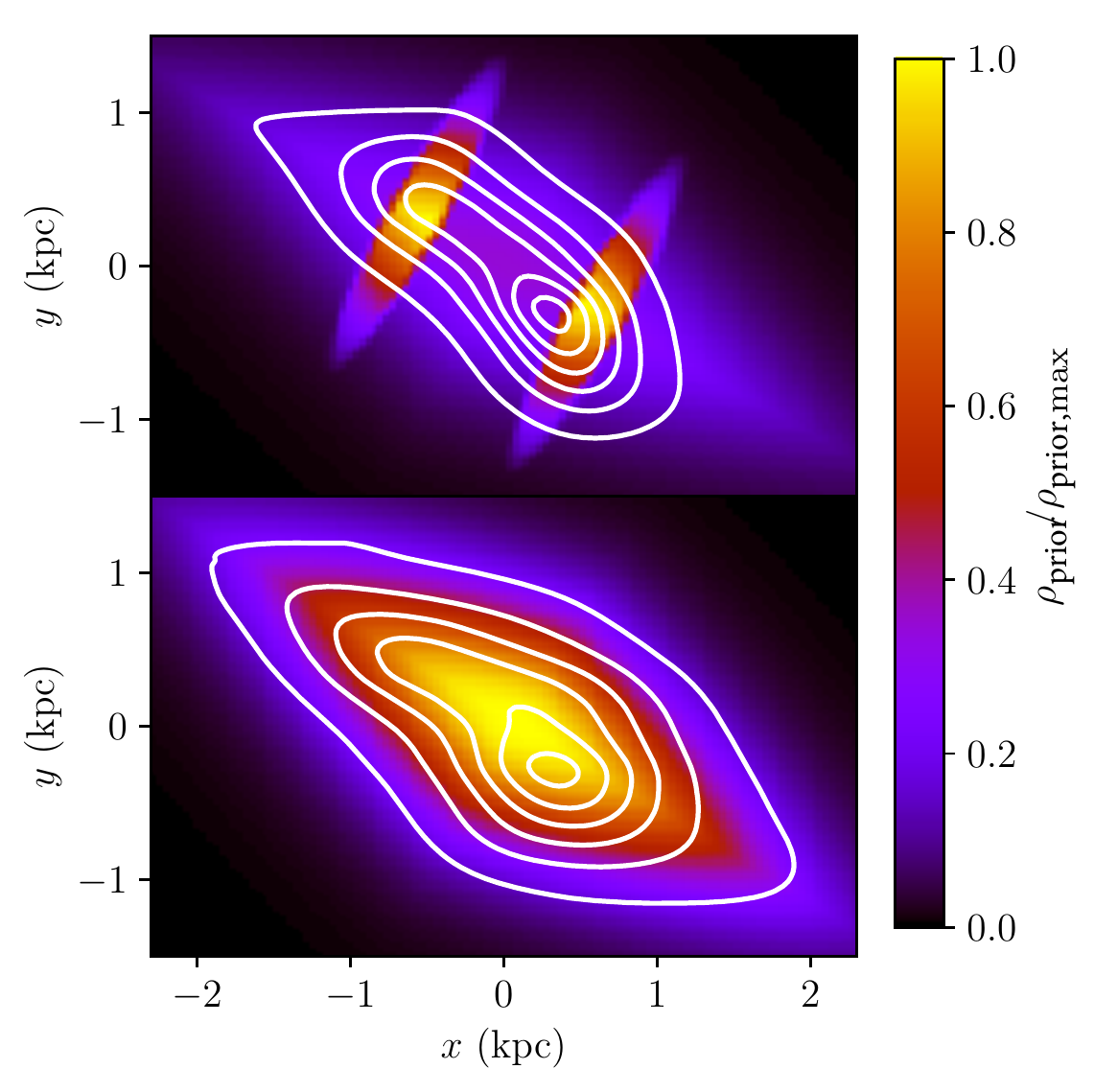}
    \caption{SX (top) and S (bottom) parametric density models at $z = 0.495\, \mathrm{ kpc}$, fitted to the VVV data using the 10 Gyr bulge Parsec derived luminosity function which has been convolved with a Gaussian with $\sigma=0.24$
     (case D). 
     They are used as the prior models for a non-parametric fit.
     The broadened luminosity function has driven the X-component to be unnaturally contrasting to the S component, which necessitates the non-parametric model (white contours) modulate significantly from the prior density. By contrast, the S-model is still largely visible in the non-parametric solution, with the modulated X-bulge arms visible at $x=\pm0.5$ 
    }
\end{figure}

S17 found that the best-fitting luminosity function was significantly broader than the luminosity function they had simulated with \textsc{galaxia} \citep{SharmaGalaxiaCodeGenerate2011}, using the same isochrones we have used in our analysis. We also tried a similarly broad luminosity function, by convolving our luminosity function \newtext{(of approximate Gaussian width 0.06 mag)} with an additional Gaussian with a standard deviation of 0.24 mag. The density slices in the "Broad $\Phi$" column of Figures~\ref{fig:datasystematics} and  \ref{fig:simulationsystematics} are consistent with the broadened luminosity function requiring a narrower and more angled bulge. A similar  relationship can be seen in Fig.~16 of S17. 
In the top panel of Fig.~\ref{fig:TS}, 
the SX parametric model with broadened luminosity function (case D) had a slightly improved TS for the data, while it was disfavoured for the simulation. However, this broader luminosity function is not consistent with recent measured intrinsic RC magnitude dispersions \newtext{in the $K_s$ band of 0.03-0.09 mag} \citep{HallTestingasteroseismologyGaia2019,ChanemphGaiaDR2parallax2019}.
Also, in Fig.~\ref{fig:parampairplotdata}, the X-shape parameters, $n$ and $x_{1}$, are anomalous for case D. The consequence of this was that the broader luminosity function fit resulted in
unnaturally narrow X-arms as depicted in Fig.~\ref{fig:broadluminosity functionsystematics}.
As can be seen in the non-parametric results shown in the bottom panel of  Fig.~\ref{fig:results}, the broader luminosity function (case D) provided a high TS for the simulations indicating a bad fit. This is to be expected as the simulations were based on our standard narrower luminosity function. The TS for the data was so high for the broad luminosity function that we could not accommodate it in Fig.~\ref{fig:results} without making the range of the plot too great to see any of the other details. This was because the non-parametric model was being heavily penalised for deviating greatly from the prior SX model, which had converged to a physically unnatural solution, shown in the top panel of Fig. \ref{fig:broadluminosity functionsystematics}. 

Since our prior for the maximum entropy deconvolution was unnatural for the broad luminosity function, we wanted to check if a different prior gave similar results. So we repeated the test, but instead we used an S-model as the prior density, shown in the bottom panel of Fig. \ref{fig:broadluminosity functionsystematics}.
As can be seen in the top panel of Fig.~\ref{fig:TS}, this S-model with a broad luminosity function (case I) was disfavoured by both the data and the simulation for the parametric case.
Also, as presented in the bottom panel of Fig.~\ref{fig:results}, 
case I did have a significant TS for the non-parametric fit in the case of the data.
This indicates that from a TS perspective, our non-parametric results disfavour a broad luminosity function.

\subsection{Metallicity Distribution Systematics}
\label{sec:metalicity}
\begin{figure}\label{fig:metallicitygradient}
\centering
    \includegraphics[width=0.9\columnwidth]{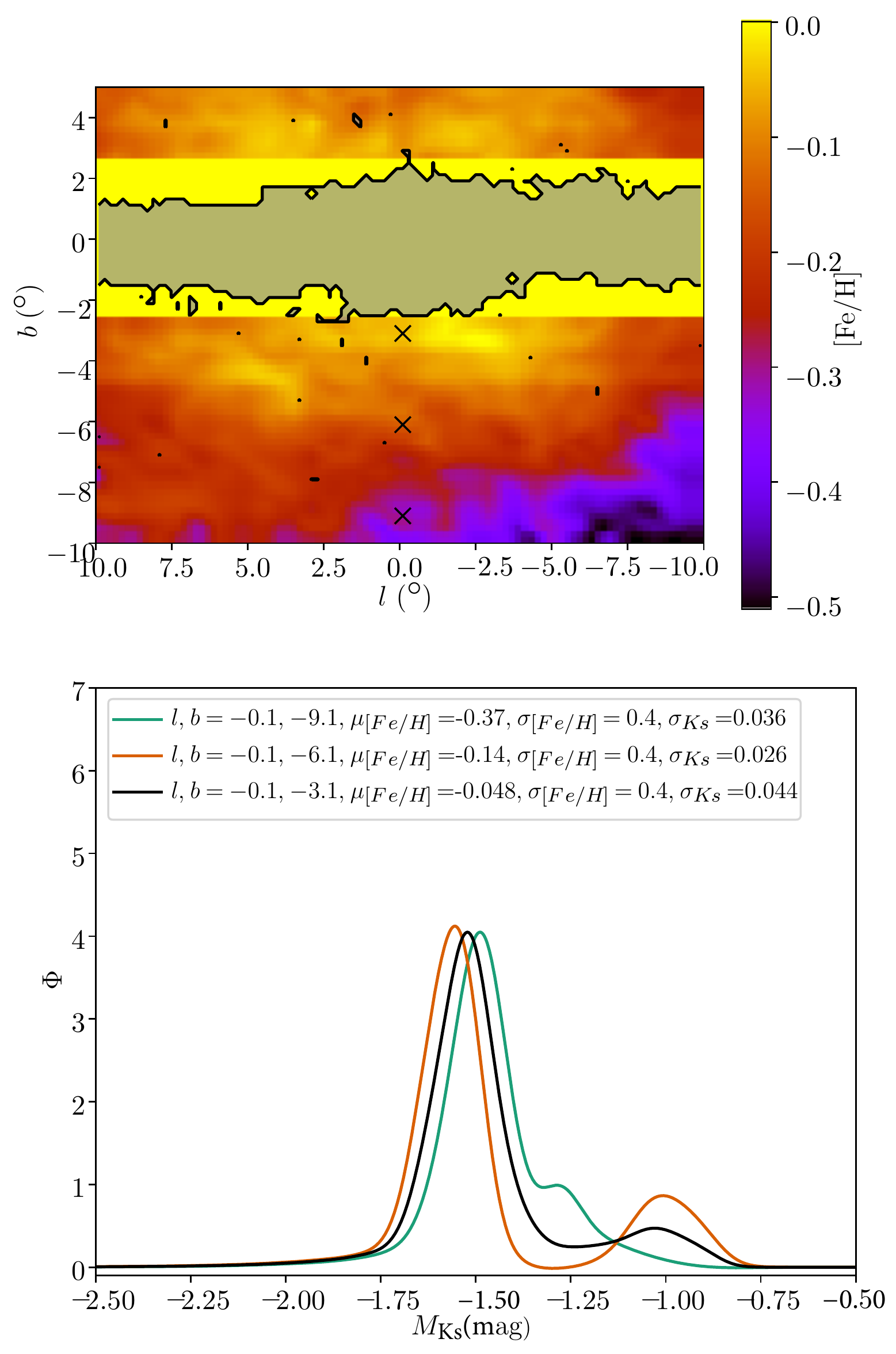}
    \caption{\textit{Top:} Mean photometric metallicity map, [Fe/H] of \protect\cite{Gonzalez2018StructureBehindBar}. Where the map does not have coverage at $|b| < 2.6$ we assume the fiducial value $[Fe/H]=0.0$. The black crosses indicate the locations of the three luminosity functions plotted in the bottom panel. \textit{Bottom:} 
    The RC+RGBB
    luminosity functions for a range of fields of view, assuming a metallicity distribution as in the above panel. They have been convolved with a Gaussian with dispersion $\sigma$, the photometric uncertainty. In order of increasing metallicity, the mean absolute magnitude of the RC is -1.49 mag, -1.51 mag and -1.52 mag. }
\end{figure}

Our base case assumed that the metallicity distribution is constant throughout the bulge. Several spectroscopic studies, e.g. \cite{Zoccali2017GIBS} and \cite{Garcia2017APOGEEBulgeMetallicity}, have observed a vertical metallicity gradient in the bulge, where stars near the Galactic midplane are on average more metal rich than stars on the periphery of the bulge. We used the photometric metallicity map generated by the BEAM-II calculator \citep{Gonzalez2018StructureBehindBar} to allow the metallicity distribution function in the computation of our semi-analytic luminosity function to have a different mean metallicity for every line-of-sight. The metallicity dispersion was kept fixed at 0.4 for this test.
Shown in Fig. \ref{fig:metallicitygradient} (top panel) is the metallicity map of \cite{Gonzalez2018StructureBehindBar}, where we have filled the missing values with [Fe/H] = 0.0. From the luminosity functions in bottom panel of Fig. \ref{fig:metallicitygradient}, it is clear that the lower metallicity line-of-sight has a fainter RC, and is naturally broader, though the difference in brightness is only 0.03 mag between $b=-9.1$ and $b=-3.1$. Some part of the broadness is from the overlapping of the RC and RGBB, since the RGBB is brighter at lower metallicities. Qualitatively, the density which was fitted using the metallicity gradient is nearly identical to the  base case as seen in the last column of  Fig. \ref{fig:datasystematics}. 
The insensitivity to the metallicity gradient 
can be seen in case E for the bottom and top panel of Fig.~\ref{fig:TS}.
The TS changes for the metallicity cases are negligible in comparison to the TS changes associated with the other systematics.
The E case does appear to have an anomalous $x_1$ in Fig.~\ref{fig:parampairplotdata}. However, as $A\approx 0$ for the $E$ case, its X-component is negligible. 
We conclude from this test that the inclusion of a simple unimodal metallicity gradient does not significantly affect our results. A more sophisticated double population model, consistent with spectroscopic observations, is necessary to properly include a metallicity gradient.

\begin{figure}\label{fig:centraloverdensity}
\centering
\includegraphics[width=0.85\columnwidth]{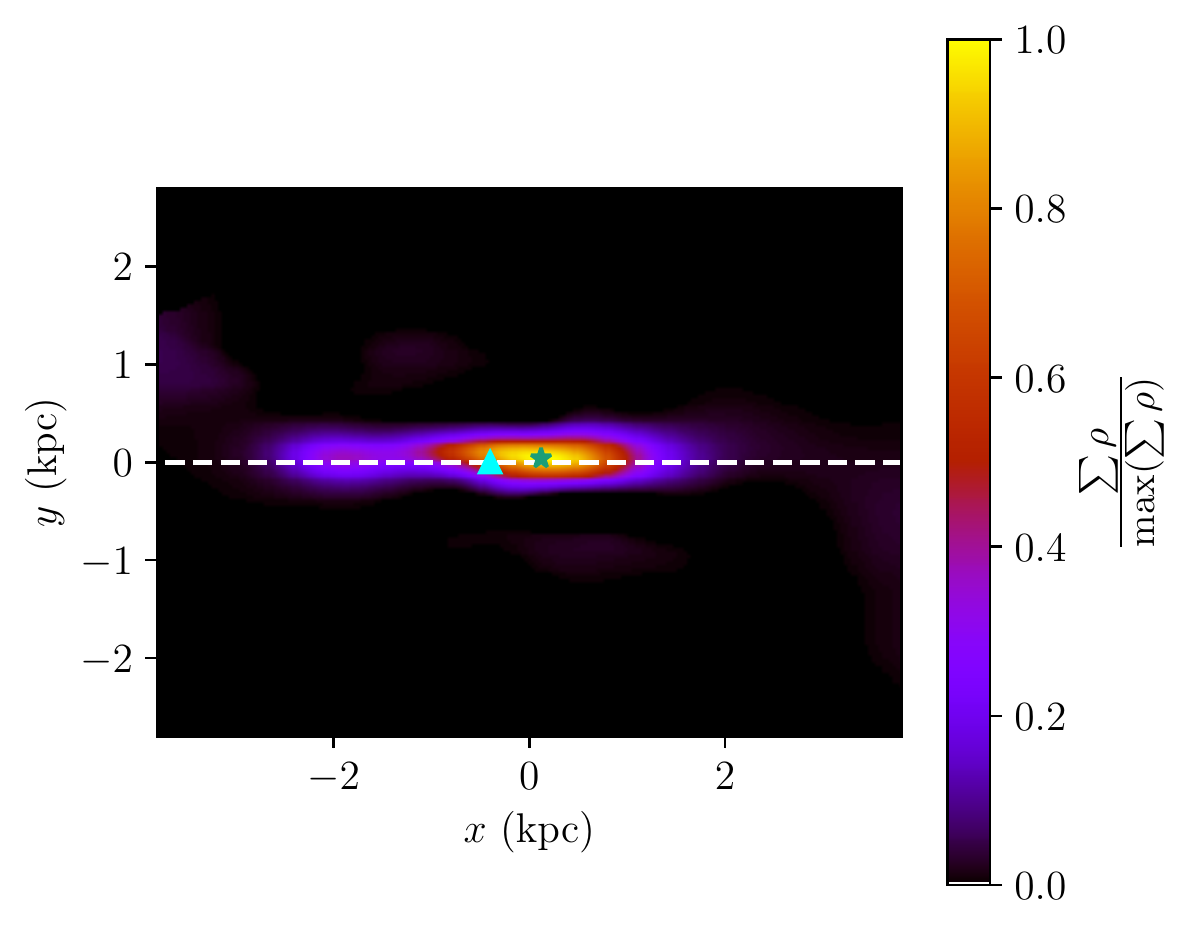}
    \caption{Difference between the deconvolved density using a crowding + extinction based mask and a extinction only mask in Cartesian co-ordinates where $x$ is aligned with the Sun-Galactic centre line. The density difference  has been summed over $|z < 1\,\mathrm{kpc}|$. The white dashed line indicates $l=0^{\circ}$. The maximum density of the difference (indicated by a green star) is 150 pc behind the maximum density location of the crowding + extinction based mask reconstructed bulge. The cyan triangle is at the expected maximum density location for a population which would have an RC 0.1 mag brighter than our PARSEC derived semi-analytic luminosity function, such as a 5 Gyr old population or a more metal rich population.}
\end{figure}

\subsection{Sun Position Systematic}

Our simulated population of stars had the Sun located at $Z_{\odot} = 15$~pc, which is different to the $Z_{\odot} = 0$~pc assumed in our base model. We tested the significance of this assumption by fitting an S-model with the Sun in the same position as in our simulation (case F). We still assumed symmetry in the maximum entropy density about $b=0^{\circ}$.
The top panel of 
Fig.~\ref{fig:TS} shows how parametric case F  provided an  improved fit to the simulation. This is to be expected as it corresponds with the model used to generate the simulation.
In the case of the VVV data, it is harder to interpret the case F result in Fig.~\ref{fig:TS} as we have changed both the position of the Sun and the parametric form of the prior density. The difference between case F and case H is the position of the Sun, where both differ from the base case by having an S-model parametric form.
The VVV data TS of case F was significantly larger than case H in the parametric case, however, there was less of a difference when fitting the parametric model to the simulation.
This confirms that the VVV data prefers $Z_{\odot}=0$~pc when fitting the parametric S-model as seen in the top panel of Fig.~\ref{fig:results}. 
When comparing the same cases, F and H, for the non-parametric method, case F had a significantly larger TS than case H for both the simulated population and the VVV data.
It is hard to interpret this result for the non-parametric model, given that it had an assumed symmetry around the $Z_{\odot}=0$~pc plane. However, we relaxed this assumption in P19 without issue.
 %


Case H is an S-model with $Z_\odot=0$~pc.
As can be seen from the top panel of Fig.~\ref{fig:TS}, for the parametric fit, the  data significantly prefer the SX model.
Also, for the parametric fit, the F case is very slightly favoured over the SX model for the simulation.
This follows in that the F case is of the same form as the model used to generate the simulation. However, case F is even more disfavoured by the data than case H.
From this we conclude that, for the parametric fit, the data favours the SX model over the S-model and this conclusion is not affected by reasonable changes in $Z_\odot$.

\subsection{Mask Systematic}
We changed the region in which the data is excluded, from the combined extinction and $K_s$-band uncertainty boundary case ($\sigma > 0.06$), to a colour excess mask $E(J-K)>0.9$.
This systematic test changes the amount of data used in the analysis, so the likelihood is not comparable to the base case. In Fig. \ref{fig:datasystematics}, the density that is reconstructed with an extinction only mask has a prominent bar-like feature at $|z| < 0.2\,\mathrm{kpc}$, that is pointed nearly directly towards the Sun. Note, that this feature is not seen in the corresponding simulation result of Fig.~\ref{fig:simulationsystematics}. We extracted this feature by subtracting the baseline case.  Plotted in Fig. \ref{fig:centraloverdensity} is the sum of the density difference for all density with $|z| < 1\, \mathrm{kpc}$.
At first glance, this apparent over-density looks similar in structure to the younger, secondary population of bulge stars in S17 (E component of the S+E model). The green star indicates the maximum density of the difference and is located at $(x,y) = (120\,\mathrm{pc},90\,\mathrm{pc})$. This is 150 pc behind the centre of the bulge ($(x,y) = (0\,\mathrm{pc},0\,\mathrm{pc})$). This suggests that the stars are unlikely to be from a significantly younger or more metal rich population than the rest of the stars in our bulge model, as they would have a brighter RC in the luminosity function than we have modelled.  A 5 Gyr old population with a similar metallicity distribution to our fiducial case has an RC which is 0.1 mag brighter, which corresponds to a difference of 400~pc closer at 8~kpc, indicated by the cyan triangle on Fig. \ref{fig:centraloverdensity}. 

We argue based on the reconstructed distance from the Sun, that the apparently over-dense region is not consistent with a different population of stars. Its orientation, which is suspiciously pointed directly towards the Sun, and is distinctly different from the majority of the bulge population also makes it inconsistent with main population of the bulge stars. This was one of our motivations in using the crowding+extinction based mask over the extinction only based mask. A combination of significant crowding and residual extinction deteriorates the quality of the star count catalogues, including the photometric zero-point.

%
%
%
%

%
%
%
%
%

%
%
%
%
%
%
%
%
%
%
%
%
%
%
%
%
%
%
%
%

%
%
%
%
%
%
%
%

%

%

%

%

%

%
%
%
%
%
%
%

%
%
%
%
%
%

%
%
%
%
%
%
%

\section{Applications}
\label{sec:applications}
\subsection{Properties of the Bulge}
\subsubsection{Mass of the bulge}
From the fitted density and IMF we can estimate the total mass of the bulge. Integrating the RC+RGBB stellar density over the entire bulge region gives us a total of $19.1\times10^{6}$ (RC + RGBB) stars. Based on our luminosity function, 0.062 \% of all stars are in either the RC or RGBB, so the total number of stars in the bulge is $N_{\rm{total}} = 30.7 \times 10^9$. Stars in the bulge with a mass >1$M_{\odot}$ have evolved into stellar remnants, so the normalisation of the IMF is then given by
\begin{equation}
    \xi_0 = \frac{N_{\rm{total}} }{\int^{1\,M_{\odot}}_{0.15\,M_{\odot}} ~\xi(m)\, {\rm d}m},
\end{equation}
where $\xi$ is the IMF and $\xi_0$ is the normalisation of the IMF. We use the Chabrier IMF, which was also used to generate our luminosity function. With the IMF correctly normalised, the mass of the bulge is then calculated by integrating the IMF multiplied by the final mass of the star, over the range $0.15 M_{\odot} < m < 150 M_{\odot}$. Stars with an initial mass $<1M_{\odot}$ have not yet evolved into remnants, so the final mass is equal to the initial mass. Stars with initial mass $1M_{\odot} < m < 8M_{\odot}$ have evolved into white dwarfs, where the final mass is related to the initial mass by $m_f = 0.48 + 0.077m_i$ \citep{Maraston1998}. To determine the final mass stars with initial mass $>8M_{\odot}$, which have evolved into neutron stars or black holes, we use the results of the numerical population synthesis code \textsc{sevn} \citep{Spera2015}. Therefore, the total stellar mass of the bulge (assuming a Chabrier log-normal IMF) is $M_{bulge} = 1.64 \times 10^{10} \,M_{\odot}$. This includes the mass of the stellar remnants, which make up $30.1\%$ of the total mass.

Parametric modelling of VVV bulge stars in S17 found a total stellar mass of the bulge assuming a Chabrier IMF of $2.36 \times 10^{10} M_{\odot}$, with the stellar remnants making up 49\% of the total mass. Both the total mass and remnant fraction of S17 are larger than we are reporting. However, if we were to have the same remnant fraction as S17, then our total mass would be $2.24\times 10^{10} M_{\odot}$ which would be consistent with S17 once our systemic uncertainties have been incorporated. 

A dynamical estimate of the bulge mass by combining the VVV bulge stellar distribution of WG13 with kinematic information from BRAVA in \cite{PortailMadetomeasuremodelsGalactic2015} found a bulge stellar mass of 1.3-1.7$\times10^{10}M_{\odot}$, which is consistent with our estimated mass. They also provide a mass-to-clump ratio, which is used to estimate the total stellar mass of the bulge from the number of RC+RGBB stars. For a Chabrier IMF, there are approximately $905M_\odot$ of bulge mass for each RC+RGBB star. So for our estimated $19.1\times10^{6}$ (RC+RGBB) stars the estimated mass was $1.73\times10^{10}M_{\odot}$. This is remarkably similar to our value, considering \cite{PortailMadetomeasuremodelsGalactic2015} used different isochrones, metallicity distribution and treatment of the compact remnants to those used in our estimation. Additionally, we list the bulge mass estimates for all of our systematic test cases in Table \ref{tab:mass}. As can be seen, the mass estimates of the simulated data encompass the mass of the model used for the simulation with a spread of a few percent.
\newtext{As the systematic error is much greater than the statistical error, we use the range of best fit bulge mass estimates for our different cases to get an estimate of the uncertainty in our mass estimate.}
The mass estimates for the bulge from the VVV data are in the range 1.33-1.71 $\times 10 ^{10} M_{\odot}$, which is in agreement with the results of \cite{PortailMadetomeasuremodelsGalactic2015}.

\begin{table}[]
\caption{Total stellar mass estimate for the Galactic bulge for all test cases. A Chabrier IMF was assumed, which gave a remnant fraction of 30.1\% 
The cases considered  are: base (A), no behind-the-bar feature subtraction (B), exponential background (C), broad luminosity function (D), metallicity gradient (E),   S-model prior with $Z_{\odot}=15$~pc (F), S-model prior and broad luminosity function with $Z_{\odot}=15$~pc (G),  S-model prior with $Z_{\odot}=0$~pc (H), S-model prior with $Z_{\odot}=0$~pc with a broad luminosity function (I),
and extinction mask (J). The mass of the simulated stellar population is Mass$^{\rm{Sim}}_{\rm{Bulge}} = 1.92 \times10^{10} M_{\odot}$.
\label{tab:mass}}
\centering
\begin{tabular}{lcc}
\toprule
Case & Mass$^{\rm{VVV}}_{\rm{Bulge}}$ ( $\times 10 ^{10} M_{\odot}$) & Mass$^{\rm{Sim}}_{\rm{Bulge}}$ ( $\times 10 ^{10} M_{\odot}$) \\ \hline
A     & 1.64 &  1.89                                   \\
B     & 1.70 & 1.92                                   \\
C     & 1.33  & 1.84                                  \\
D     & 1.61  &1.90                                  \\
E     & 1.63 & 1.89                                   \\
F     & 1.52 & 1.91                                    \\
G     & 1.58 & 1.93                                    \\
H     & 1.53 & 1.92                                   \\
I     & 1.57  & 1.93                                  \\
J     & 1.71 & 1.90 \\ \bottomrule
\end{tabular}
\end{table}

\subsubsection{Distance to the Galactic centre}
As mentioned previously, we associate the Galactic centre with the location of the maximum density of the bulge. In all cases we examined, this maximum bulge density was in the same location for the parametric and non-parametric fit. According to our base non-parametric model, the distance from the Sun to the Galactic centre is $7.9$~kpc, where the assumed mean absolute magnitude of the RC is $\mu_{M_{Ks},RC}=-1.53$. 
WG13 found the main effect of changing $\mu_{M_{Ks},RC}$ was to change the distance to the Galactic centre.
If we had instead used the observed local RC mean magnitude of $\mu_{M_{Ks},RC} = -1.62$ \citep{ChanemphGaiaDR2parallax2019,HallTestingasteroseismologyGaia2019}, then all distances would be increased by a factor of 1.04. With the brighter RC, the distance to the Galactic centre would then be $8.24$~kpc, which is consistent with the recent measurement of $8.18 \pm 0.04$~kpc calculated using parallax observations of Sgr A*  \citep{GravityCollabDistance2019}. 

\subsubsection{Estimating the X-component proportion}

The X component was obtained by setting the $1$ in $(1+A)$ from the SX model definition in Eq.~\ref{eq:s_x_model} to 0.
The X-component proportion was then computed by integrating  the X component and SX model over all coordinates and then taking the ratio of them. These ratios are listed in Table \ref{tab:x-ratios}.

\begin{table}
	\caption{Ratios given by the X component of each corresponding model integrated in all directions down to a scalar divided by overall integrated SX model, for data and simulation fits.}\label{tab:x-ratios}
	\begin{tabular}{|c|c|c|c|c|c|c|}
		\hline
		&  A   &    B    &   C    &   D   &   E   &   J   \\ \hline
		Data     & 0.23 &  0.23   &  0.18  & 0.25  & 0.24  & 0.92  \\ 
		Simulations & 0.20  & -0.0062 & -0.048 & 0.012 & 0.018 & 0.016 \\ \hline
	\end{tabular} 
\end{table}

\begin{figure}
	\centering
	\includegraphics[width=\linewidth]{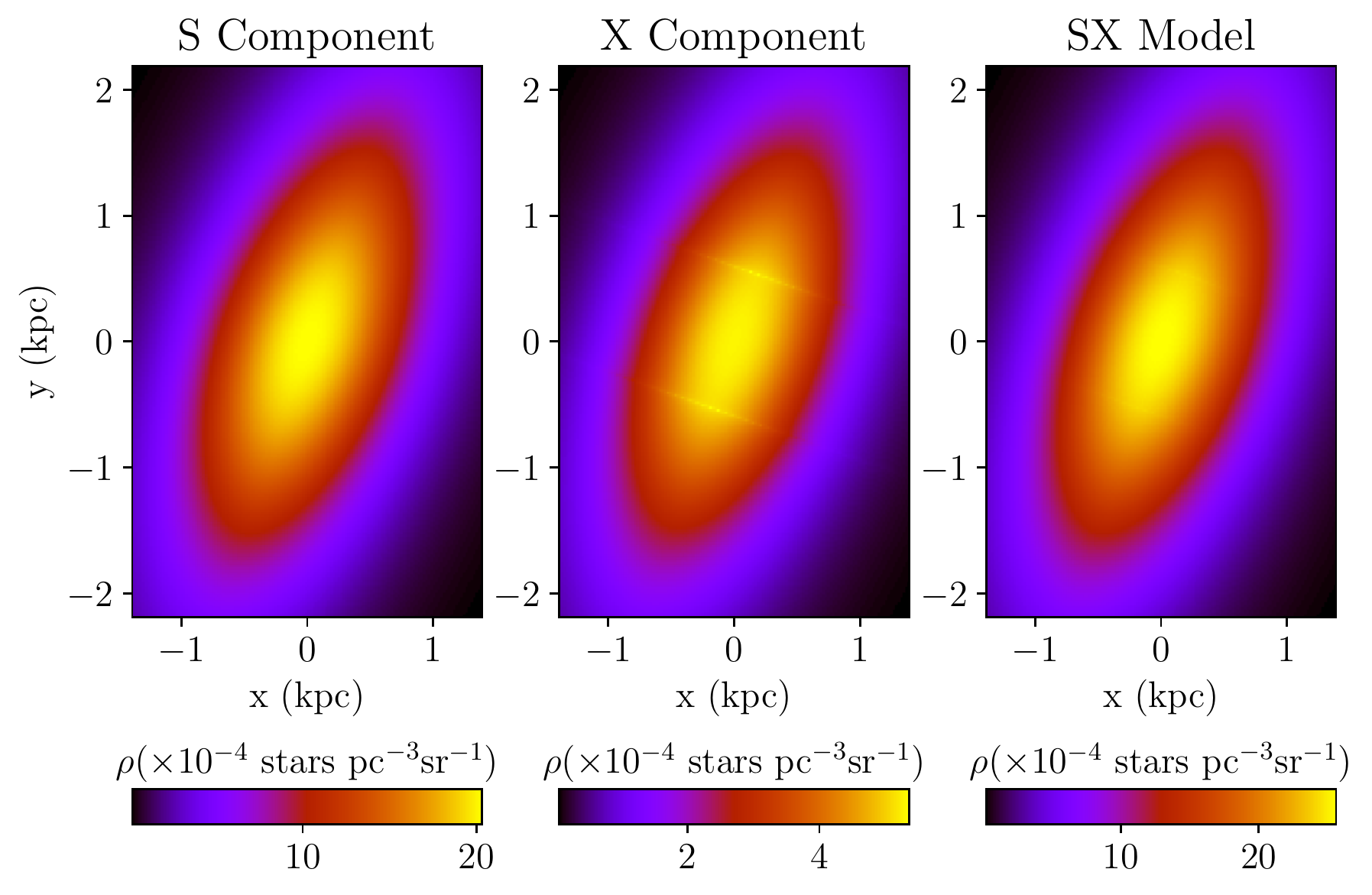}
	\caption{Sample slice at $z=310 pc$ of the parametric model in case A, fitted to simulations.
	A simple ratio of the X component to the full SX model can imply there is a significant X-arm component when there isn't one.
	Due to the very small exponent $n \sim 0.02$, the X component has effectively the same shape as the S component only with small cusps at the origins of the exponential functions.}
	\label{fig:ParDegeneracy}
\end{figure}

A partial degeneracy in the SX model, due to allowing the X-arm power law exponent ($n$) to vary, turns up in our extinction mask parametric fit (case J) to the data. The additional density unveiled by the extinction mask depicted in Fig.~\ref{fig:centraloverdensity} may be the main driving factor in this behaviour which only showed up in that model case.
The result of this is visible in Fig.~\ref{fig:parampairplotdata}, where the J case is an outlier in the $A$ and $n$ parameters.
With an exponent, $n$, less than 1, the X-arms become very broad. This case is not shown in Fig.~\ref{fig:TS} because it involves a different amount of data, so the change of likelihood will be on a different scale to that in the other cases.
Another case of $A$ and $n$ replacing the bulk of the S component of the SX model is in parametric case A on the simulations.
A slice near the edge of the Galactic plane data mask, at 310 pc, is displayed in Fig. \ref{fig:ParDegeneracy}.

As the parameter $n$ approaches 0, the perturbation tends towards a constant with a cusp at the X-arm origins from the exponential term.
Although this model can appear to have a strong X component, the fact we have $n\ll 1$ tells us that this component is near constant, so it is effectively adding to the normalisation of the S component rather than giving an X shaped perturbation.
This result could in principle have come out for any of the simulation cases, so this behaviour is not particular to the A model, just the random model initialisation that resulted in a convergence to a model that has the X component trace the bulge rather than, for example, fall below the mask by having a large X-arm parting factor $C$.

Based on the above arguments we  discard the A case parametric estimate for the simulation and the J case parametric result for the data in Table~\ref{tab:x-ratios}. It follows that our simulation results are consistent with a negligable X-component which is correct as the model used to generate the simulation had no X-component. Additionally, we can conclude that our parametric fit to the data has the X-component contributing  a range of 18\% to  25\% to the bulge mass.
This estimate of 
 the X-bulge component contribution is consistent with that found for the WG13 model by \cite{PortailMadetomeasuremodelsGalactic2015} which was 24\%.

\subsubsection{Bulge angle}
As can be seen from Table~\ref{tab:Par_Sims_Table} our bulge angles  with respect to the Sun-Galactic centre line
($\alpha$) for the simulation ranged from $19.1^\circ$ to $29.3^\circ$ which encompasses the simulated value of $\alpha=19.2^\circ$. As can be seen from Table~\ref{tab:Par_Data_Table} our parametric fit of the VVV data had bulge angles in the range of  $18^\circ$ to $32^\circ$. This is consistent with previous estimates. E.g. WG13 obtained a best fit of $27^\circ$ and S17 obtained a best fit of $20^\circ$.
\newtext{The dependence of the viewing angle on the intrinsic RC luminosity dispersion for triaxial features was observed by \cite{StanekModellingGalacticBar1997} and S17. As $\sigma_{RC}$ broadens, the depth of the bar needs to decrease along each line of sight. For a triaxial density, an increase in angle relative to the Sun-Galactic Centre position will directly lead to a smaller depth through the bar for each line of sight.}

\subsection{Gamma-Ray Galactic Centre Excess}\label{sec:fermiresults}
\begin{figure}
\centering
    \includegraphics[width=\columnwidth]{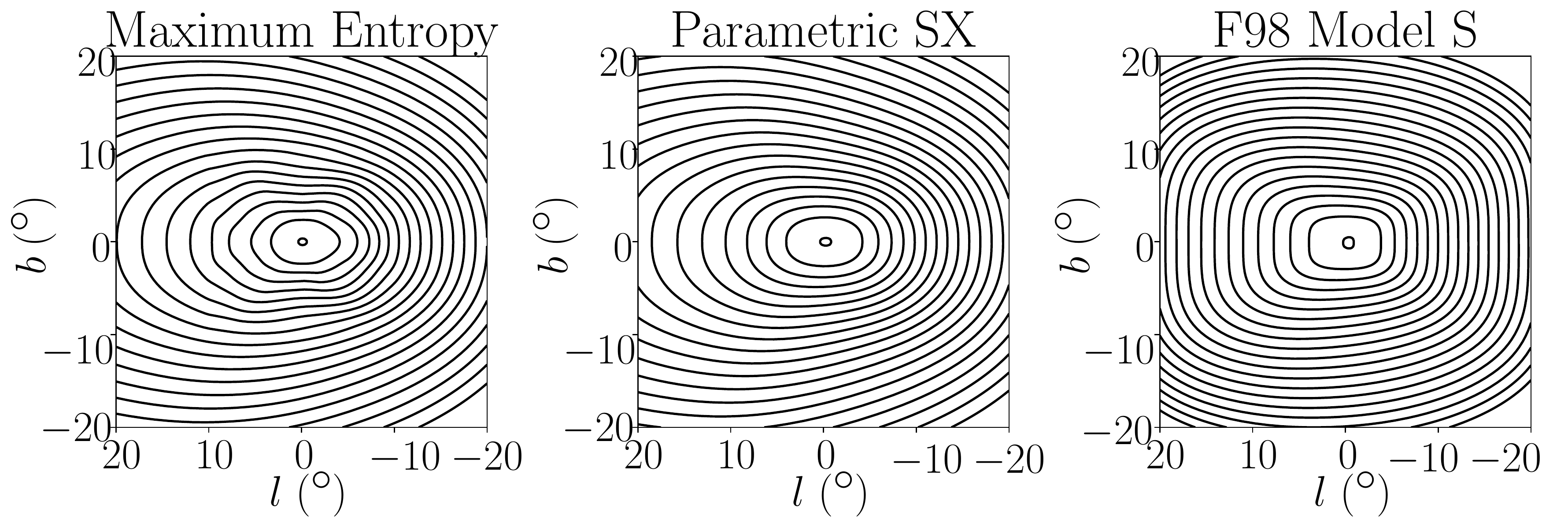}
    \caption{
    Integrated density, $T(l,b) = \int \rho(s,l,b) \, {\rm d} s$  for the maximum entropy deconvolution, the parametric SX prior density for the deconvolution and the parametric S-model of F98. 
    \label{fig:templates}
    }
\end{figure}

\begin{figure}
	\centering
	\includegraphics[width=0.8\columnwidth]{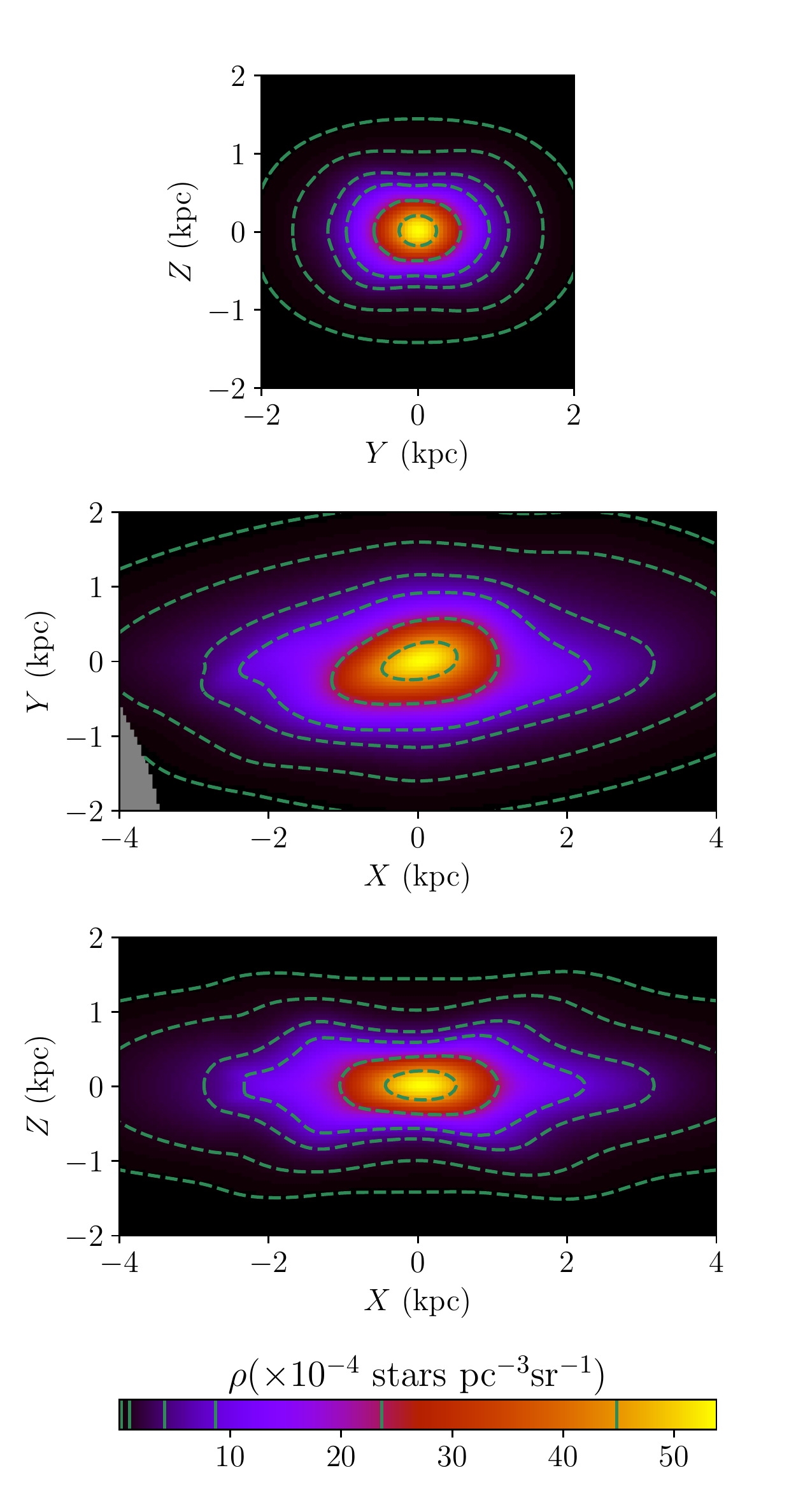}
	\caption{Slices at the Galactic centre of the stellar density across different axis slices for our base non-parametric model. The 3 perpendicular axes are aligned along the bulge angle and centre using $\alpha$ and $\Delta R_0$ from our best fitting parametric model for the base case. Where $X$ is along the main axis of the bar and $Z$ is perpendicular to the Galactic plane.
	}
	\label{fig:contourplotbasemodel}
\end{figure}

The work of \cite{MaciasStrongEvidencethat2019} found the S-bulge model (denoted by F98S hereafter) from \cite{FreudenreichCOBEModelGalactic1998} provided the best fit to the {\em Fermi\/} GCE in a template fitting analysis.
We created a template from our base parametric model and our non-parametric model fitted to the VVV data for comparison with the quality of the F98S template fit. We assumed that the density of MSPs is spatially correlated with the RC stellar density.
The template ($T$) for the {\em Fermi--LAT\/} analysis needs to be proportional to the expected flux of the MSPs, so it was constructed using:
\begin{equation}
T(l,b)=\int_s \rho(s,l,b)\, {\rm d}s
\end{equation}
where $\rho$ is, as before,  the RC+RGBB stellar density of the bulge.
\newtext{Note that an extra factor of $s^2$ is not necessary as this is the flux so whilst the number density is increasing as $s^2$ the observed flux is falling as $s^2$.}
We show a comparison between the F98S template and templates generated from our parametric and non-parametric fits in Fig.~\ref{fig:templates}. Our non-parametric template has a noticeable ``peanut'' like morphology. This may at first seem 
in contrast to the X-shaped morphology apparent from Fig.~\ref{fig:datasystematics} for example.
However, in that figure each slice in $z$ is normalized by the maximum density in that slice.
As is well known, when no such normalization is done the bulge has a more peanut-like morphology as can be seen from the third panel of the cross-sections in Fig.~\ref{fig:contourplotbasemodel}.


In fitting to the {\em Fermi\/}-LAT data, we  followed the same method as  \cite{MaciasStrongEvidencethat2019}. The bulge template was fitted simultaneously with the resolved point sources, gas correlated templates, inverse Compton templates (ICS-F98SA50) \citep{Porter:2017vaa}, {\em Fermi\/}--bubbles templates, and Sun/Moon templates.
\newtext{The unresolved MSP Galactic disk component has been found to have an undetectable contribution \citep{BartelsFermiLATGeVexcess2018} and so we did not include it.}
The energy range of the photons used in the {\em Fermi--LAT\/} analysis was 667 MeV to 158 GeV, distributed over 15 logarithmically spaced energy bins. 
A $40^\circ \times 40^\circ$ region around the Galactic centre was used with $0.5^\circ \times 0.5^\circ$ pixels. 
This large region of interest was necessary to be able to constrain the background components.
Also, no mask was used in the {\em Fermi--LAT\/} analysis. This made our non-parametric method of estimating the bulge from the VVV data particularly suitable as it allowed us to obtain an estimate of the bulge morphology over a $40^\circ \times 40^\circ$ area  with no masked regions.

We evaluated the improvement to the fit to the {\em Fermi--LAT\/} data  by working out 
${\rm TS}_{\rm Fermi}=2\log{\cal L}_{\rm null}-2\log{\cal L}_{\rm bulge}$ where ${\cal L}_{\rm null}$ is the maximum likelihood with all the above mentioned templates' normalisations treated as free parameters in each of the 15 energy bands.
$\cal {L}_{\rm bulge}$ is the maximum likelihood estimate using all the above mentioned templates and the the bulge template where the template normalisations were all fitted simultaneously.
As discussed by \cite{MaciasStrongEvidencethat2019}, a TS$_{\rm Fermi}\ge 34.8$ corresponds to a 4$\sigma$ detection of a new extended source.
In Table~\ref{tab:fermi}, we list the change in
TS$_{\rm Fermi}$ for the different
 bulge templates\footnote{\url{https://github.com/chrisgordon1/galactic_bulge_templates}} we considered.
The non-parametric template was preferred by the {\em Fermi--LAT\/} data, with $\Delta{\rm TS}_{\rm Fermi}=177$ compared to the previous best-fitting template, F98S.
A similar values was obtained when using a S-model fitted to the VVV data instead of F98S.
Compared to our parametric SX template, our non-parametric template had $\Delta{\rm TS}_{\rm Fermi}=65$.
\newtext{Our $40^\circ \times 40^\circ$ templates were significantly larger than the area covered by the VVV data. The extrapolated regions of the templates accounted for around half of the magnitude of the TS values listed in Table~\ref{tab:fermi}.}
Each successive enhancement in our bulge model, from S to SX to non-parametric, resulted in a steady improvement in the quality of fit to the {\em Fermi\/} data. This provides further evidence that 
the GCE traces the stellar content of the Galactic bulge. We found that the inferred gamma-ray energy spectra of the bulge was not very sensitive to the bulge morphology and was similar to previous analysis 
\citep{MaciasStrongEvidencethat2019}.

\begin{figure*}
\centering
    \includegraphics[width=0.9\textwidth]{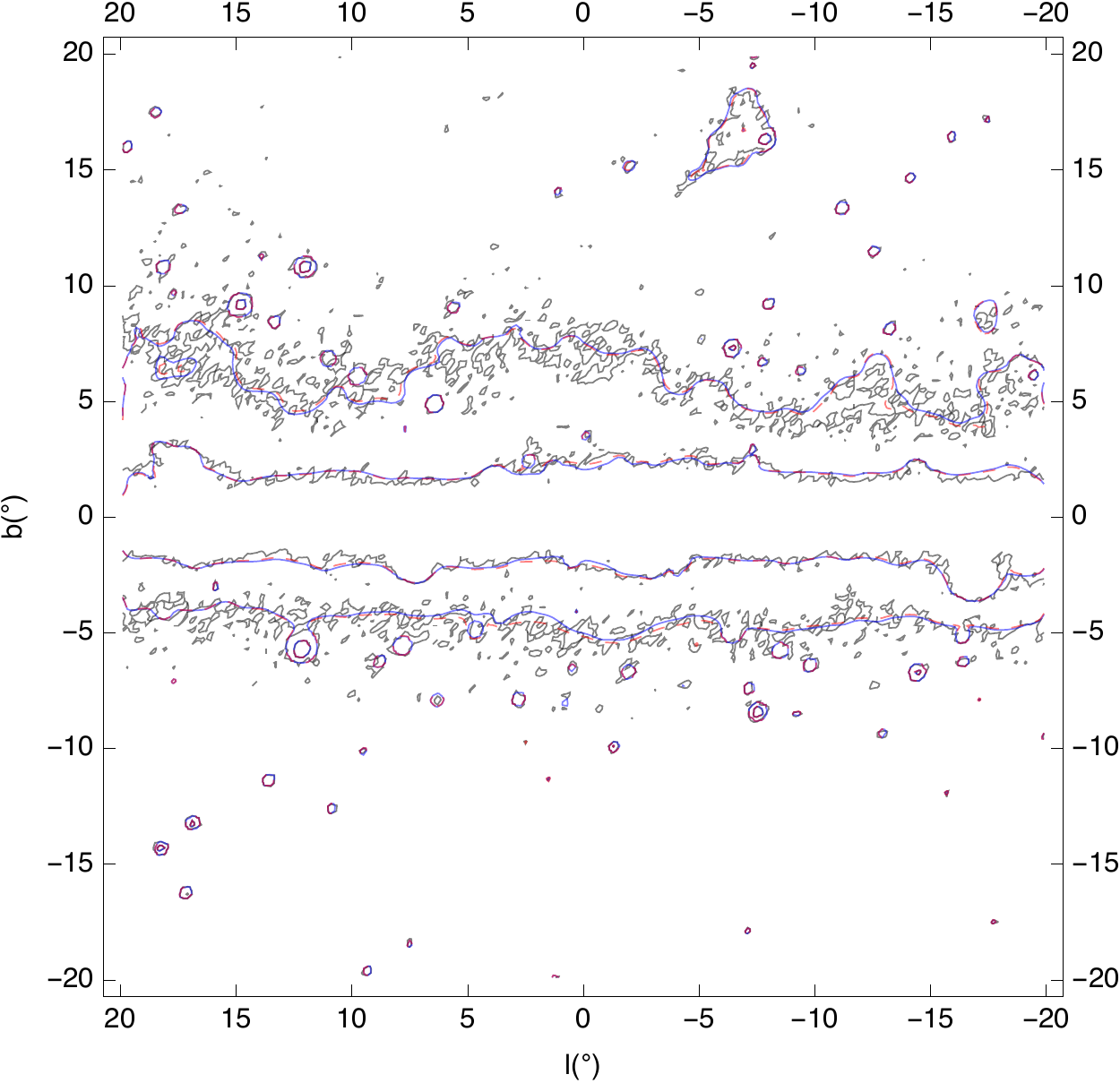}
   \caption{Contours of the {\em Fermi--LAT\/} data (black), a model without a Galactic bulge (blue), and model with our non-parametric Galactic bulge (red, dashed). The energy range is 1.1 to 2.8 GeV and the contour levels are 750 and 2000 in units of photons per square degree.}
    \label{fig:FermiContours}
\end{figure*}

\begin{figure}
\centering
    \includegraphics[width=\columnwidth]{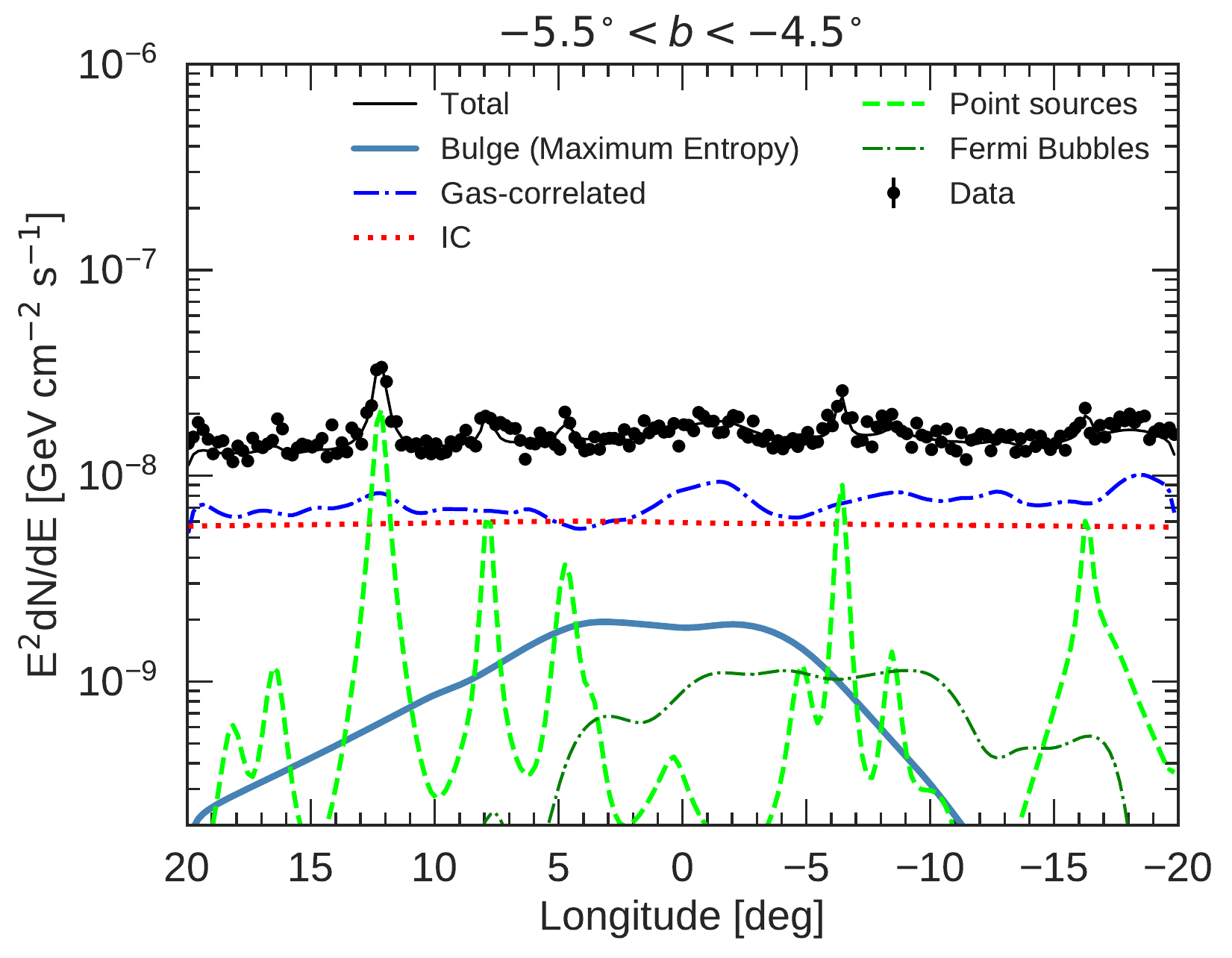}
   \caption{Spatial distribution of the main model components included in the {\em Fermi--LAT\/} fit. The flux profiles in the energy range $[1.1,2.8]$ GeV are displayed. Black dots represent the data and the continuous black line the total best-fitting model. Other components not shown here (\textit{e.g.,} isotropic, Sun, Moon and Loop I) are $\sim \mathcal{O}(1)$ less bright in the region used to construct the profile. }
    \label{fig:FermiProfile}
\end{figure}

Contour plots of the data and two alternative models are shown in Fig.~\ref{fig:FermiContours}. The improvement of the fit when the Galactic bulge component is included is particularly noticeable around $(l,b)=(5^\circ,-5^\circ)$.  
The contribution of the Galactic bulge to the {\em Fermi--LAT\/} model fit is shown in Fig.~\ref{fig:FermiProfile}. 
The peanut nature for the bulge shape is evident in this figure, even after accounting for the PSF smoothing of the {\em Fermi--LAT\/} instrument. Around the $l=5^\circ$ region there is a larger ratio of bulge to total signal than in other longitudes displayed. 
This helps in explaining why that area has one of the most noticeable improvements in fitting to the gamma-ray data presented in Fig.~\ref{fig:FermiContours}.
Also, this figure shows how typically the bulge component is an order of magnitude smaller than the overall signal. This makes it hard to assign a statistical significance to the difference in $\Delta{\rm TS}_{\rm Fermi}$ values seen in  Table~\ref{tab:fermi}, as small errors in the larger components could cause one template to be preferred over the other. One alternative method to account for this complication may be to use a maximum entropy non-parametric approach to modulate the larger components as handled by the SkyFACT method \citep{StormSkyFACTHighdimensionalmodeling2017a}, which also found a preference for a boxy bulge model of the GCE in the {\em Fermi--LAT\/} data \citep{BartelsFermiLATGeVexcess2018}. %
\begin{table}
\caption{\label{tab:fermi} A comparison of the different bulge templates ability to explain the {\em Fermi--LAT\/} GCE. Where for model $i$, we list $\Delta {\rm TS}_{\rm Fermi}= 2\ln{\cal L}_{\rm nonparam}/{\cal L}_i$.
}
\begin{center}
\begin{tabular}{l|r}
Model&$\Delta{\rm TS}_{\rm Fermi}$ \\ \hline
Non-parametric bulge&0          \\
SX bulge    &65        \\
S-bulge           &177        
\end{tabular}
\end{center}
\end{table}

\section{Conclusions}\label{sec:Conclusion}

We have used a non-parametric method incorporating maximum entropy and smoothness regularisation to deconvolve the density distribution of bulge stars in the VVV MW-BULGE-PSFPHOT catalogue. We have also proposed a maximum entropy method for determining the background non-RC+RGBB stars, based on prior estimates using parametric models. 
Reasonable values for the regularisation parameters were found by testing the deconvolution method on a simulated stellar population of the galaxy made of a 10~Gyr old eight-fold symmetric bulge, thin disc, and thick disc. Testing our maximum entropy deconvolution and background fitting method on a simulated population, we were able to nearly perfectly reconstruct the density even in the heavily extincted and crowded regions which had been masked in the analysis.

Applying the deconvolution method to the VVV data we found many of the features previously observed in the literature, including the X-shaped bulge from the split RC peak, 
the dependence of the viewing angle on the intrinsic RC luminosity dispersion, and the  feature  behind the bar.
The $R_0$ gradient was not clearly seen in the MW-BULGE-PSFPHOT star counts when using the modified Richardson-Lucy deconvolution method assuming eight-fold symmetry.

We performed extensive systematic tests of the maximum entropy deconvolution method to test our assumptions regarding the choice of background model, metallicity distribution, intrinsic dispersion of the RC, position of the Sun above the Galactic mid-plane, and the deconvolution method itself. 

The maximum entropy background was significantly preferred over the widely used exponential background by both the parametric models we fitted and the maximum entropy deconvolution method. Future studies of bulge star counts should be wary using the exponential background, as we have shown it has a tendency to over estimate the background star counts at the bright end of the luminosity function, causing the density of stars to be significantly underestimated at nearby distances.

A broad, unimodal metallicity distribution with spatially varying mean metallicity did not significantly effect the bulge stellar density. A bi-modal metallicity distribution is likely needed, which will become possible as the coverage of bulge spectroscopic surveys grows.

Qualitatively our results were broadly consistent with the modified Richardson-Lucy deconvolution of WG13. However, we were able to obtain  less noisy and higher resolution reconstructions with our maximum entropy method when using the narrow RC dispersion which recent observations with {\em Gaia\/} have favoured \citep{HallTestingasteroseismologyGaia2019,ChanemphGaiaDR2parallax2019}. This resulted in somewhat less dense X-arms. Our method inpainted regions where the data was masked. This meant that we did not need to assume eight-fold symmetry to obtain a reconstruction of the whole bulge area.

From our fits to several different model cases, we found
our bulge angle was in the range $[18^\circ,  32^\circ]$ , our bulge mass was in the range  $[1.3\times10^{10}$, $1.7\times10^{10}]M_\odot$, and our X-bulge contribution to the bulge was in the range $[18,25]\%$. These are all compatible with other recent bulge estimates using the VVV data.

Our non-parametric method allowed us to 
inpaint masked regions and smoothly join onto a parametric model outside the region of the VVV data. This made it suitable for 
providing a template to be used in fitting the {\em Fermi--LAT\/} GeV Galactic centre excess.
We found our non-parametric template  provided a better fit than the previously implemented parametric S-model (F98S) and our parametric fits to the VVV data. This further supports the unresolved population of millisecond pulsars interpretation of the GeV Galactic centre excess, traced by the Galactic bulge stellar population.

\section*{Acknowledgements}
We thank Iulia Simion for helpful discussions related to this work. Also, we are grateful to Elena Valenti for giving us early access to the MW-BULGE-PSFPHOT VVV catalogue. O.M. was supported by World Premier International Research Center Initiative (WPI Initiative), MEXT, Japan and by the Japan Society for the Promotion of Science under Grant Numbers KAKENHI-JP17H04836,-JP18H04340 and -JP18H04578.
This work was made possible by the use of the Research Compute Cluster (RCC) facilities at the University of Canterbury.
The following software packages were used in this work:  \textsc{astropy}, \textsc{Fermi Science Tools}, \textsc{matplotlib}, \textsc{numpy}, \textsc{pylbfgs}, and \textsc{scipy}.

\bibliographystyle{mnras}
\setcitestyle{authoryear,open={((},close={))}}
\bibliography{references} %

\appendix

\section{Results Tables}
The best-fiting likelihood values we obtained for our parametric and non-parametric fits are listed in Table~\ref{tab:TSmaxentandparam}. The best fit parameter values are listed in Tables \ref{tab:Par_Data_Table} and \ref{tab:Par_Sims_Table}.
\begin{table}
\caption{Minimum values of $-2\ln \mathcal{L}$ for the parametric and non-parametric models.
The base case (A) values of $(-1.36968, -2.35631, -1.4102015,	-2.32068)\times10^8$ have been subtracted from columns one to four respectively.
The non-base cases considered  are:  no behind-the-bar feature subtraction (B), exponential background (C), broad luminosity function (D), metallicity gradient (E),   S-model prior with $Z_{\odot}=15$~pc (F), S-model prior and broad luminosity function with $Z_{\odot}=15$~pc (G),  S-model prior with $Z_{\odot}=0$~pc (H), S-model prior with $Z_{\odot}=0$~pc with a broad luminosity function (I), extinction mask (J).
\newtext{Note that as case J has a different amount of data, it's $-2\ln {\cal L}$ value cannot be compared directly with the other cases.}
}\label{tab:TSmaxentandparam}
{
\centering
\begin{tabular}{lrrrr}
             &   & VVV Data &   & Simulation  \\ \hline 
\textbf{Case} & Param.\ & Non-param.\ & Param.\ & Non-Param.\ \\ \hline
A & 0         & 0         & 0         & 0         \\
B & 17086     & 974       & 733       & 307      \\
C & 65507     & 60554     & 55654     & 69758     \\
D & -1793     & 2917614   & 13797     & 76778     \\
E & 266       & 184       & 109       & -1641     \\
F & 38934     & 241421    & -5523     & 176708    \\
G & 21665     & 209841    & 15475     & 161736    \\
H & 19723     & 1361      & 640       & 95        \\
I & 15107     & 25589     & 22740     & 6252      \\
J & $-2\times 10^7$ & $-4\times 10^7$ & $-2\times 10^7$ & $-3\times 10^7$ \\ \bottomrule
                                
\end{tabular}
}

\end{table}

\begin{table*} %

	\setlength\tabcolsep{2pt}

	\caption{Parametric SX and S-models fitted to VVV data used as priors in Table \ref{tab:TSmaxentandparam}. The best fits and 68\% errors are given for each case on alternating lines.
	}
	\label{tab:Par_Data_Table} 
	\scriptsize{
		\begin{tabular}{lrrrrrrrrrrrrr}
			\toprule
			\small Label                                  & \small $c_{\perp}$ & \small $c_{\parallel}$ &  \small $x_0$ &   \small $y_0$ &   \small $z_0$ & \small $\rho_0 \times 10^6$ & \small $\alpha$ & \small $\Delta R_0$ &   \small $C$ &    \small $A$ & \small $x_1$ & \small $y_1$ &    \small $n$ \\ \midrule
			\small A) Base case                           &       1.581 &           2.359 &  1.853 &   0.672 &  0.4605 &                0.123 &    20.12 &      -0.0968 & 1.386 &   0.69 & 0.731 & 1.090 &   2.31 \\
			\small                                        &       0.008 &           0.009 &  0.006 &   0.001 &  0.0004 &                0.002 &     0.03 &       0.0009 & 0.005 &   0.02 & 0.004 & 0.005 &   0.09 \\ \midrule
			\small B) No feature behind the bar           &       1.856 &           2.319 &   1.88 &   0.664 &  0.4544 &                0.119 &     18.0 &       -0.198 & 1.359 &   0.68 & 0.781 &  1.11 &    2.2 \\
			\small ~~~incorporated into background        &       0.007 &           0.008 &   0.02 &   0.002 &  0.0007 &                0.003 &      0.2 &        0.001 & 0.004 &   0.05 & 0.007 &  0.02 &    0.2 \\ \midrule
			\small C) Exponential background              &       1.309 &           3.177 &  1.641 &  0.7105 &  0.4798 &               0.1158 &    23.55 &      -0.0386 & 1.346 & 0.6246 & 0.621 & 0.734 &  1.981 \\
			\small ~~~instead of MaxEnt background        &       0.001 &           0.002 &  0.001 &  0.0007 &  0.0003 &               0.0001 &    0.002 &       0.0005 & 0.002 & 0.0009 & 0.001 & 0.001 &  0.001 \\ \midrule
			\small D) Broad luminosity function           &       1.172 &           2.124 &  1.735 &   0.610 &  0.4658 &               0.1788 &    28.88 &      -0.0711 & 1.356 &   2.13 & 0.170 & 1.135 &   18.0 \\
			\small                                        &       0.007 &           0.009 &  0.008 &   0.002 &  0.0007 &               0.0009 &     0.06 &       0.0009 & 0.003 &   0.04 & 0.003 & 0.008 &    0.4 \\ \midrule
			\small E) Metallicity gradient                &       1.546 &           2.383 &  1.884 &  0.6802 &  0.4582 &               0.1193 &   19.863 &      -0.1127 & 1.389 &  0.727 & 0.729 & 1.057 &  2.244 \\
			\small ~~~accounted for                       &       0.002 &           0.002 &  0.002 &  0.0003 &  0.0002 &               0.0002 &    0.001 &       0.0007 & 0.001 &  0.001 & 0.002 & 0.001 &  0.002 \\ \midrule
			\small F) S-model prior                       &       1.677 &           2.616 & 1.3812 & 0.58753 &    0.42 &               0.2322 &  19.7886 &      -0.0724 &     - &      - &     - &     - &      - \\
			\small ~~~with $Z_{\odot}=15$~pc              &      0.0003 &          0.0002 & 0.0002 & 0.00012 &  0.0003 &               0.0004 &   0.0003 &       0.0003 &     - &      - &     - &     - &      - \\ \midrule
			\small G) S-model prior and broad luminosity  &       1.242 &           2.779 & 1.2332 &  0.4819 & 0.40921 &               0.3687 &   31.945 &      -0.0698 &     - &      - &     - &     - &      - \\
			\small ~~~function with $Z_{\odot}=15$~pc     &       0.001 &           0.003 & 0.0013 &  0.0004 & 0.00018 &               0.0005 &    0.005 &       0.0008 &     - &      - &     - &     - &      - \\ \midrule
			\small H) S-model prior with $Z_{\odot}=0$~pc &      1.6734 &           2.592 & 1.3921 &  0.5915 &  0.4271 &               0.2269 &  19.8241 &      -0.0767 &     - &      - &     - &     - &      - \\
			\small                                        &      0.0008 &           0.003 & 0.0009 &  0.0004 &  0.0002 &               0.0002 &   0.0003 &       0.0008 &     - &      - &     - &     - &      - \\ \midrule
			\small I) S-model prior with $Z_{\odot}=0$~pc &       1.221 &           2.733 &  1.253 &  0.4884 & 0.41672 &               0.3596 &   31.851 &      -0.0712 &     - &      - &     - &     - &      - \\
			\small ~~~\& broad luminosity function        &       0.003 &           0.004 & 0.0012 &  0.0004 & 0.00016 &               0.0004 &    0.006 &       0.0006 &     - &      - &     - &     - &      - \\ \midrule
			\small J) Extinction mask                     &       0.970 &           2.691 & 26.442 &  0.7440 &  0.4786 &             0.004990 &   18.768 &      -0.1018 & 1.302 & 38.903 & 0.815 & 0.891 & 0.8855 \\
			\small                                        &       0.002 &           0.001 &  0.002 &  0.0007 &  0.0002 &             0.000005 &    0.002 &       0.0006 & 0.001 &  0.002 & 0.001 & 0.001 & 0.0009 \\ \bottomrule
		\end{tabular}
	}

	\caption{Parametric SX and S-models, fitted to an S-model simulation.
		The best fits and 68\% errors are given for each case on alternating lines. 
	}\label{tab:Par_Sims_Table}
	\begin{tabular}{lrrrrrrrrrrrrr}
		\toprule
		\small Label                                  & \small $c_{\perp}$ & \small $c_{\parallel}$ &   \small $x_0$ &   \small $y_0$ &  \small $z_0$ & \small $\rho_0 \times 10^6$ & \small $\alpha$ & \small $\Delta R_0$ &    \small $C$ &    \small $A$ &  \small $x_1$ & \small $y_1$ &   \small $n$ \\ \midrule
		\small A) Base case                           &       1.864 &           2.464 &   1.608 &  0.6851 & 0.4845 &               0.1492 &   19.414 &      -0.0031 & 1.8136 &   0.42 & 0.0003 & 0.409 & 0.022 \\
		\small                                        &       0.004 &           0.003 &   0.001 &  0.0006 & 0.0002 &               0.0007 &    0.006 &       0.0003 & 0.0006 &   0.01 & 0.0002 & 0.005 & 0.001 \\ \midrule
		\small B) No feature behind the bar           &       1.864 &           2.467 &   1.600 &  0.6846 & 0.4835 &               0.1897 &   19.405 &      -0.0023 &  1.092 & -0.016 &  0.050 & 7.538 & 0.178 \\
		\small ~~~incorporated into background        &       0.003 &           0.004 &   0.001 &  0.0004 & 0.0002 &               0.0003 &    0.003 &       0.0006 &  0.003 &  0.003 &  0.001 & 0.005 & 0.002 \\ \midrule
		\small C) Exponential background              &       1.733 &           2.481 &   1.545 &  0.7116 & 0.4943 &               0.1932 &    21.17 &       0.0638 & 0.6724 & -0.205 &  0.020 &  2.10 & 0.222 \\
		\small ~~~instead of MaxEnt background        &       0.004 &           0.005 &   0.002 &  0.0006 & 0.0003 &               0.0007 &     0.02 &       0.0004 & 0.0006 &  0.006 &  0.002 &  0.04 & 0.007 \\ \midrule
		\small D) Broad luminosity function           &       1.893 &           2.545 &   1.377 &  0.6043 & 0.4785 &               0.2386 &    26.90 &       0.0460 &  2.659 &  0.402 &  0.011 & 1.954 &  0.40 \\
		\small                                        &       0.008 &           0.007 &   0.002 &  0.0006 & 0.0004 &               0.0005 &     0.03 &       0.0007 &  0.002 &  0.001 &  0.001 & 0.007 &  0.02 \\ \midrule
		\small E) Metallicity gradient                &       1.852 &           2.523 &   1.601 &  0.6864 & 0.4843 &               0.1817 &    19.10 &      -0.0178 &  7.483 &  0.019 &  2.779 &  4.51 & 8.308 \\
		\small ~~~accounted for                       &       0.004 &           0.005 &   0.001 &  0.0005 & 0.0003 &               0.0003 &     0.01 &       0.0008 &  0.005 &  0.001 &  0.008 &  0.01 & 0.006 \\ \midrule
		\small F) S-model prior                       &       1.868 &           2.506 &   1.586 &  0.6790 & 0.4746 &               0.1930 &    19.49 &      -0.0003 &      - &      - &      - &     - &     - \\
		\small ~~~with $Z_{\odot}=15$~pc              &       0.003 &           0.004 &   0.001 &  0.0004 & 0.0002 &               0.0002 &     0.04 &       0.0007 &      - &      - &      - &     - &     - \\ \midrule
		\small G) S-model prior and broad luminosity  &      1.9941 &          2.6591 & 1.30221 & 0.56743 & 0.4640 &               0.2677 &  29.2638 &       0.0548 &      - &      - &      - &     - &     - \\
		\small ~~~function with $Z_{\odot}=15$~pc     &      0.0002 &          0.0002 & 0.00008 & 0.00005 & 0.0001 &               0.0002 &   0.0001 &       0.0003 &      - &      - &      - &     - &     - \\ \midrule
		\small H) S-model prior with $Z_{\odot}=0$~pc &       1.861 &           2.476 &   1.599 &  0.6841 & 0.4840 &               0.1886 &   19.552 &      -0.0065 &      - &      - &      - &     - &     - \\
		\small                                        &       0.003 &           0.003 &   0.001 &  0.0005 & 0.0002 &               0.0002 &    0.004 &       0.0006 &      - &      - &      - &     - &     - \\ \midrule
		\small I) S-model prior with $Z_{\odot}=0$~pc &       1.954 &           2.604 &  1.3187 &  0.5733 & 0.4740 &               0.2616 &  29.2719 &       0.0514 &      - &      - &      - &     - &     - \\
		\small ~~~\& Broad luminosity function        &       0.001 &           0.002 &  0.0006 &  0.0003 & 0.0002 &               0.0002 &   0.0009 &       0.0006 &      - &      - &      - &     - &     - \\ \midrule
		\small J) Extinction mask                     &       1.839 &           2.513 &   1.582 &  0.6844 & 0.4861 &               0.1851 &    19.84 &      -0.0164 &   6.76 &  0.041 &   0.98 &  2.23 &  0.82 \\
		\small                                        &       0.005 &           0.006 &   0.002 &  0.0006 & 0.0004 &               0.0004 &     0.02 &       0.0007 &   0.05 &  0.003 &   0.01 &  0.07 &  0.03 \\ \bottomrule
	\end{tabular}

\end{table*}

\subsection{Deconvolution Method Systematic}\label{sec:Richardson-Lucy}

Since our data differ from previous 3-D RC bulge studies in its photometry and completeness, we investigated how these changes are reflected in past methods applied to view the VVV RC.
Given our semi-analytic formulation of a $K_s$-band luminosity function, we compare the results of past methods using different luminosity functions and backgrounds to our maximum entropy non-parametric density model.
We continued to use the semi-analytic luminosity function derived in P19 (abbreviated here as the PARSEC luminosity function). We also used the parametric function fitted to Monte Carlo simulations of WG13 (abbreviated as the BaSTI luminosity function).
The WG13 luminosity function construction involved random draws of star masses from a Salpeter IMF and metallicity from the Baade's window metallicity distribution measured by \cite{Zoccali2008BulgeMetalContent}.
Then, the $K_s$ absolute magnitude was obtained from interpolated $\alpha$ enhanced BaSTI isochrones \citep{PietrinferniLargeStellarEvolution2004} assuming an age of 10 Gyr.
The parametrisation of the WG13 BaSTI based luminosity function takes the form of the sum of two Gaussians corresponding to the RC and RGBB with parameters $\mu_{\rm M_{K_s,RC}}=-1.72$, $\sigma_{\rm RC} = 0.18$, $\mu_{M_{\rm K_s,\rm RGBB}}=-0.91$, $\sigma_{\rm RGBB}=0.19$ and relative fraction $f_{\rm RGBB}=0.20$ ($\mu$ and $\sigma$ taking their typical meanings in a Gaussian distribution).
A notable difference here is that the RC dispersion is 3 times the width of our semi-analytic form, which is approximately 0.06 when fitting a Gaussian to the RC component.

As in WG13, we fitted a background of the form
\begin{equation}\label{eq:exponbackground}
B(K_{s}) = \exp(a+b(K_{s}-13)+c(K_{s}-13)^2)
\end{equation}
to the magnitude ranges $11 \leq K_s\leq 11.9$ mag and $14.3\leq K_s\leq 15$ mag for each line-of-sight.
Several adjustments they recommended were retained for this background fit. Higher extinction and crowding in fields with $|b|<2^{\circ}$ were accommodated by setting the second order coefficient, $c$, to 0 and restricting the upper fitted magnitude range to 14.5 mag. The bright latitude end magnitude range for regions where $l\geq5.5^{\circ}$ was reduced down to $11 \leq K_s\leq 11.7$.
The star count model for each field of view takes the form of Eq.~\ref{eq:stellarstatistics}, converted to the form of a background plus a linear convolution via the transform of line-of-sight distance ($s$) to distance modulus ($\mu$).
The luminosity function was convolved with the mean combined photometric and systematic uncertainty for each $K_s$ along each line-of-sight to account for their effects.
The VVV data was re-discretised into $\sim 1.5^{\circ}\times \sim 0.5^{\circ}$ spatial bins over 0.05 mag $K_s$ bins.
For each line-of-sight, the density distribution was initialised to a Hann window function over a distance modulus of 11.2 to 17, renormalised to the observed counts.
We then applied the modified Richardson-Lucy procedure of WG13, retaining their stopping criteria, for both the BaSTI and PARSEC luminosity functions.
This produced an estimate of the bulge density 
which depended on $\mu$ which 
we mapped onto a density which depends on 
$s$.
We then reprojected the bulge density to Cartesian form using linear interpolation. For the low resolution data, step sizes of ($\Delta$x $\times$ $\Delta$y  $\times$ $\Delta$z) = (0.15  $\times$ 0.1  $\times$ 0.075) kpc were used.
This simple reprojection only produced a noisy unsymmetrised view of the density model. For a view of the deconvolved bulge density assuming eight-fold symmetry, the appropriate frame needs to be found.

We applied a process of finding the maximally eight-fold symmetric frame following WG13.
For each slice in the $z$ direction, we carried out a simple grid search over distance to the Galactic centre $R_0$ and bulge angle $\alpha$, in steps of 0.02 kpc and 0.5 deg.
For each $\alpha$ fixed, we shifted the bulge centre to some value of $R_0$ and computed the symmetrised density
\begin{equation}\label{rhobar}
\bar{\rho}(x,y,z) = \frac{1}{N} \left[ \rho(x,y,z) + \rho(-x,y,z) + \mbox{6 other octants} \right] 
\end{equation}
where octant positions without matching densities in the $(l,b,s)$ projection were ignored from the computation. Parameter $N$ is the number of octants with non-masked densities.
Then the quantity 
\begin{equation}\label{eq:RLsymmetrisation}
\frac{1}{\rm N_{z}} \sum_{z=0.4\rm kpc}^{1\rm kpc}  \frac{\left<\rho_{{\rm rms}}\right>_z}{\left<\rho\right>_z}
\end{equation}
was minimised, where $N_{z}$ is the number of slices between 0.4 and 0.8 kpc in the chosen cartesian grid, so the quantity is comparable between resolutions.
The parameter $\rho_{{\rm rms}}$ denotes the root mean square deviation between each octant's density in the symmetrisation and the average density, $\bar{\rho}$, of those points, which was then averaged across all points in each z-slice.

Rather than minimising Eq.~\ref{eq:RLsymmetrisation} directly, $\left<\rho_{{\rm rms}}\right>_z/\left<\rho\right>_z$ was minimised over individual slices of $z$ for our $R_0$ grid search.
This was an intermediary step in the bulge angle selection process to account for potential magnitude shifts in the model resulting from factors such as metallicity gradients, on top of the required shift in finding the maximally eight-fold symmetric frame.

This process was then repeated for $0.2^\circ \times 0.2^\circ$ spatial bins using our maximum entropy derived background, described in Section \ref{subsec:background}, and Cartesian grid spacing adjusted to ($\Delta$x $\times$ $\Delta$y  $\times$ $\Delta$z) = (0.04 $\times$ 0.04 $\times$ 0.03)~kpc, to accommodate the finer data resolution.

\begin{figure}
	\centering
	\includegraphics[width=\columnwidth]{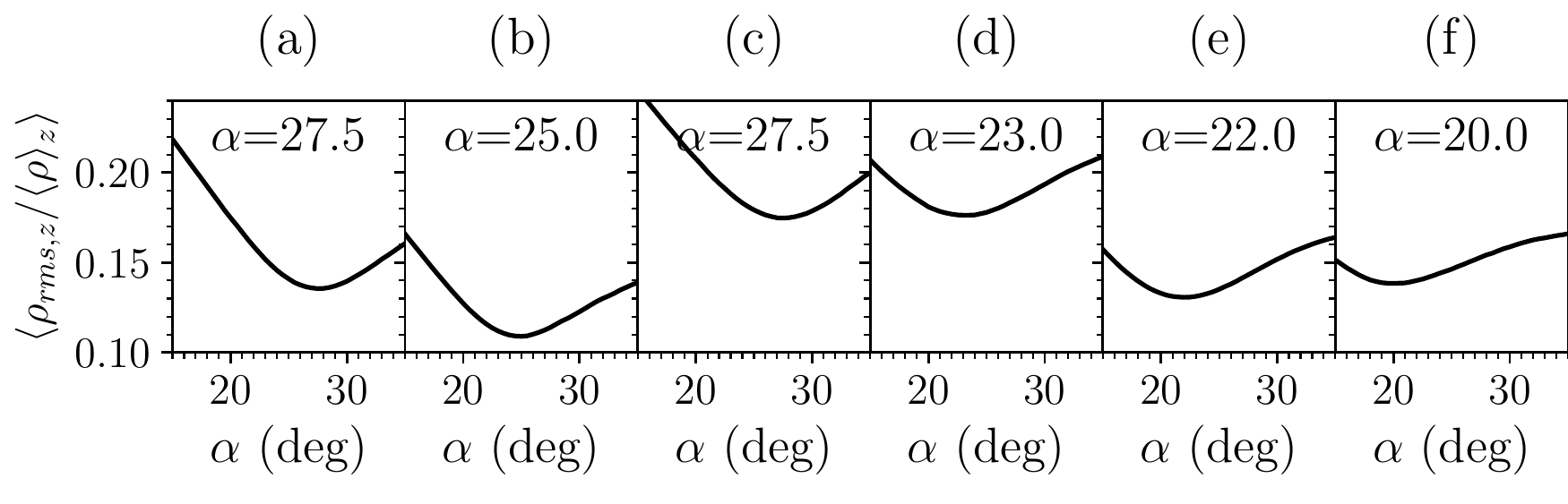}
	\includegraphics[width=\columnwidth]{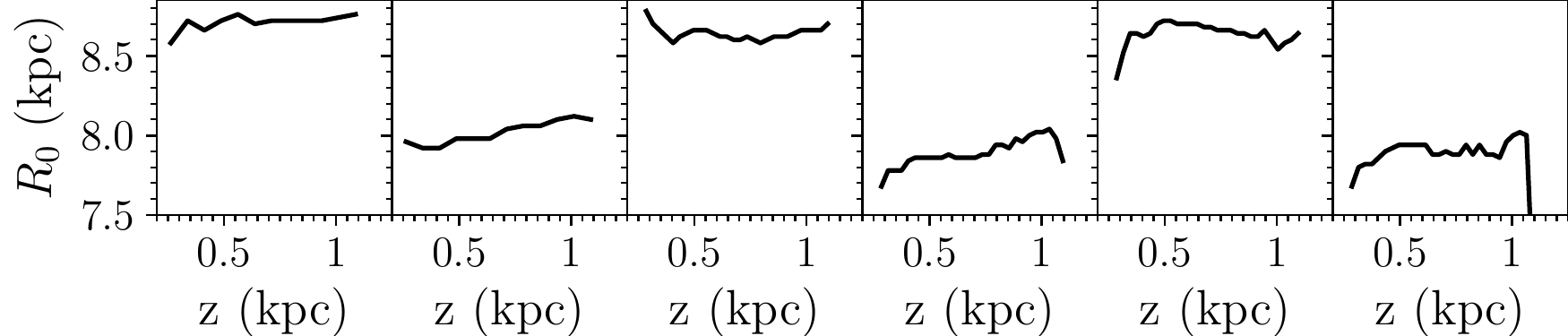}
	\caption{Maximally eight-fold symmetric angle (top) and $R_0$ (bottom) orientation of modified Richardson-Lucy deprojected data. From left to right: (a) BaSTI luminosity function on low resolution data (b) PARSEC luminosity function on low resolution data (c) BaSTI luminosity function on high resolution data (d) PARSEC luminosity function on high resolution data (e) BaSTI luminosity function on simulated data (f) PARSEC luminosity function on simulated data. }
	\label{fig:eight-fold symmetrymaximisationbestangle}
\end{figure}

\begin{figure}
	\centering
	\includegraphics[width=\columnwidth]{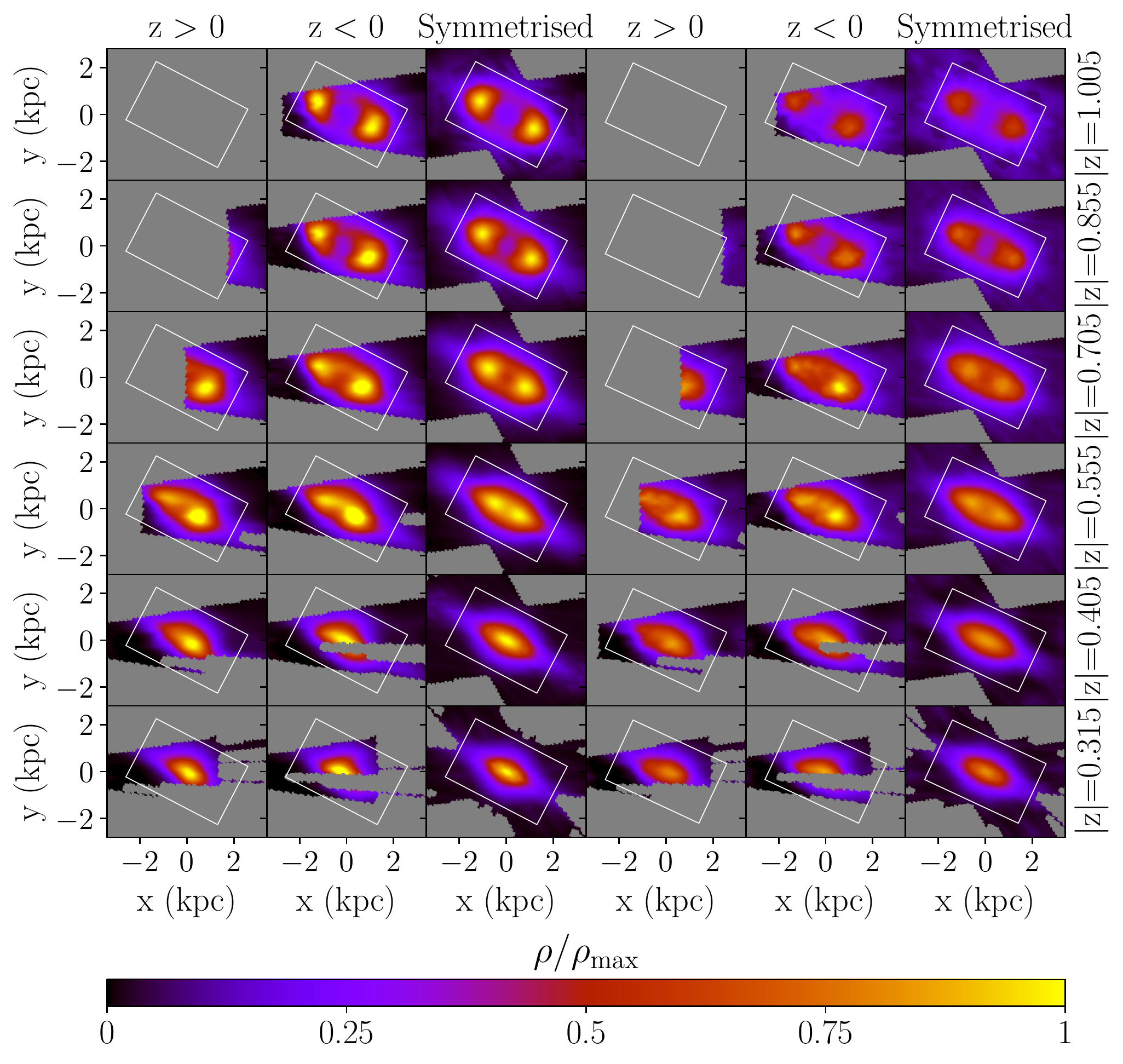}
	\caption{Three dimensional reconstruction of low resolution VVV data. Columns 1-3 using BaSTI luminosity function and 4-6 using PARSEC luminosity function. Slices of $|z|$ (measured in kpc) normalised by the maximum of the BaSTI symmetrised model.}
	\label{fig:eight-fold symmetrymaximisationdensity0}
\end{figure}

\begin{figure}
	\centering
	\includegraphics[width=\columnwidth]{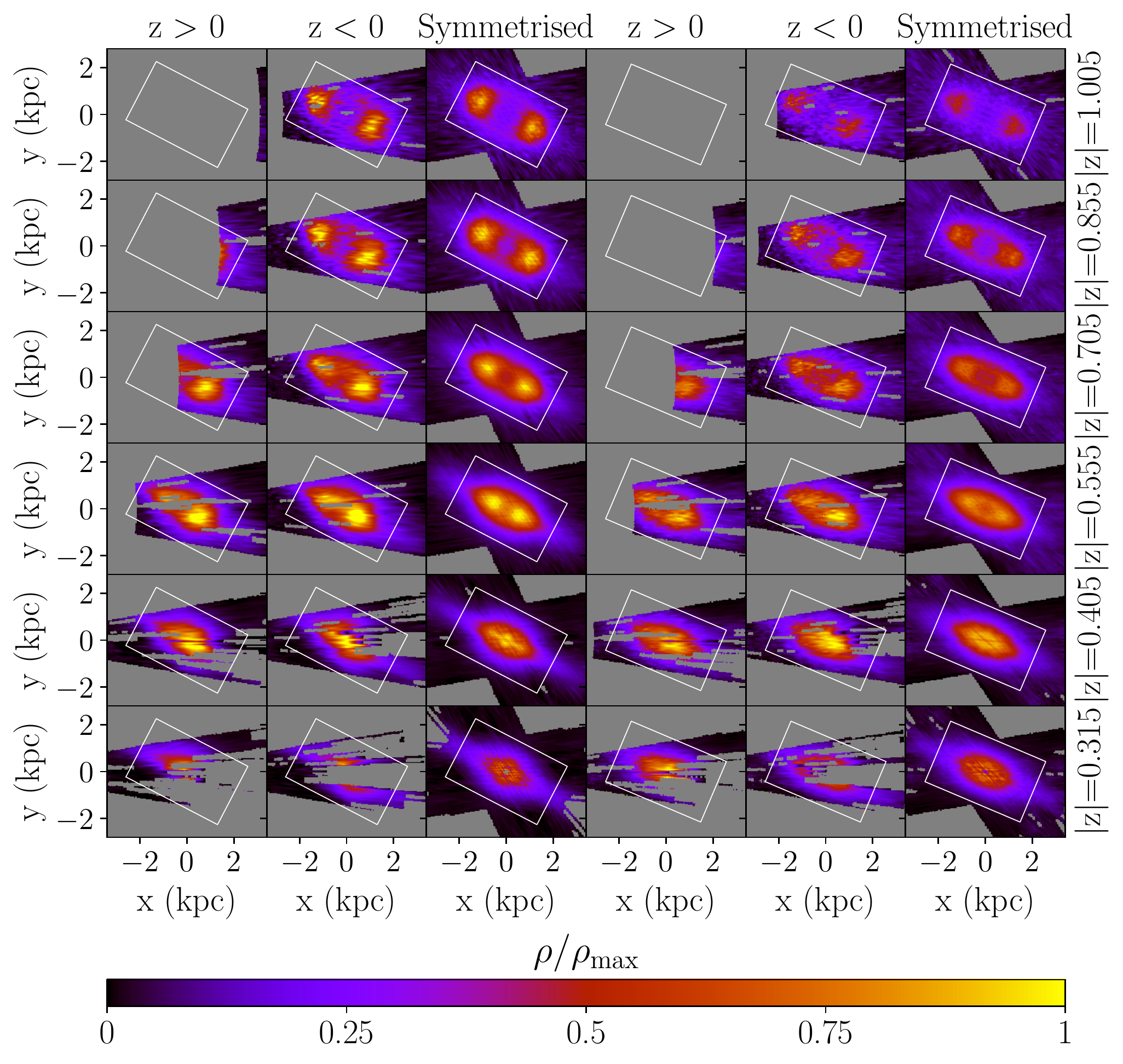}
	\caption{Three dimensional reconstruction of high resolution VVV data. Columns 1-3 using BaSTI luminosity function and 4-6 using PARSEC luminosity function.
	Slices of $|z|$ are measured in kpc.
	}
	\label{fig:eight-fold symmetrymaximisationdensity1}
\end{figure}
\begin{figure}
    \centering
    \includegraphics[width=\columnwidth]{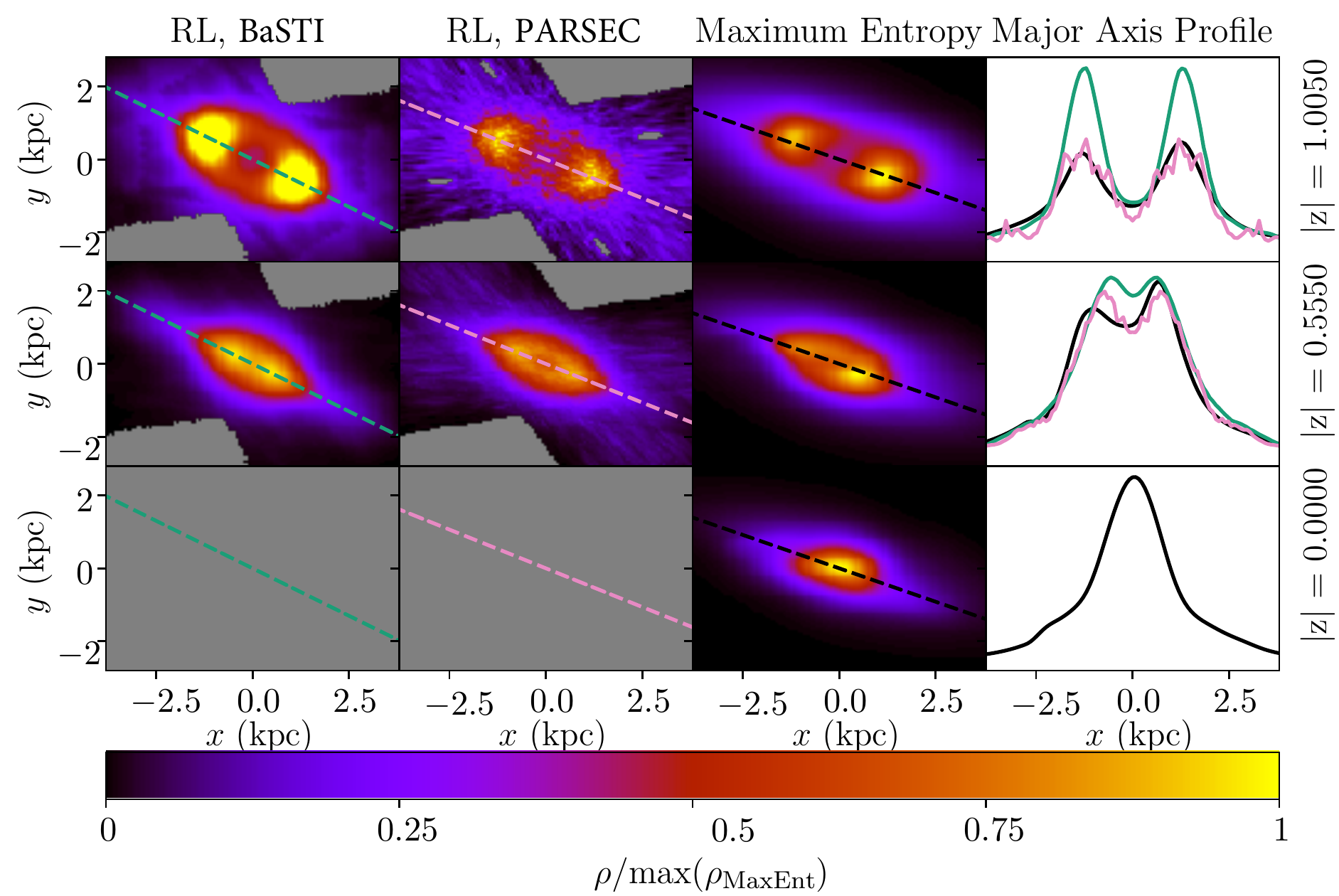}
    \caption{
    Comparison between the modified Richardson-Lucy (RL) deconvolution and maximum entropy deconvolution. The left column implements the same method and resolution as WG13 except on our updated data set. The middle column is constructed in the same way as the left column except that the narrower PARSEC luminosity function is used instead of the BaSTI luminosity function used by WG13.
    Density slices have been normalised to the maximum value in the corresponding maximum entropy slice.
    The green, pink, and black profile plots in the fourth column are along the lines shown in column one, two,  and three respectively. Slices of $|z|$ are measured in kpc.
    }
    \label{fig:comparisontoWG13}
\end{figure}
\begin{figure}
	\centering
	\includegraphics[width=\columnwidth]{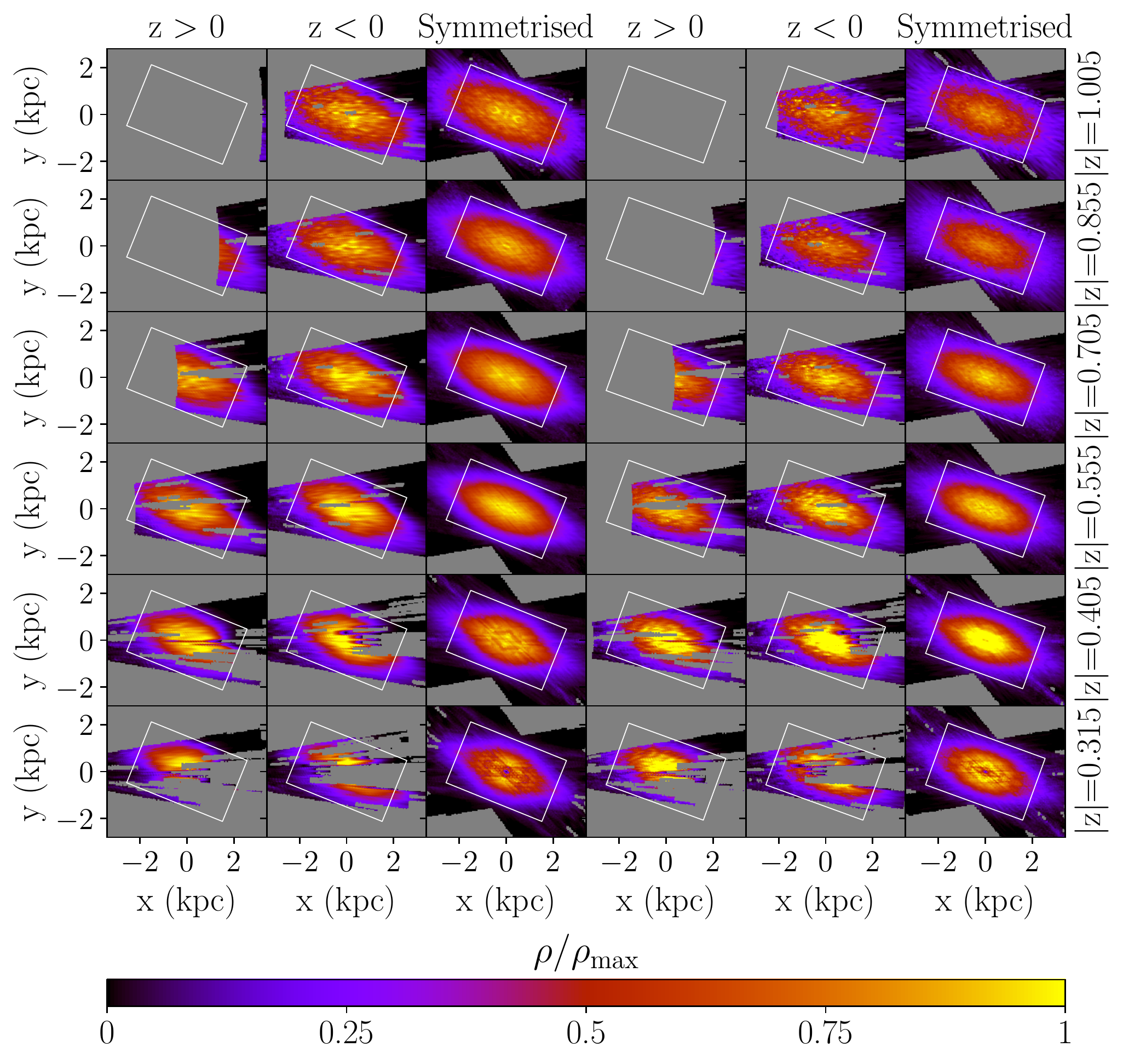}
	\caption{Three dimensional reconstruction of S-model simulations. Columns 1-3 using BaSTI luminosity function and 4-6 using the PARSEC luminosity function. Slices of $|z|$ are measured in kpc.}
	\label{fig:eight-fold symmetrymaximisationdensity2}
\end{figure}

In Fig.~\ref{fig:eight-fold symmetrymaximisationbestangle}, we recovered the relation observed in S17, in which the broader BaSTI luminosity function 
results in a larger bulge angle in comparison to the narrower PARSEC luminosity function.
We note how the shift in $R_0$ for each slice to maximise eight-fold symmetry is nearly flat with a constant shift in the BaSTI cases and a much shallower gradient than found by WG13 in our semi analytic PARSEC luminosity function cases.
 Figures \ref{fig:eight-fold symmetrymaximisationdensity0} and \ref{fig:eight-fold symmetrymaximisationdensity1} show our density deconvolutions on the data using the BaSTI and PARSEC luminosity functions  across the two different resolutions we considered. The region used in the maximisation of eight-fold symmetry,  compatible with WG13, is bounded by a white rectangle.
The X-bulge structure and features seen in WG13, such as the near-far RC density asymmetry, are visibly recovered.
The $K$- and $K_s$-band RC magnitude widths being observed using {\em Gaia\/} DR2 of 0.03-0.09 mag \citep{HallTestingasteroseismologyGaia2019,ChanemphGaiaDR2parallax2019} are consistent with the PARSEC luminosity function which is narrower than the BaSTI luminosity function.

In Fig.~\ref{fig:comparisontoWG13} we show a comparison between the modified Richardson-Lucy deconvolution and our non-parametric method. As can be seen from the profile plot in the right most panel, the modified Richardson-Lucy deconvolution with the BaSTI luminosity function has significantly denser X-arms at high $|z|$. However, this is primarily due to the use of the BaSTI luminosity function rather than the PARSEC  luminosity  function. If the PARSEC luminosity function is used with the modified Richardson-Lucy deconvolution (as in the second column) then the peaks are similar to our non-parametric deconvolution.
But, as can be seen from the second column, of the figure, when the PARSEC luminosity function is used with the modified Richardson-Lucy deconvolution, a much noisier reconstruction is obtained even though the low resolution case is being used. The PARSEC luminosity function has an intrinsic RC dispersion that is more consistent with observations (as mentioned above). It is distinct advantage that  our non-parametric model can give non-noisy reconstructions with the narrower PARSEC  luminosity function at higher resolution.
We checked the method against simulations for the finer resolution to examine possible shortcomings in that regime independently of the actual data.

In Fig.~\ref{fig:eight-fold symmetrymaximisationdensity2} we 
show the results of
the deconvolution and symmetrisation of the simulated data
with our standard $0.2^\circ \times 0.2^\circ$ resolution.
The bulge angle was effectively recovered using $0.5^\circ$ steps in a grid search for the PARSEC luminosity function case and a larger angle using the broader BaSTI luminosity function as 
seen in our earlier results and also
by S17.
The shift in $R_0$ is mostly flat across $z$ slices in both cases with a slight negative gradient in the BaSTI case.
Comparing to the gradient in the data fits, it is not apparent whether or not these comparably shallow gradients are spurious.
The $R_0$ eight-fold symmetric maximisation on the data results in a very flat shift in $R_0$ across $z$ slices between 400 and 800 pc. Above 800 pc the counts are very low at this resolution, causing excessively noisy features and below 400 pc our mask starts interfering substantially with the symmetrisation procedure.
We find a negligible gradient using the broader BaSTI derived luminosity functions.
It is not clear within this method how one might interpret the apparent magnitude-shift gradient depending on the broadness of the luminosity function here and how much of it is an artefact of the symmetrisation, when there is a persistent asymmetry at odds with the assumption of eight-fold symmetry.
Our metallicity distribution systematic in Section \ref{sec:metalicity} for comparison, found unimodal corrections driven by observation were negligible.

\bsp	%
\label{lastpage}
\end{document}